\documentclass[conference]{IEEEtran}

\ifCLASSINFOpdf
\else
\fi
\usepackage{epsfig,endnotes,xspace,url,diagbox}

\usepackage{amsmath, amsthm, amsfonts}
\usepackage{bbm}
\usepackage{enumerate}
\usepackage{varioref}
\usepackage{graphicx}
\usepackage{subfigure}
\usepackage[hidelinks]{hyperref}
\usepackage{pifont}

\usepackage{caption}
\usepackage{float}
\usepackage{cleveref}
\usepackage{thmtools}
\usepackage{thm-restate}

\usepackage{epstopdf}
\usepackage{xcolor}
\usepackage{colortbl} 

\usepackage{multirow}
\usepackage{comment}
\usepackage{color}
\usepackage[utf8]{inputenc}
\usepackage{mathtools}

\usepackage{wrapfig}
\usepackage{tabularx}
\usepackage{xspace}
\usepackage{algorithm}
\usepackage{algorithmic}
\usepackage{cite}
\usepackage{booktabs}
\usepackage{tcolorbox}
\usepackage{enumitem}
\usepackage{url}
\theoremstyle{plain}
\newtheorem{definition}{Definition}

\setlist[itemize]{leftmargin=*,noitemsep, topsep=0pt}

\newcommand{\ie}{\emph{i.e., }}
\newcommand{\eg}{\emph{e.g., }}
\newcommand{\aka}{\emph{aka. }}
\newcommand{\mypara}[1]{\smallskip\noindent\textbf{#1.} \xspace}
\newcommand{\mysubpara}[1]{\noindent\textit{#1.} \xspace}

\newcommand{\algcomment}[1]{\hfill {\color{blue} $\triangleright$ \emph{\small{#1}}}}

\newcommand{\myfullcomment}[1]{\STATE {\textcolor{gray}{\small\textit{\# #1}}}}
\newcommand{\IfThen}[2]{%
  \STATE \textbf{if} #1\ \textbf{then} #2
}

\newcommand{\ElseIfThen}[2]{%
  \STATE \textbf{else if} #1\ \textbf{then} #2
}

\newcommand{\myonline}{\ensuremath{\mathsf{CMIA}}\xspace}
\newcommand{\myoffline}{\ensuremath{\mathsf{PMIA}}\xspace}

\newcommand{\myonlineopt}{\ensuremath{\mathsf{CMIA}}$_\text{opt}$\xspace}

\newcommand{\myonlineloss}{\ensuremath{\mathsf{CMIA}}$_\text{loss}$\xspace}

\newcommand{\myonlinegibbs}{\ensuremath{\mathsf{CMIA}}$_\text{Gibbs}$\xspace}

\newcommand{\revision}[1]{\textcolor{red}{#1}}

\begin{document}
%
\title{Cascading and Proxy Membership Inference Attacks}


\author{
\IEEEauthorblockN{
Yuntao Du\IEEEauthorrefmark{1},
Jiacheng Li\IEEEauthorrefmark{1},
Yuetian Chen\IEEEauthorrefmark{1},
Kaiyuan Zhang\IEEEauthorrefmark{1},
Zhizhen Yuan\IEEEauthorrefmark{1},
\\
Hanshen Xiao\IEEEauthorrefmark{1}\IEEEauthorrefmark{2},
Bruno Ribeiro\IEEEauthorrefmark{1}, and
Ninghui Li\IEEEauthorrefmark{1}
}
\IEEEauthorblockA{\IEEEauthorrefmark{1}Purdue University, \IEEEauthorrefmark{2}NVIDIA Research}
\IEEEauthorblockA{\small \{ytdu, li2829, yuetian, zhan4057, yuan462, hsxiao, ribeirob, ninghui\}@purdue.edu}
}
	

%


\IEEEoverridecommandlockouts
\makeatletter\def\@IEEEpubidpullup{6.5\baselineskip}\makeatother
\IEEEpubid{\parbox{\columnwidth}{
		Network and Distributed System Security (NDSS) Symposium 2026\\
		23 - 27 February 2026 , San Diego, CA, USA\\
		ISBN 979-8-9919276-8-0\\  
		https://dx.doi.org/10.14722/ndss.2026.230661\\
		www.ndss-symposium.org
}
\hspace{\columnsep}\makebox[\columnwidth]{}}

\maketitle

\begin{abstract}
A Membership Inference Attack (MIA) assesses how much a trained machine learning model reveals about its training data by determining whether specific query instances were included in the dataset. 
We classify existing MIAs into adaptive or non-adaptive, depending on whether the adversary is allowed to train shadow models on membership queries.
In the adaptive setting, where the adversary can train shadow models after accessing query instances, we highlight the importance of exploiting membership dependencies between instances and propose an attack-agnostic framework called Cascading Membership Inference Attack (\myonline), which incorporates membership dependencies via conditional shadow training to boost membership inference performance.

In the non-adaptive setting, where the adversary is restricted to training shadow models before obtaining membership queries, we introduce Proxy Membership Inference Attack (\myoffline). 
\myoffline employs a proxy selection strategy that identifies samples with similar behaviors to the query instance and uses their behaviors in shadow models to perform a membership posterior odds test for membership inference.
We provide theoretical analyses for both attacks, and extensive experimental results demonstrate that \myonline and \myoffline substantially outperform existing MIAs in both settings, particularly in the low false-positive regime, which is crucial for evaluating privacy risks\footnote{Our code is available at \href{https://github.com/zealscott/MIA}{https://github.com/zealscott/MIA}.}.

\end{abstract}


%
\IEEEpeerreviewmaketitle

\section{Introduction}
\label{sec:intro}

Machine learning (ML) has advanced rapidly over the past decade, with models such as neural networks increasingly being trained on sensitive datasets. 
This raises a critical need to ensure that these trained models are privacy-preserving.
Membership inference attacks (MIAs)~\cite{sp17miashokri} 
quantify the degree to which a model leaks information by predicting whether some instances were part of its training set. 
Closely connected with Differential Privacy (DP)~\cite{dwork_dp}, MIAs have become a widely adopted approach for empirical privacy auditing of ML models~\cite{arxiv20privacy_meter,tensorflow_mia}, and serve as crucial components for more sophisticated attacks~\cite{usenix21llm_extract,usenix23diffusion_extract}.

Generally speaking, MIAs exploit discrepancies in a target model's behavior between training samples (\ie members) and non-training samples (\ie non-members). 
These discrepancies are often captured through signals such as output losses.  
To effectively leverage these signals for membership inference, a prevalent approach is the \textit{shadow training} technique \cite{sp17miashokri}, which involves training multiple shadow models on datasets drawn from the same distribution as the target model’s training data. 
These shadow models provide insights into how a model’s output on an instance varies depending on whether the instance is included in the training dataset.
Shadow-based MIAs~\cite{iclr23canary,ccs24rapid} have achieved state-of-the-art performance, particularly when evaluated using the increasingly recommended metric~\cite{sp22lira}: True-Positive Rate (TPR) at a low False-Positive Rate (FPR).


We classify MIAs into two categories: adaptive and non-adaptive (see~\Cref{sec:definition} for a formal definition).
In the adaptive setting, which has been extensively studied in prior work~\cite{ sp19comprehensive,iclr23canary,codaspy21mia,sp22lira},
the adversary can train shadow models \textit{after} knowing the query instances to be inferred.
For each query instance, the adversary can train shadow \textit{in} models (trained with the instance) and shadow \textit{out} models (trained without the instance), enabling the computation of a membership score that takes advantage of both shadow in and shadow out models. 
In the non-adaptive setting, which has garnered attention in recent studies~\cite{nips23quantile, ccs24seqmia, icml24rmia, ccs24rapid}, the adversary can only train shadow models \textit{before} learning the query instances. 
As a result, only shadow \textit{out} models are available to compute the membership scores for query instances.

In this paper, we study MIAs in both settings. 
For the adaptive setting, we highlight the importance of exploiting membership dependencies between instances for attack, which is largely overlooked by existing adaptive MIAs~\cite{sp22lira,iclr23canary}. 
Building on the theoretical analysis of joint membership estimation using Gibbs sampling, we introduce Cascading Membership Inference Attack (\myonline), an attack-agnostic framework designed to enhance attack performance through conditional shadow training. 
\myonline operates by iteratively running a base shadow-based attack to identify highly probable samples, then using their inferred membership to train new shadow models to improve the inference of remaining instances. 

In the non-adaptive setting, 
we propose Proxy Membership Inference Attack (\myoffline), which approximates the likelihood that a query instance in the training set using the behaviors of the shadow models' training data as proxies. 
We also introduce several methods for selecting proxy data at different granularities, including global, class, and instance levels. 


We compare the proposed attacks to a wide range of state-of-the-art MIAs and conduct extensive experiments across six benchmark datasets and five model architectures. 
The experimental results demonstrate that \myonline consistently improves attack performance across all evaluated attack algorithms and datasets, including those previously considered difficult to attack. 
For instance, \myonline improves LiRA~\cite{sp22lira} by more than $\mathbf{5\times}$ in the true-positive rate at a 0.001\% false-positive rate on MNIST, showing a significant performance boost.
Additionally, \myoffline significantly outperforms all existing MIAs in the non-adaptive setting while maintaining high efficiency.
We also conduct comprehensive ablation studies to evaluate the impact of various components in \myonline and \myoffline and assess the effectiveness of existing defenses against the proposed attacks. 
In summary, we make the following contributions:

\begin{itemize}
    \item We provide a new formulation of the MIA game in~\Cref{sec:definition}, 
    which allows the clear differentiation of the adaptive and non-adaptive settings for MIA.
    \item In the adaptive setting, we highlight the importance of modeling membership dependencies and present a theoretical analysis of joint membership estimation using approximate Gibbs sampling. 
    We also introduce \myonline, an attack-agnostic framework that enhances attack performance via conditional shadow training.
    \item In the non-adaptive setting,  we introduce \myoffline, a new attack that approximates the membership posterior odds ratio test using proxy data.
    \item We conduct extensive experiments and demonstrate that \myonline and \myoffline consistently outperform all state-of-the-art MIAs in various attack scenarios.
\end{itemize}

\mypara{Organization}
The rest of this paper is organized as follows. 
~\Cref{sec:definition} introduces the definitions and threat models for membership inference attacks. 
We then describe the proposed adaptive attack (\ie \myonline) in~\Cref{sec:cmia} and the non-adaptive attack (\ie \myoffline) in~\Cref{sec:pmia}. ~\Cref{sec:experiments} presents the experimental results for both attacks. Related work is discussed in ~\Cref{sec:related}, and the paper concludes in ~\Cref{sec:conclusion}. 

\section{Problem Definition and Threat Models}
\label{sec:definition}

The goal of a membership inference attack (MIA)~\cite{sp17miashokri} is to determine whether some instances were included in the training data of a given trained model $f_\theta$.
In this paper, we consider the model to be a neural network classifier $f_\theta: \mathcal{X} \to \Delta^m$, where $f_\theta$ is a learned function that maps an input data sample $x \in \mathcal{X}$ to a probability distribution over $m$ classes, where $\Delta^m$ denotes the $m$-dimensional simplex.
Given a dataset $D$, we use $f_\theta \gets \mathcal{T}(D)$ to denote that the neural network $f_\theta$, parameterized by weights $\theta$, is learned by applying the training algorithm $\mathcal{T}$ on the training set $D$.

In the literature~\cite{csf18privacy,arixv20revisiting,sp22lira}, membership inference has been defined via a security game in which the adversary is asked to determine the membership of a single instance.  
However, there exists a disconnection between this definition and the experiments, which train shadow models and determine memberships for a set of query instances (referred to as the membership query set).  
More importantly, using such a single-instance game definition for MIA, it is difficult to clearly differentiate adaptive versus non-adaptive MIA settings, as will be detailed later.
To address these challenges, we define membership inference through the following security game:

\begin{definition}[Membership Inference Security Game]
\label{def:game} 
The following game is between a challenger and an adversary that both have access to a data distribution $\mathbb{D}$: 
\begin{enumerate}[leftmargin=*]
    \item The challenger samples a training dataset $D \sim \mathbb{D}$, trains a target model $f_\theta \gets \mathcal{T}(D)$ on the dataset $D$, and grants the adversary query access to the model $f_\theta$.

    \item The challenger randomly selects a subset $D_\mathrm{a} \subseteq D$ and samples a set $D_\mathrm{b}$ from $\mathbb{D}$. These two sets are combined to create a query set: $D_\mathrm{query} = D_\mathrm{a} \cup D_\mathrm{b}$, which the challenger then sends to the adversary.

    
    \item The adversary responds with a set $D_\mathrm{g} \subseteq D_\mathrm{query}$, which represents that the adversary guesses that instances in $D_\mathrm{g}$ are used when training $f_\theta$, and instances in $D_\mathrm{query}  \setminus D_\mathrm{g}$ are not used when training $f_\theta$.

\end{enumerate}
\end{definition}

In this paper, we focus on membership inference attacks in black-box scenarios, where the adversary is granted oracle access to the target model $f_\theta$, but is not given the model parameters.  
That is, the adversary can obtain the output softmax probabilities for any input instance $x$.



State-of-the-art MIAs ~\cite{sp22lira,iclr23canary,ccs24rapid,icml24rmia} leverage shadow models~\cite{sp17miashokri} to analyze how the model's outputs depend on whether specific instances are used in training or not.
We assume the adversary constructs a dataset and samples subsets from it to train the shadow models.
The above game allows the adversary to access the data distribution $\mathbb{D}$, which they can sample when constructing the dataset.
Depending on when the shadow models are trained, we consider two threat models.

\mypara{Adaptive Setting}
In this setting, the adversary is allowed to train shadow models \textit{after} receiving the query set $D_\text{query}$ (\ie after step 2 in~\Cref{def:game}). 
Thus, the adversary samples $D^\prime$ from $\mathbb{D}$ and combines it with $D_\text{query}$ to create a dataset for shadow training, \ie $D_\text{adv}^{\text{adapt}} \gets D^\prime\cup D_\text{query}$. 
This enables the adversary to train shadow \textit{in} models and shadow \textit{out} models for each query instance, and exploit the behavioral discrepancies to mount an attack.

This setting models several attack situations. 
One situation is that the adversary is willing to spend substantial computation to train shadow models for specific query instances~\cite{sp22lira,iclr23canary}.
Another situation is that the target model's training set is drawn from a dataset that the adversary also possesses, and the adversary tries to learn which specific instances are used~\cite{sp19comprehensive,codaspy21mia}. 
MIAs in the adaptive setting are also advocated for empirical privacy auditing~\cite{arxiv20privacy_meter,tensorflow_mia} of ML models, where the goal is to assess the extent of worst-case privacy leakage.

\mypara{Non-Adaptive Setting}
In contrast, under the non-adaptive setting, the adversary is only allowed to train shadow models \textit{before} the adversary learns the query set $D_\text{query}$ (\ie before step 2~\Cref{def:game}).  
Thus, the adversary constructs $D_\text{adv}^\text{non-adapt}$ only by sampling from $\mathbb{D}$, \ie $D_\text{adv}^\text{non-adapt} \sim \mathbb{D}$.
For most practical classification tasks, the sizes of $D_\text{adv}^\text{non-adapt}$ and $D_\text{query}$ are relatively small compared to the entire data distribution $\mathbb{D}$, meaning the probability of each instance belonging to both datasets is quite low.
As a result, the adversary can only observe a model's behavior when the query instance is not part of the training set, thus having access only to shadow \textit{out} models but not shadow \textit{in} models.

This setting models the situation where the adversary is given a sequence of membership queries and needs to answer them without paying the cost of retraining shadow models for each query.    
Recent studies~\cite{nips23quantile, ccs24seqmia, icml24rmia, ccs24rapid} focus on this setting, as it presents a more efficient attack scenario. 




\mypara{Discussion}
It is worth noting that some studies~\cite{sp22lira, ccs24rapid, icml24rmia} refer to the adaptive and non-adaptive settings as ``online'' and ``offline'', respectively.
This terminology may lead to confusion, as the offline setting models queries being processed without training new shadow models, which is similar in spirit to ``online algorithms''~\cite{albers2003online}.
Thus, we adopt the terms ``adaptive'' and ``non-adaptive'', which are more aligned with well-established concepts in the security and privacy domain (\eg the (adaptive) chosen-ciphertext attacks~\cite{luby96pseudorandomness, 98adaptive}).
We use these terms throughout this paper for clarity.

\mypara{Differences from Previous MIA Game}
Notably, our membership inference security game, as defined in~\Cref{def:game}, goes beyond existing studies~\cite{sp22lira,csf18privacy,arixv20revisiting} by using a query set rather than a single instance. This modification offers a few advantages. 
Firstly, it more accurately reflects the experimental methodologies employed in existing research, where performance evaluations (\eg TPR at low FPR) are typically conducted using a query set, rather than isolating assessments to each instance. 
Secondly, this query-set-based approach enables a clear distinction between adaptive and non-adaptive settings, achieved by controlling the timing of shadow model training relative to the receipt of the query set.

\mypara{Missed Opportunities of Existing MIAs}
Our paper is motivated by the observations that existing MIAs \revision{do} not fully take advantage of the available information in both adaptive and non-adaptive settings. 

\mysubpara{$\bullet$  Membership Dependencies in the Adaptive Setting}
Existing adaptive MIAs~\cite{sp22lira,iclr23canary,sp19comprehensive} typically predict the membership of each query instance \textit{independently}. They thus fail to take advantage 
of the conditional membership dependencies among instances, leading to suboptimal performance.

\mysubpara{$\bullet$  Proxy Shadowing in the Non-Adaptive Setting}
In the non-adaptive setting, the adversary can construct only shadow out models, and lacks the knowledge of models trained with a specific query instance behave on that instance.  
However, we observe that it is possible to find instances that are similar to the query instances and observe their behavior.


\section{Cascading Membership Inference Attack}
\label{sec:cmia}

In this section, we describe the methodology of Cascading Membership Inference Attacks (\myonline), which is an attack-agnostic framework designed for the adaptive setting. 
We first present a novel attack paradigm, denoted {\em Joint MIA}, which seeks to {\em jointly} estimate the membership of {\em all} query instances. 
We then provide a detailed description of \myonline. 

\begin{figure}
    \centering
    \includegraphics[width=0.6\linewidth]{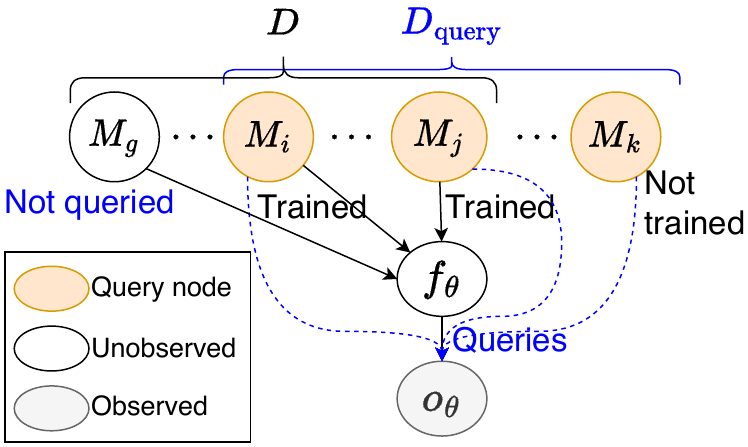}
    \caption{Statistical dependencies of joint MIA shows that
conditioning on output $o_\theta$ creates a collider dependency between the membership indicators $M_i$ and $M_j$, \ie $M_i \not \!\!\bot\!\!\!\bot M_j \mid o_\theta$.}
    \vspace{-4mm}
    \label{fig:DAG}
\end{figure}

\subsection{Theoretical Intuition of Joint MIA}
\label{sec:adaptive_theory}


We assume that instances in the size-$n$ query set $D_\text{query}$ are ordered in a canonical order, \ie $D_\text{query}=\left\{(x_i,y_i) \mid 1\le i\le n \right\}$.  
Given target model $f_\theta$ trained using dataset $D$, we define the (unobserved) vector of membership variables as $\mathbf{M} = (M_1, M_2, \ldots, M_n)$, where each $M_i$ is a Bernoulli variable, $M_i := \mathbbm{1}[(x_i, y_i) \in D]$. 
We use $o_\theta=\{(x_i,f_\theta(x_i))|\forall (x_i,y_i) \in D_\text{query}\}$ to denote the output softmax probabilities that the adversary can obtain from $D_\text{query}$.

Existing MIAs~\cite{sp22lira,iclr23canary,icml24rmia} typically treat each query instance $(x_i,y_i)$ in isolation, computing individual membership probabilities $\Pr(M_i = 1 | o_\theta)$ independently. At first glance, this approach appears reasonable, as the data instances in the training set $D$ could be assumed to be independently and identically distributed (i.i.d.), implying that the membership indicators $M_i$ and $M_j$ are marginally independent, \ie $M_i \bot\!\!\!\bot M_j$, for all $i \neq j$. However, this independence assumption is violated when conditioning on the model's output $o_\theta$ over the entire query dataset $D_\text{query}$. In this conditional setting, the membership indicators, although independent marginally, become dependent due to the collider effect induced by $o_\theta$, as illustrated in \Cref{fig:DAG}. Specifically, conditioning on $o_\theta$ renders $M_i$ and $M_j$ conditionally dependent, i.e., $M_i \not \!\!\bot\!\!\!\bot M_j \mid o_\theta$.

The practical significance of this dependence becomes evident in the adaptive setting, where the adversary can train shadow models after knowing $D_\text{query}$, which overlaps with the target's training dataset $D$.
Existing adaptive MIAs~\cite{sp17miashokri, sp19comprehensive,codaspy21mia, sp22lira, iclr23canary} \textbf{do not explore this conditional joint dependence}.
By leveraging the joint membership distributions rather than relying on marginal probabilities, an adversary can mount more effective membership inference attacks. We first formulate the definition of Joint MIA via Gibbs sampling.



\mypara{Joint MIA with Gibbs Sampling (accurate but prohibitively expensive)}
Sampling from the joint distribution $\Pr(\mathbf{M} | o_\theta)$ requires capturing the dependencies between the membership indicators introduced by conditioning on $o_\theta$.
%
The traditional Gibbs sampling procedure updates each variable sequentially~\cite{gelfand1990sampling}, with each update based on the conditional distribution of the variable given the current states of all other variables. This iterative process can be expressed formally at any iteration $t \geq 0$ as:
\begin{equation}
\label{eq:Gibbs}
M_i^{(t+1)} \sim \Pr(M_i \mid \mathbf{M}_{-i}^{(t+1,t)}, o_\theta), \: \forall i,
\end{equation}
where $\mathbf{M}_{-i}^{(t+1,t)} = M_1^{(t+1)}, \ldots, M_{i-1}^{(t+1)}, M_{i+1}^{(t)}, \ldots, M_n^{(t)}$.

In~\Cref{appendix:converge_gibbs}, we provide a theorem to show that it is possible to perform this joint sampling and converge to the true target stationary distribution $\Pr(\mathbf{M}|o_\theta)$, as well as the almost sure convergence of any metric of success of the attack.


However, using Gibbs sampling for Joint MIA is computationally prohibitive, as convergence usually requires many iterations. 
The challenge now lies in devising a practical and effective method for performing a joint membership inference attack, which benefits from jointly inferring all instances in $D_\text{query}$ without paying the high cost of full Gibbs sampling.


\subsection{\myonline: A Fast Approximation of Joint MIA}
\label{sec:jMIA}

We now introduce \myonline, a heuristic that significantly speeds up the Gibbs sampling procedure by approximating it and performing just a single joint sampling step.

Notably, an important hyperparameter of the Gibbs sampling process is the ordering in which variables in $D_\text{query}$ are sampled, referred to as the {\em scan order}. The ordering in~\Cref{eq:Gibbs} is known as the systematic scan (also known as deterministic or sequential scan). 
Interestingly, this ordering can be modified, with significant implications for the convergence rate of Gibbs sampling~\cite{roberts2015surprising,he2016scan}. The most common strategy is actually random ordering, which shuffles $D_\text{query}$ after each iteration.  In adaptive Gibbs sampling~\cite{latuszynski2013adaptive}, the scan order is adaptively determined. 

Our approach (\myonline) also uses adaptive ordering, but we focus exclusively on identifying a single, (highly likely) joint membership sample. 
This allows for optimizations that would be incorrect or inefficient in standard Gibbs sampling contexts, where the goal is to explore the entire distribution, including less likely observations.
More specifically, a key innovation in \myonline is the dynamic reordering of the instances $(x_i, y_i)$ within $D_\text{query}$ to \textit{prioritize} those with higher membership probabilities (instances we will call them {\bf anchors}). 
For the initial sample, the instances are rearranged to satisfy $\Pr(M_1 = 1 | M_{-1}^{(0,0)}, o_\theta) \geq \Pr(M_{i} = 1 | M_{-i}^{(0,0)}, o_\theta), \forall i$, with ties resolved arbitrarily (in practice, ties and near-ties are sampled jointly as shown in 
\Cref{sec:cmia_framework}).
To illustrate the effect of this reordering, consider a scenario where $\Pr(M_1 = 1 | M_{-1}^{(0,0)}, o_\theta) \approx 1$, indicating near certainty that $(x_1, y_1)$ is a member of $D$. In this context, the reordered $D_\text{query}$ sets the stage for a cascading effect, where subsequent sampling decisions are informed by this initial high confidence. By leveraging this certainty, the process streamlines the exploration of the sample space, focusing on the most promising regions first.


The sampling process proceeds iteratively, with each step refining the ordering of the remaining instances to prioritize those with the highest membership probabilities. Specifically, for the $i$-th sample, the instances are reordered such that $\Pr(M_i = 1 | M_{-i}^{(1,0)}, o_\theta) \geq \Pr(M_j = 1 | M_{-j}^{(1,0)}, o_\theta)$ for $i < j$, where $i=2,\ldots,n$ denotes the current iteration. Ties are again resolved arbitrarily.
At each iteration, the process checks if the membership probability for the next instance exceeds a predetermined threshold, 
in which case the instance is deemed a member, and its membership status is set to $M_i = 1$. Otherwise, the process terminates, and the remaining instances are deemed non-members.

This greedy approach identifies a subset of highly probable members while avoiding unnecessary computations for less likely candidates.
As such, it can be viewed as a computationally efficient heuristic for Maximum a Posteriori (MAP) estimation, which aims to find the single most likely point from the distribution.
Empirically, as we see in~\Cref{sec:exp_online}, joint MIA achieves significantly higher attack performance compared to independent inference of prior work.


\subsection{Implementation Details of \myonline}
\label{sec:cmia_framework}

In this section, we describe the implementation details of \myonline.
\myonline approximates the sampling procedure outlined in the previous section, and estimates conditional membership probabilities by performing a base shadow-based MIA using conditional shadow models.
Specifically, in each iteration, \myonline identifies multiple highly probable instances (\ie anchors), by conducting membership inference using a base shadow-based MIA.
It then utilizes their inferred membership to generate conditional shadow models, which enhance the inference of the remaining instances.
This process can be repeated until no additional anchors can be reliably identified.
We begin by providing a definition for shadow-based MIA, which will serve as the base attack in \myonline.

\mypara{Shadow-based MIAs}
Shadow-based MIAs~\cite{sp17miashokri,sp22lira} train shadow models to imitate the target model's behavior. 
Formally, shadow-based MIAs are defined as follows:
\begin{definition} [Shadow-based MIA]
\label{def:shadowMIA}
Let $D_\mathrm{adv}^\mathrm{adapt}$ be the adversary's dataset in the adaptive setting.
A shadow-based MIA constructs multiple shadow dataset-model pairs $\mathcal{P}_\mathrm{shadow} = \{(D_\mathrm{shadow}^j, f_\mathrm{shadow}^j)\}$, where each pair consists of a shadow dataset $D_\mathrm{shadow}^j$ sampled from $D_\mathrm{adv}^\mathrm{adapt}$ and the corresponding shadow model $f_\mathrm{shadow}^j$ trained on the dataset, i.e., $f_\mathrm{shadow}^j \gets \mathcal{T}(D^j_\mathrm{shadow})$.
The attack model $\mathcal{M}$ (e.g., LiRA~\cite{sp22lira}) then utilizes these shadow dataset-model pairs to generate a continuous membership score to predict the membership of instance $(x_i, y_i)$ for the target model $f_\theta$:
\begin{equation*}
    s(x_i,y_i) = \mathcal{M}\left(f_\theta, (x_i, y_i\right),  \mathcal{P}_\mathrm{shadow}).
\end{equation*}
\end{definition}



Note that different MIAs may employ different training algorithms $\mathcal{T}$ to train shadow models.
In \myonline, we use these shadow-based MIAs by only modifying the input shadow dataset, while both the training algorithm $\mathcal{T}$ and the attack model $\mathcal{M}$ remain unchanged.

\begin{figure}[t]
    \centering
    \includegraphics[width=0.499\textwidth]{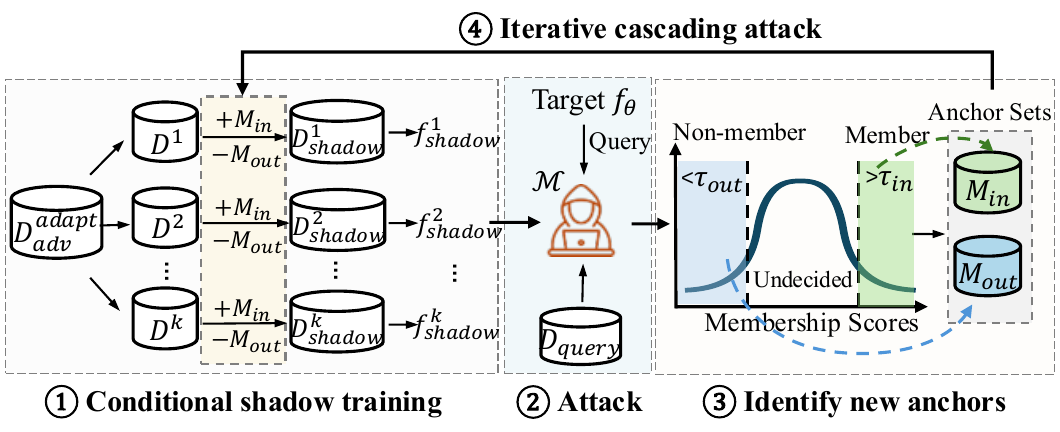}
    \vspace{-5mm}
    \caption{Demonstration of \myonline. The adversary \ding{172} constructs conditional shadow datasets by sampling from $D_\text{adv}^{\text{adapt}}$ and incorporating the membership of anchors, \ding{173} performs the base attack $\mathcal{M}$, \ding{174} uses the computed membership scores to identify new anchors ($M_\text{in}/M_\text{out}$), \ding{175} repeats the above processes to enhance the inference of the remaining instances. }
    \label{fig:cmia_framework}
    \vspace{-4mm}
\end{figure}

\mypara{Framework Overview}
The pipeline of \myonline is illustrated in~\Cref{fig:cmia_framework} and outlined in~\Cref{alg:cmia}. 
First, the maximum number of cascading iterations $K$ is set. 
In each iteration $k$, the adversary constructs shadow datasets by sampling from $D_\text{adv}^{\text{adapt}}$ (defined in~\Cref{sec:definition}) and incorporating with anchor sets ($M_\text{in}/M_\text{out}$).
Specifically, samples from $M^{k-1}_\text{in}$ are included while those from $M^{k-1}_\text{out}$ are excluded from the sampled set to construct the conditional shadow dataset (line 7). 
This procedure repeats $N$ times to generate $N$ conditioned shadow dataset-model pairs.
Then, the base attack $\mathcal{M}$ is executed using these dataset-model pairs to compute membership scores for all instances in $D_\text{query}$ (lines 11-15). 
New anchors are selected based on their membership scores relative to learned decision thresholds (lines 19–23, detailed below).
The iterations stop when either the maximum number of iterations is reached or the number of new anchors falls below $\delta$ (line 24). 
Finally, shadow datasets/models from all iterations are used to perform the final attack on all query instances (lines 29-33).

\mypara{Thresholds Selection}
Anchors (\ie highly probable instances) are selected by comparing membership scores against two decision thresholds: $\tau_\text{in}$ for members and $\tau_\text{out}$ for non-members.
To determine these thresholds, for each iteration, we perform the attack on a ground-truthed shadow model and analyze the relationship between membership scores and actual membership.
Specifically, we randomly select a shadow model and treat it as the target, then apply the base attack on this model to compute membership scores for instances in $D_\text{adv}^\text{adapt}$.
Since the adversary knows the ground truth about the members (\ie the training set) of the selected shadow model, we set the decision thresholds by ordering the membership scores and setting: 
(i) $\tau_\text{in}$ as the highest score among non-members, and
(ii) $\tau_\text{out}$ as the 10th lowest score among members.
In other words, $\tau_\text{in}$ is chosen to avoid false positives when predicting members, while $\tau_\text{out}$ is selected to tolerate less than 10 false negatives when predicting non-members.
This distinction is made because the adversary typically focuses on reliably identifying members rather than non-members~\cite{sp22lira}.
We analyze the impact of thresholds in~\Cref{sec:exp_ablation}.

\mypara{Implementation Details} 
We set the number of the cascading iterations $K=10$ for all experiments, as we observe that all evaluated attacks exhibit significant improvements within only a few iterations (as shown in~\Cref{sec:exp_online}).
The stopping criterion $\delta$ is set to $30$, meaning that the iteration halts if fewer than $30$ new anchors can be selected.
Although the threshold selection procedure can be repeated by using different ground-truthed shadow models to obtain an average threshold estimate, we find that it remains robust even with a single randomly chosen shadow model.
All above hyperparameters are kept fixed across all evaluated attacks and datasets to demonstrate the robustness of the framework.

\begin{algorithm}[t]
\caption{\textbf{Cascading Membership Inference Attack.} 
}
\label{alg:cmia}
\begin{algorithmic}[1]
\REQUIRE Target model $f_\theta$, adversary's dataset $D_\text{adv}^\text{adapt}$, training algorithm $\mathcal{T}$, base MIA algorithm $\mathcal{M}$, membership query set $D_\text{query}$, number of iterations $K$, stopping criterion $\delta$
\STATE $M^{0}_{\text{in}} \gets \{\},\ M^{0}_{\text{out}} \gets \{\}$ \algcomment{initialize anchor sets}
\FOR{$k = 1$ \textbf{to} $K$} 
    \myfullcomment{Step 1: Train conditional shadow models}
    \STATE $\mathcal{P}^k_{\text{shadow}} \gets \{\}$ \algcomment{initialize shadow dataset-model set}
    \FOR{$N$ \text{times}}
        \STATE $D_{\text{tmp}} \sim D_\text{adv}^\text{adapt}$ 
        \algcomment{sample a dataset}
        \STATE $D_{\text{shadow}} \gets (D_\text{tmp} \setminus M^{k-1}_{\text{out}}) \cup M^{k-1}_{\text{in}}$ 
        \STATE $\mathcal{P}^k_{\text{shadow}} \gets \mathcal{P}^k_{\text{shadow}} \cup \{(D_{\text{shadow}}, \mathcal{T}(D_{\text{shadow}}))\}$
    \ENDFOR
    \myfullcomment{Step 2: Attack with conditional shadow models}
    \STATE $\mathcal{S}^k \gets \{\}$  \algcomment{initialize membership scores set}
    \FOR{each $(x, y) \in D_\text{query}$} 
        \myfullcomment{compute membership score, defined in~\Cref{def:shadowMIA}}
        \STATE $\mathcal{S}^k \gets \mathcal{S}^k\cup \{\mathcal{M}\left(f_\theta, (x, y\right),  \mathcal{P}^k_\text{shadow})\}$
    \ENDFOR
    \myfullcomment{Step 3: Identify anchor samples using membership scores}
    \STATE $M^k_{\text{in}} = M^{k-1}_{\text{in}},\ M^k_{\text{out}} = M^{k-1}_{\text{out}}$
    \myfullcomment{thresholds selection is detailed in Section~\ref{sec:cmia_framework}}
    \STATE $\tau^k_{\text{in}}, \tau^k_{\text{out}} \gets \texttt{SelectThresholds}(\mathcal{M}, \mathcal{P}^k_\text{shadow},D_\text{adv}^\text{adapt})$
    \FOR{each $s^k(x, y) \in \mathcal{S}^k$}
        \IfThen{$s^k(x, y) > \tau^k_{\text{in}}$}{$M^k_{\text{in}} \gets M^k_{\text{in}} \cup \{(x,y)\}$}
        \ElseIfThen{$s^k(x, y) < \tau^k_{\text{out}}$}{$M^k_{\text{out}} \gets M^k_{\text{out}} \cup \{(x,y)\}$}
    \ENDFOR
    \IF{$|M^{t}_{\text{in}}| - |M^{k-1}_{\text{in}}| < \delta \textbf{ and } |M^{t}_{\text{out}}| - |M^{k-1}_{\text{out}}| < \delta$}
        \STATE \textbf{break}
    \ENDIF
\ENDFOR
\myfullcomment{Perform final attack using all shadow dataset-model pairs}
\STATE $\mathcal{P}_\text{shadow} \gets \bigcup_{k} \mathcal{P}^k_\text{shadow},\ \mathcal{S} \gets \{\}$ 
\FOR{each $(x, y) \in D_\text{query}$} 
    \STATE $\mathcal{S} \gets \mathcal{S}\cup \{\mathcal{M}\left(f_\theta, (x, y\right),  \mathcal{P}_\text{shadow})\}$
\ENDFOR
\RETURN $\mathcal{S}$ 
\end{algorithmic}
\end{algorithm}

\section{Proxy Membership Inference Attack}
\label{sec:pmia}

While \myonline shows impressive performance, it requires 
knowledge of query instances, and cannot be applied in the non‑adaptive setting. 
In this section, we propose Proxy Membership Inference Attack (\myoffline), a new non-adaptive MIA that can respond to arbitrary membership queries without training additional shadow models.
We begin by presenting the theoretical intuition behind \myoffline and then describe the attack procedure in detail.

\subsection{Theoretical Intuition of Marginal MIA}
\label{sec:nonadaptive_theory}

In the non-adaptive setting, the adversary is given a sequence of membership queries and needs to answer them \textit{independently}, \ie for instance $(x_i,y_i) \in D_\text{query}$ we predict its membership status: $M_i | e_\theta$, where $e_\theta \gets \mathcal{Q}(f_\theta) $ denotes the information the adversary can observe from the model $f_\theta$ with the available data. 
We refer to this type of attack as \textbf{Marginal MIA}, as our predictions are only based on $\Pr(M_i| e_\theta)$ regardless of the membership status of other instances in $D_\text{query}\setminus \{(x_i,y_i)\}$.
We follow~\cite{icml19white} and frame the marginal MIA as a Bayesian binary hypothesis testing problem.
In the black-box scenario we study, $e_\theta$ is the output of $f_\theta$.
We now show how to calculate the membership posterior odds ratio test for marginal MIA.

\begin{restatable}[A Membership Posterior Odds Test for Marginal MIA]{theorem}{marginalMIA}
\label{thm:marginalMIA}
Let $\mathcal{S}^+_{(x_i,y_i)}$ and $\mathcal{S}^-_{(x_i,y_i)}$ be the sets of all subsets in the support domain $\mathbb{D}$ that include or exclude the query instance $(x_i,y_i)$, respectively.
Let $M_i = \mathbbm{1}[(x_i,y_i) \in D]$ and $\mathcal{L}(D,e_\theta)= \Pr(\mathcal{Q}(\mathcal{T}(D))=e_\theta)$ denote the likelihood function, where $\mathcal{T}(D)$ represents the model is trained with dataset $D$.
Given the adversary's observation $e_\theta$, the Bayesian posterior odds test for marginal MIA is defined as:
\begin{equation*}
    \mathcal{A}_\mathrm{odds}(x_i,y_i) = \mathbbm{1}\left[\frac{\Pr\left(M_i = 1  \mid \mathcal{Q}(f_\theta)=e_\theta \right)}{\Pr\left(M_i = 0  \mid \mathcal{Q}(f_\theta)=e_\theta \right)} > 1 \right],
\end{equation*}
which can be obtained from
\begin{equation} 
\label{equ:marginal_mia_optimal}
\!\!    \mathcal{A}_\mathrm{odds}(x_i,y_i)\!\! =\!\! \mathbbm{1}
    \!\left[\frac{\mathbb{E}_{D' \sim \mathcal{S}^+_{(x_i,y_i)}}\!\!\!\mathcal{L}(D',e_\theta)}{\mathbb{E}_{D' \sim \mathcal{S}^-_{(x_i,y_i)}}\!\!\!\mathcal{L}(D',e_\theta)} > \frac{\Pr(M_i = 0)}{\Pr(M_i = 1)} \right]\!\!.
\end{equation}
\end{restatable}
A detailed proof based on Bayes' rule is provided in~\Cref{appendix:proof_optimal}.
Note that the expectations in the formula are taken with respect to datasets sampled uniformly at random from $\mathcal{S}^+_{(x_i,y_i)}$ and $\mathcal{S}^-_{(x_i,y_i)}$, which are the countable collections of all possible datasets that include or exclude the query instance $(x,y)$.
The theorem establishes that the membership posterior odds test $\mathcal{A}_\mathrm{odds}$ for an adversary involves comparing a special likelihood ratio to a threshold derived from prior probabilities.

\mypara{Connecting with Evaluation Metrics}
For most practical classifiers, the training set $D$ is relative small compared to the entire data distribution, so the prior probability $\Pr(M_i = 1)$ is low for most data points $(x_i, y_i)$, making the threshold $\frac{\Pr(M_i = 0)}{\Pr(M_i = 1)}$ fairly large. 
Thus, the adversary must have a significantly high likelihood ratio to infer membership correctly. 
This statistical analysis helps explain why recent studies~\cite{sp22lira, icml24rmia} emphasize evaluating MIAs on the FPR at low FPR.

\mypara{Connecting with SOTA MIAs}
We observe that most state-of-the-art MIAs, e.g.,~\cite{sp22lira,icml24rmia}, adhere to the principles of the posterior odds ratio test outlined in~\Cref{equ:marginal_mia_optimal}. 
For instance, LiRA~\cite{sp22lira} estimates the likelihood ratio for a query instance by utilizing the (scaled) losses of its shadow \textit{in} and shadow \textit{out} models.
However, in the non-adaptive setting, computing the likelihood ratio becomes challenging because the adversary only has access to the shadow \textit{out} models for query instances. 
As a result, LiRA adopts a one-sided hypothesis test and estimates only the likelihood that the instance is not in the training set.
This deviation leads to suboptimal performance, as evidenced by recent studies~\cite{ccs24seqmia, ccs24rapid}.

\mypara{\myoffline: Better Approximating the Posterior Odds Test}
Motivated by these insights, we propose a new attack, \myoffline, which better aligns with the membership posterior odds test presented above while avoiding the need to train additional shadow models for queries. 
The key idea is to approximate the likelihood that a query instance is in the training set using the behaviors of the shadow models' training data as proxies. 
We also explore various proxy-finding strategies at different levels of granularity to further improve the attack's performance.



\begin{algorithm}[t]
\caption{\textbf{Proxy Membership Inference Attack.}
}
\label{alg:pmia}
\begin{algorithmic}[1]
\REQUIRE Target model $f_\theta$, adversary's dataset $D_\text{adv}^{\text{non-adapt}}$, training algorithm $\mathcal{T}$, query instance $(x, y)\in D_\text{query}$
\STATE \textcolor{gray}{\texttt{\# Prepare Phase: Shadow Model Training}}
\STATE $\mathcal{D}_{\text{shadow}} \gets \{\},\ \mathcal{F}_{\text{shadow}} \gets \{\}$
\FOR{$N$ \text{times}}
    \STATE $D_{\text{shadow}} \sim D_\text{adv}^{\text{non-adapt}}$ \algcomment{sample a shadow dataset}
    \STATE $\mathcal{D}_{\text{shadow}} \gets \mathcal{D}_{\text{shadow}} \cup \{D_{\text{shadow}}\}$
    \STATE $\mathcal{F}_\text{shadow} \gets \mathcal{F}_\text{shadow} \cup \{\mathcal{T}(D_{\text{shadow}})\}$
\ENDFOR

\vspace{0.5em}
\STATE \textcolor{gray}{\texttt{\# Inference Phase: Query on $(x, y)$}}
\STATE $\text{confs}_{\text{in}} \gets \{\},\ \text{confs}_{\text{out}} \gets \{\}$
\myfullcomment{collect out confidence scores}
\FOR{each $f_\text{shadow} \in \mathcal{F}_{\text{shadow}}$}
    \STATE $\text{confs}_{\text{out}} \gets \text{confs}_{\text{out}} \cup \{\phi(f_\text{shadow}(x)_y)\}$
\ENDFOR
\myfullcomment{collect in confidence scores via proxies, see~\Cref{sec:pmia_framework}}
\STATE $D_\text{proxy} \gets \texttt{FindProxy}(D_{\text{adv}}^{\text{non-adapt}}, (x, y))$ \algcomment{find a proxy set}

\FOR{each $D^{i}_\text{shadow} \in \mathcal{D}_\text{shadow}$}
    \FOR{each $(u,v) \in D_\text{proxy}$}
        \IF{$(u,v) \in D^{i}_\text{shadow}$}
            \STATE $\text{confs}_{\text{in}} \gets \text{confs}_{\text{in}} \cup \{\phi(f^{i}_\text{shadow}(u)_v)\}$
        \ENDIF
    \ENDFOR
\ENDFOR

\STATE Compute mean $\tilde{\mu}_{\text{in}}$, and variance $\tilde{\sigma}_{\text{in}}^2$ from $\text{confs}_{\text{in}}$
\STATE Compute mean $\mu_{\text{out}}$, and variance $\sigma_{\text{out}}^2$ from $\text{confs}_{\text{out}}$

\STATE $\text{conf}_{\text{obs}} \gets \phi(f_\theta(x)_y)$ \algcomment{query target model}
\STATE \textbf{return} $\displaystyle \tilde{\Lambda} = \frac{p(\text{conf}_{\text{obs}} \mid \mathcal{N}(\tilde{\mu}_{\text{in}}, \tilde{\sigma}^2_{\text{in}}))}{p(\text{conf}_{\text{obs}} \mid \mathcal{N}(\mu_{\text{out}}, \sigma^2_{\text{out}}))}$
\end{algorithmic}
\end{algorithm}

\subsection{Attack Method}
\label{sec:pmia_framework}


\mypara{Likelihood Estimators}
Building on the analysis in the previous section, our goal is to estimate the likelihood ratio of the query instance for inference. 
We first follow LiRA~\cite{sp22lira} and define the output of query instance $(x,y)$ on model $f$ as:
\begin{equation*}
    \phi(f(x)_y) = \log \left(\frac{f(x)_y}{1- f(x)_y}\right),
\end{equation*}
where $f(x)_y$ denotes the softmax probability (\aka confidence score) of the model $f$ on the instance $(x,y)$. 
This scaling transformation stabilizes the distribution and, empirically, allows it to be well-approximated by a normal distribution. 
We then adopt the same likelihood estimators used in LiRA~\cite{sp22lira}:
\begin{equation*}
\label{equ:lira_gaussian}
\begin{aligned}
    \mathbb{E}_{D^\prime \sim \mathcal{S}^+_{(x,y)}}\Tilde{\mathcal{L}}(D^\prime,e_\theta) &= p\left(\phi(f_\theta(x)_y) \mid \mathcal{N}(\mu_{\text{in}}, \sigma^2_{\text{in}})\right), \\ 
    \mathbb{E}_{D^\prime \sim \mathcal{S}^-_{(x,y)}}\Tilde{\mathcal{L}}(D^\prime,e_\theta) &= p\left(\phi(f_\theta(x)_y) \mid \mathcal{N}(\mu_{\text{out}}, \sigma^2_{\text{out}})\right),
\end{aligned}
\end{equation*}
where $p\left(\phi(f_\theta(x)_y) \mid \mathcal{N}(\mu, \sigma^2)\right)$ is the probability density function over $\phi(f_\theta(x)_y)$ under a normal distribution with mean $\mu$ and variance $\sigma^2$.
In the adaptive setting, these two normal distributions can be estimated by training shadow \textit{in} models and shadow \textit{out} models on the query instance $(x,y)$ and retrieving the scaled confidence scores from shadow models.
However, in the non-adaptive setting, the adversary is not allowed to train shadow models on query instances, meaning that the distribution $\mathcal{N}(\mu_{\text{in}}, \sigma^2_{\text{in}})$ is unavailable.
Therefore, we aim to find proxy data $D_\text{proxy}$ from the adversary’s dataset $D_\text{adv}^\text{non-adapt}$ and use their behaviors (\ie $\mathcal{N}(\Tilde{\mu}_{\text{in}}, \tilde{\sigma}_{\text{in}})$) to approximate the behavior of a query instance $\mathcal{N} (\ie (\mu_{\text{in}}, \sigma^2_{\text{in}})$).

\mypara{Overview}
The workflow of \myoffline is outlined in~\Cref{alg:pmia}, consisting of two phases: the prepare and inference phases. 
In the prepare phase, we follow the standard shadow training technique~\cite{sp22lira,sp17miashokri} and train $N$ shadow models by randomly sampling shadow datasets from $D_\text{adv}^\text{non-adapt}$ (lines 3–7).
In the inference phase, we first collect the distribution of confidence scores for the target instance $(x, y)$ when it is \textit{not} in the model's training set (lines 11–13). 
Next, we select proxy data from $D_\text{adv}^\text{non-adapt}$ for the query instance and gather their confidence score distributions when they are \textit{in} the shadow models' training sets (lines 15–22).
Finally, we estimate the mean and variance for both collected confidence distributions (lines 23–24), query the target model $f_\theta$ on $(x,y)$ to output a parametric likelihood ratio $\Tilde{\Lambda}$ as the membership score.

\mypara{Finding Proxy Data}
The effectiveness of \myoffline relies on selecting appropriate proxy data from $D_\text{adv}^\text{non-adapt}$ and using their confidence distribution to approximate the likelihood of the query instance being in the training set. 
We propose three strategies for selecting these proxies at different granularities:

\begin{itemize}
    \item \textit{Global-level}. We use all instances in the attacker’s dataset as the proxy set $D_\text{proxy}$ and collect their confidence scores when they are part of the shadow model’s training set. This produces a global \textit{in} confidence distribution that is used to compute the likelihood for all query instances.
    \item \textit{Class-level}. Since the adversary knows the label of the query instance, we select samples from $D_\text{adv}^\text{non-adapt}$ that belong to the same class as the query instance to form $D_\text{proxy}$.
    \item \textit{Instance-level}. We retrieve the top-10 similar samples from $D_\text{adv}^\text{non-adapt}$ using some similarity measurements (\eg cosine similarity in embedding space), and use them as $D_\text{proxy}$.
\end{itemize}

We implement all the above strategies and analyze their effectiveness in~\Cref{sec:exp_ablation}. 
Despite their simplicity, we find that using these proxy data achieves state-of-the-art performance. 
We thus leave the exploration of more advanced proxy selection strategies for future work.

\mypara{Implementation Details}
We follow LiRA and train 256 shadow models during the prepare phase.
We also apply standard data augmentations to fit $n$-dimensional spherical Gaussians, which are collected by querying the shadow models $n$ times per sample ($n$ is set as 9 for all experiments). 
The final membership score, $\Tilde{\Lambda}$, is then computed using the likelihood ratio between two multivariate normal distributions.
When using the instance-level strategy to select proxy data, we adopt different approaches depending on the data modality.
For image datasets, we first embed each image into a 512-dimensional vector using a pretrained CLIP~\cite{icml21clip} encoder. We then use the Faiss library~\cite{24faiss} to retrieve the top-10 most similar samples from the attacker's dataset, based on cosine similarity in this embedding space. 
For non-image datasets where a powerful pretrained encoder is unavailable, we use Wasserstein distance on the raw data for proxy selection.




\section{Evaluation}
\label{sec:experiments}

We conduct a comprehensive evaluation of our adaptive (\myonline) and non-adaptive (\myoffline) attacks across various datasets and attack settings.
We first describe the experimental setup in~\Cref{sec:exp_setup}. 
Next, we evaluate the attack performance in adaptive and non-adaptive settings in~\Cref{sec:exp_online} and~\Cref{sec:exp_offline}, respectively. 
We also analyze the impact of attack components in~\Cref{sec:exp_ablation} and perform additional analyses and assess their performance against defenses in~\Cref{sec:exp_add}.

\subsection{Experimental Setup}
\label{sec:exp_setup}

\mypara{Datasets}
We select four image benchmark datasets (\ie MNIST~\cite{mnist}, Fashion-MNIST~\cite{fmnist}, CIFAR-10~\cite{cifar10}, and CIFAR-100~\cite{cifar10}) for our main experiments.
The results on two non-image datasets that are commonly used in MIAs (\ie Purchase and Texas~\cite{sp17miashokri}) are reported in~\Cref{sec:exp_add}.
A detailed dataset description is provided in~\Cref{appendix:data_description}.

\mypara{Network Architecture}
We consider four widely used neural network architectures for image classification: ResNet50~\cite{cvpr16resnet}, VGG16~\cite{vgg}, DenseNet121~\cite{densenet}, and MobileNetV2~\cite{mobilenetv2}. 
To reduce overfitting, we follow the settings of previous studies~\cite{sp22lira, iclr23canary} when training the target models. 
Specifically, we use the SGD algorithm with a learning rate of 0.1, momentum set to 0.9, and weight decay~\cite{krogh1991simple} set to $5\times 10^{-4}$. 
Additionally, we employ a cosine learning rate schedule~\cite{loshchilov16sgdr} for optimization, and apply data augmentation~\cite{aaai20random} during the training of the target models.
The train and validation accuracy of the target model in both settings are reported in~\Cref{appendix:acc_model}.

\begin{table*}[t]
\centering
\caption{Performance comparison of \textit{adaptive} attacks using \myonline on ResNet50 trained on four image datasets (\ie MNIST, Fashion-MNIST, CIFAR-10, and CIFAR-100). 
For each method, results are provided for both the original attack and the enhanced version using \myonline. 
\%Imp. denotes the relative improvement of \myonline over the baseline.
The best result is in bold.
}
\vspace{-1mm}
\label{tab:cmia_main_resnet}
\resizebox{0.9999\textwidth}{!}{
\begin{tabular}{ll|*{12}{c}}
\toprule
\multirow{2}{*}{\textbf{Method}} & \multirow{2}{*}{} & \multicolumn{4}{c}{\textbf{TPR @ 0.001\% FPR}} & \multicolumn{4}{c}{\textbf{TPR @ 0.1\% FPR}} & \multicolumn{4}{c}{\textbf{Balanced Accuracy}} \\
\cmidrule(lr){3-6}\cmidrule(lr){7-10}\cmidrule(lr){11-14}
& & MNIST & FMNIST & C-10 & C-100 & MNIST & FMNIST & C-10 & C-100 & MNIST & FMNIST & C-10 & C-100 \\
\midrule
\multirow{3}{*}{Calibration} & Base & 0.01\% & 0.52\% & 0.28\% & 1.48\% & 0.19\% & 2.23\% & 1.02\% & 5.51\% & 51.05\% & 54.21\% & 54.62\% & 61.18\% \\
& \textbf{\myonline} & \textbf{0.08\%} & \textbf{1.24\%} & \textbf{0.59\%} & \textbf{3.81\%} & \textbf{0.55\%} & \textbf{4.72\%} & \textbf{3.65\%} & \textbf{8.52\%} & \textbf{52.21\%} & \textbf{55.37\%} & \textbf{56.13\%} & \textbf{64.09\%} \\
& \cellcolor[gray]{0.9}\%Imp.  & \cellcolor[gray]{0.9}700.00\% & \cellcolor[gray]{0.9}138.46\% & \cellcolor[gray]{0.9}110.71\% & \cellcolor[gray]{0.9}157.43\% & \cellcolor[gray]{0.9}189.47\% & \cellcolor[gray]{0.9}111.66\% & \cellcolor[gray]{0.9}257.84\% & \cellcolor[gray]{0.9}54.63\% & \cellcolor[gray]{0.9}2.27\% & \cellcolor[gray]{0.9}2.14\% & \cellcolor[gray]{0.9}2.76\% & \cellcolor[gray]{0.9}4.76\%  \\
\midrule
\multirow{3}{*}{Attack-R} & Base & 0.00\% & 0.00\% & 0.21\% & 1.40\% & 0.10\% & 0.00\% & 1.30\% & 4.82\% & 52.15\% & 57.83\% & 54.26\% & 62.13\% \\
& \textbf{\myonline} & \textbf{0.00\%} & \textbf{0.00\%} & \textbf{0.45\%} & \textbf{2.01\%} & \textbf{0.37\%} & \textbf{0.00\%} & \textbf{1.95\%} & \textbf{6.04\%} & \textbf{52.95\%} & \textbf{58.48\%} & \textbf{55.48\%} & \textbf{63.93\%} \\
& \cellcolor[gray]{0.9}\%Imp.  & \cellcolor[gray]{0.9}- & \cellcolor[gray]{0.9}- & \cellcolor[gray]{0.9}114.29\% & \cellcolor[gray]{0.9}43.57\% & \cellcolor[gray]{0.9}270.00\% & \cellcolor[gray]{0.9}- & \cellcolor[gray]{0.9}50.00\% & \cellcolor[gray]{0.9}25.31\% & \cellcolor[gray]{0.9}1.53\% & \cellcolor[gray]{0.9}1.12\% & \cellcolor[gray]{0.9}2.25\% & \cellcolor[gray]{0.9}2.90\%  \\
\midrule
\multirow{3}{*}{LiRA} & Base & 0.12\% & 2.72\% & 2.64\% & 23.15\% & 1.23\% & 6.28\% & 8.45\% & 37.62\% & 51.26\% & 58.28\% & 62.52\% & 82.05\% \\
& \textbf{\myonline} & \textbf{0.77\%} & \textbf{4.42\%} & \textbf{3.86\%} & \textbf{36.74\%} & \textbf{2.10\%} & \textbf{8.34\%} & \textbf{9.71\%} & \textbf{45.37\%} & \textbf{52.67\%} & \textbf{60.91\%} & \textbf{63.83\%} & \textbf{84.89\%} \\
& \cellcolor[gray]{0.9}\%Imp.  & \cellcolor[gray]{0.9}541.67\% & \cellcolor[gray]{0.9}62.50\% & \cellcolor[gray]{0.9}46.21\% & \cellcolor[gray]{0.9}58.70\% & \cellcolor[gray]{0.9}70.73\% & \cellcolor[gray]{0.9}32.80\% & \cellcolor[gray]{0.9}14.91\% & \cellcolor[gray]{0.9}20.60\% & \cellcolor[gray]{0.9}2.75\% & \cellcolor[gray]{0.9}4.51\% &
 \cellcolor[gray]{0.9}2.10\% & \cellcolor[gray]{0.9}3.46\%  \\
\midrule
\multirow{3}{*}{Canary} & Base & 0.15\% & 2.95\% & 2.36\% & 25.78\% & 1.28\% & 6.65\% & 8.12\% & 38.25\% & 53.76\% & 58.94\% & 62.60\% & 83.11\% \\
& \textbf{\myonline} & \textbf{0.84\%} & \textbf{4.73\%} & \textbf{3.61\%} & \textbf{37.85\%} & \textbf{2.48\%} & \textbf{8.47\%} & \textbf{9.02\%} & \textbf{45.96\%} & \textbf{55.60\%} & \textbf{61.07\%} & \textbf{63.81\%} & \textbf{84.72\%} \\
& \cellcolor[gray]{0.9}\%Imp.  & \cellcolor[gray]{0.9}460.00\% & \cellcolor[gray]{0.9}60.34\% & \cellcolor[gray]{0.9}52.97\% & \cellcolor[gray]{0.9}46.82\% & \cellcolor[gray]{0.9}93.75\% & \cellcolor[gray]{0.9}27.37\% & \cellcolor[gray]{0.9}11.08\% & \cellcolor[gray]{0.9}20.16\% & \cellcolor[gray]{0.9}3.42\% & \cellcolor[gray]{0.9}3.61\% & \cellcolor[gray]{0.9}1.93\% & \cellcolor[gray]{0.9}1.94\%  \\
\midrule
\multirow{3}{*}{RMIA} & Base & 0.21\% & 2.05\% & 1.43\% & 10.72\% & 0.96\% & 4.71\% & 5.24\% & 30.13\% & 52.99\% & 58.16\% & 62.05\% & 80.64\% \\
& \textbf{\myonline} & \textbf{0.52\%} & \textbf{3.56\%} & \textbf{2.05\%} & \textbf{14.67\%} & \textbf{1.62\%} & \textbf{5.81\%} & \textbf{6.05\%} & \textbf{37.51\%} & \textbf{53.51\%} & \textbf{60.90\%} & \textbf{62.49\%} & \textbf{82.53\%} \\
& \cellcolor[gray]{0.9}\%Imp.  & \cellcolor[gray]{0.9}147.62\% & \cellcolor[gray]{0.9}73.66\% & \cellcolor[gray]{0.9}43.36\% & \cellcolor[gray]{0.9}36.85\% & \cellcolor[gray]{0.9}68.75\% & \cellcolor[gray]{0.9}23.35\% & \cellcolor[gray]{0.9}15.46\% & \cellcolor[gray]{0.9}24.49\% & \cellcolor[gray]{0.9}0.98\% & \cellcolor[gray]{0.9}4.71\% & \cellcolor[gray]{0.9}0.71\% & \cellcolor[gray]{0.9}2.34\%  \\
\midrule
\multirow{3}{*}{RAPID} & Base & 0.23\% & 1.31\% & 0.56\% & 9.83\% & 0.79\% & 3.44\% & 3.12\% & 21.69\% & 52.44\% & 58.40\% & 59.58\% & 75.83\% \\
& \textbf{\myonline} & \textbf{0.48\%} & \textbf{2.45\%} & \textbf{0.94\%} & \textbf{11.83\%} & \textbf{1.24\%} & \textbf{4.73\%} & \textbf{4.75\%} & \textbf{25.90\%} & \textbf{52.97\%} & \textbf{58.51\%} & \textbf{59.77\%} & \textbf{78.52\%} \\
& \cellcolor[gray]{0.9}\%Imp.  & \cellcolor[gray]{0.9}108.70\% & \cellcolor[gray]{0.9}87.02\% & \cellcolor[gray]{0.9}67.86\% & \cellcolor[gray]{0.9}20.35\% & \cellcolor[gray]{0.9}56.96\% & \cellcolor[gray]{0.9}37.50\% & \cellcolor[gray]{0.9}52.24\% & \cellcolor[gray]{0.9}19.41\% & \cellcolor[gray]{0.9}1.01\% & \cellcolor[gray]{0.9}0.19\% & \cellcolor[gray]{0.9}0.32\% & \cellcolor[gray]{0.9}3.55\%  \\
\bottomrule
\end{tabular}}
\end{table*}


\mypara{Attack Baselines}
We compare our attacks against a broad range of state-of-the-art MIAs in our experiments:

\begin{itemize}
    \item \textit{Calibration}~\cite{iclr22calibrate} employs a technique called difficulty calibration, which adjusts the loss of the query instance by calibrating its loss on shadow models as membership scores.
    \item \textit{Attack-R}~\cite{ccs22enhanced} compare the loss of query instance on the target model with its loss on shadow out models. The membership score is based on the ratio where the loss on the target model is smaller than the loss on the shadow models.
    \item \textit{LiRA}~\cite{sp22lira} exploits behavioral discrepancies by modeling its loss distribution as a Gaussian estimate and uses a likelihood ratio test to compute the membership score. 
    \item \textit{Canary}~\cite{iclr23canary} enhances LiRA by using adversarial learning to optimize the query instance for inference.
    \item \textit{RMIA}~\cite{icml24rmia} calculates the membership score based on the success ratio of pairwise likelihood ratio tests between the query instance and random instances from the population.
    \item \textit{RAPID}~\cite{ccs24rapid} combines both the loss and the calibrated loss~\cite{iclr22calibrate} to train a neural network for membership inference.
\end{itemize}

LiRA and Canary use different strategies for adaptive and non-adaptive settings. 
We distinguish between their versions based on the experimental context.
In the non-adaptive setting, in addition to these attacks, we also compare \myoffline with the following attacks that do not rely on shadow training:

\begin{itemize}
    \item \textit{LOSS}~\cite{csf18privacy} uses the loss of the query instance as the score.
    \item \textit{Entropy}~\cite{usenix21systematic} leverages a modified prediction entropy estimation as the membership score.
\end{itemize}

\mypara{Evaluation Procedures}
For the adaptive setting, we follow~\cite{sp22lira,iclr23canary} split each dataset into two disjoint subsets: $D_1$ and $D_2$. We randomly sample 50\% of $D_1$ to train the target model, and $D_2$ is used as the validation set to prevent overfitting during training.
The adversary is provided with the same $D_1$ to prepare their attack, and the query set $D_\text{query}$ is also set to be the same (\ie $D^\text{adapt}_\text{adv} = D_\text{query} = D_1$).
For the non-adaptive setting, we follow~\cite{ccs24rapid,ccs24seqmia} and divide each dataset into two disjoint subsets, one as the query set and the other as the adversary's dataset, ensuring that the adversary cannot access queries when training shadow models.
Each of these subsets is further split into training and validation sets to train the target model and shadow models. 
The details about the data split are in~\Cref{tab:dataset_split_online} and~\Cref{tab:dataset_split_offline}.

\mypara{Evaluation Metrics}
Following previous studies~\cite{sp22lira,icml24rmia}, we focus on evaluating the attack performance on the low false positive rate regime. 
Specifically, we use the following evaluation metrics for our experiments:
\begin{itemize}
    \item \textit{TPR@0.001\%FPR.} 
    It directly reflects the extent of privacy leakage of the model by allowing only one (or a few) false positives to compute the true positive rate. 
    \item \textit{TPR@0.1\%FPR.} 
    This is a relaxed version of the previous metric, allowing more false-positive samples for evaluation. 
    \item \textit{Balanced Accuracy.} 
    This metric measures how often an attack correctly predicts membership (average case).
\end{itemize}

\mypara{Attack Setup}
We use the same techniques as LiRA~\cite{sp22lira} to train shadow models, ensuring that each instance appears in exactly half of the training sets for the shadow models. 
The performance of shadow-based MIAs depends on the number of shadow models used.  
Through experimentation, we find that LiRA and Canary benefit from training a larger number (\eg 256) of shadow models, while other methods plateau after a smaller number of shadow models.  
Thus, we identify the optimal number of shadow models for each attack method individually.
For \myonline, we train the same number of shadow models as the base attack for each cascading iteration.  
For \myoffline, we train a fixed set of 256 shadow models.
The hyperparameters are set according to the original implementations, and we report the average metrics over five runs with different random seeds as the final results.


\begin{table*}[t]
\centering
\caption{Performance comparison of \textit{non-adaptive} attacks on ResNet50 trained on four image datasets.
The \%Imp. indicates the relative improvement of \myoffline compared to the strongest baseline (underlined).
}
\vspace{-1mm}
\label{tab:pmia_main_resnet}
\resizebox{0.98\textwidth}{!}{
\begin{tabular}{l|*{12}{c}}
\toprule
\multirow{2}{*}{\textbf{Method}} & \multicolumn{4}{c}{\textbf{TPR @ 0.001\% FPR}} & \multicolumn{4}{c}{\textbf{TPR @ 0.1\% FPR}} & \multicolumn{4}{c}{\textbf{Balanced Accuracy}} \\
\cmidrule(lr){
2-5}\cmidrule(lr){6-9}\cmidrule(lr){10-13}
&  MNIST & FMNIST & C-10 & C-100 & MNIST & FMNIST & C-10 & C-100 & MNIST & FMNIST & C-10 & C-100 \\
\midrule
LOSS & 0.01\% & 0.01\% & 0.00\% & 0.00\% & 0.08\% & 0.09\% & 0.00\% & 0.00\% & \underline{52.81\%} & \underline{61.51\%} & 63.35\% & 78.20\% \\
Entropy & 0.01\% & 0.01\% & 0.00\% & 0.00\% & 0.08\% & 0.10\% & 0.00\% & 0.21\% & 52.80\% & 61.16\% & 63.08\% & 78.05\% \\
Calibration & 0.05\% & 0.06\% & 0.08\% & 1.08\% & 0.34\% & 0.45\% & 1.03\% & 2.83\% & 52.51\% & 55.10\% & 57.96\% & 66.10\% \\
Attack-R & 0.00\% & 0.00\% & 0.00\% & 0.00\% & 0.00\% & 0.00\% & 0.00\% & 0.00\% & 52.62\% & 58.47\% & \underline{63.46\%} & 77.36\% \\
LiRA & 0.09\% & 0.05\% & 0.03\% & 0.98\% & 0.30\% & 0.67\% & 0.78\% & \underline{8.56\%} & 50.54\% & 53.11\% & 58.97\% & 73.25\% \\
Canary & 0.11\% & 0.08\% & 0.03\% & 1.78\% & 0.30\% & 1.02\% & 0.77\% & 7.35\% & 51.01\% & 53.79\% & 58.77\% & 73.93\% \\
RMIA & \underline{0.17\%} & 0.05\% & \underline{0.41\%} & \underline{2.73\%} & \underline{0.51\%} & \underline{1.25\%} & \underline{2.60\%} & 6.64 & 52.78\% & 58.96\% & 62.72\% & 77.53\% \\
RAPID & 0.09\% & \underline{0.15\%} & 0.15\% & 1.16\% & 0.45\% & 0.44\% & 1.34\% & 3.14\% & 52.05\% & 58.42\% & 61.39\% & \underline{78.49\%} \\
\midrule
\textbf{\myoffline} & \textbf{0.31\%} & \textbf{0.17\%} & \textbf{1.20\%} & \textbf{5.90\%} & \textbf{1.01\%} & \textbf{2.80\%} & \textbf{3.29\%} & \textbf{11.5\%} & \textbf{52.87\%} & \textbf{61.56\%} & \textbf{64.34\%} & \textbf{80.4\%} \\
\cellcolor[gray]{0.9}\%Imp.  & \cellcolor[gray]{0.9}82.35\% & \cellcolor[gray]{0.9}13.33\% & \cellcolor[gray]{0.9}192.68\% & \cellcolor[gray]{0.9}116.12\% & \cellcolor[gray]
{0.9}98.04\% & \cellcolor[gray]{0.9}124.00\% & \cellcolor[gray]{0.9}26.54\% & \cellcolor[gray]{0.9}34.35\% & \cellcolor[gray]{0.9}0.11\% & \cellcolor[gray]{0.9}0.08\% & \cellcolor[gray]{0.9}1.39\% & \cellcolor[gray]{0.9}2.43\%  \\
\bottomrule
\end{tabular}}
\end{table*}

\subsection{Evaluation of \myonline}
\label{sec:exp_online}

\mypara{Main Results}
We apply \myonline with six SOTA MIAs across four image datasets. 
As shown in~\Cref{tab:cmia_main_resnet}, \myonline consistently boosts all attacks across datasets, particularly in the low false-positive rate regime.
For example, \myonline enhances the performance of LiRA by $5\times$ at a false-positive rate of 0.001\%
and boosts the Calibration attack by $7\times$ at the same FPR on MNIST. 
These results represent a substantial advancement, as such datasets were previously considered difficult to attack. 
In addition, we find that \myonline can elevate previously weak attacks to a level comparable to strong ones. 
For instance, \myonline boosts the Calibration attack from 0.19\% to 0.55\% at a false-positive rate of 0.1\% on MNIST, nearly matching the performance of LiRA (\ie 0.56\% at the same FPR).
The Receiver Operating Characteristic (ROC) curve~\cite{nature09genomic} and results on other model architectures are reported in~\Cref{appendix:cmia_other_models}.

\begin{figure}[t]
    \centering
    \vspace{-5mm}
    \subfigure[MNIST]
    {
    \includegraphics[width=0.44\linewidth]{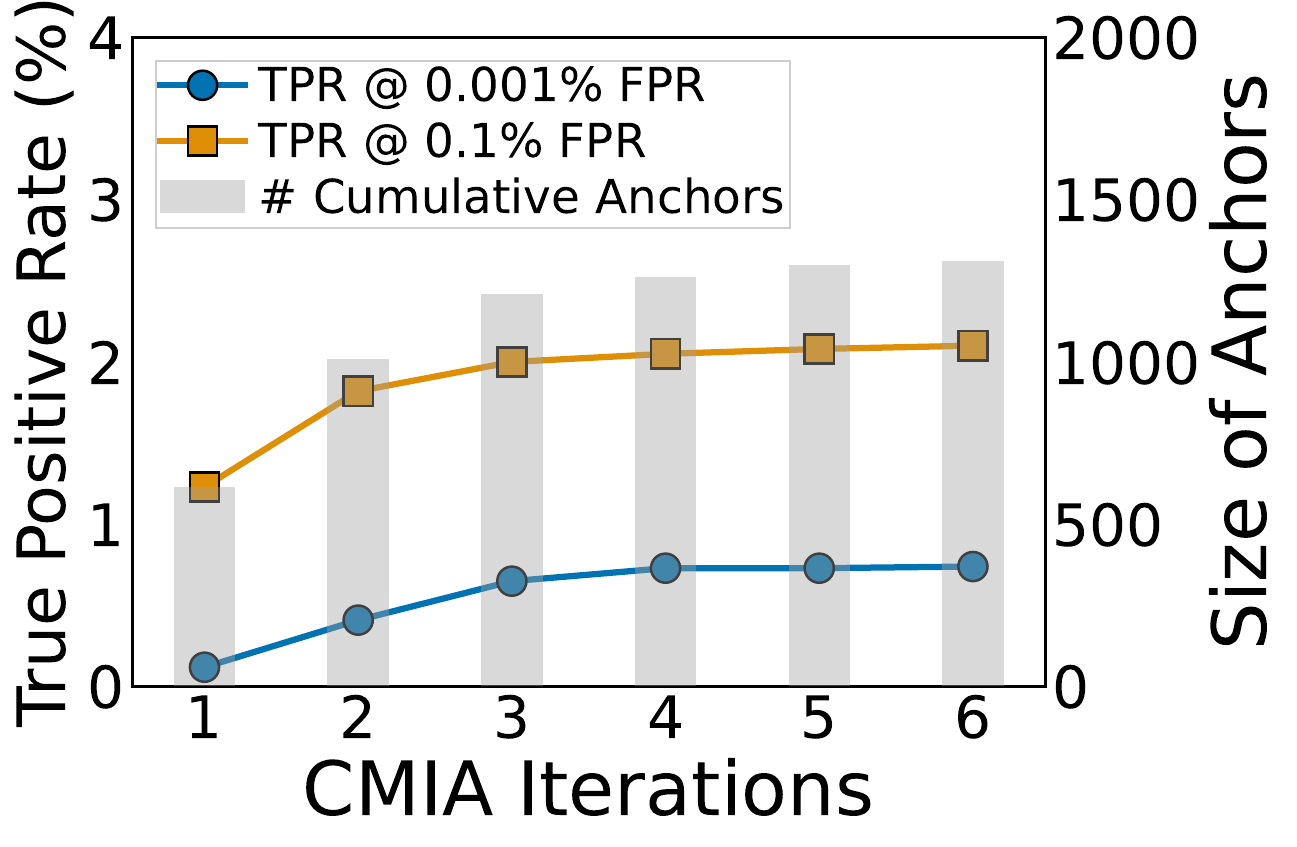}
    \label{fig:cmia_iter_mnist}
    }
    \subfigure[CIFAR-100]
    {
    \includegraphics[width=0.44\linewidth]{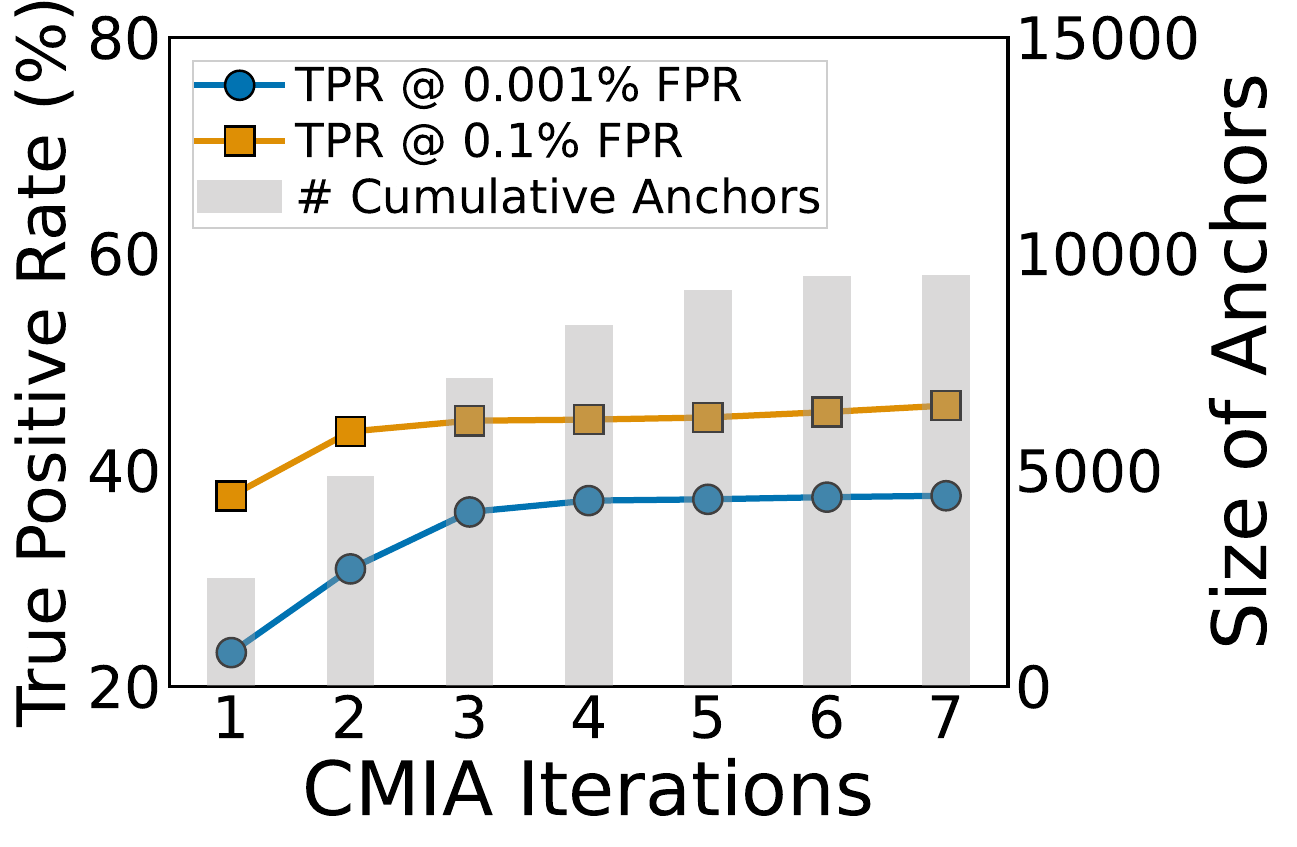}
    \label{fig:cmia_iter_cifar100}
    }
    \vspace{-3mm}
    \caption{The impact of number of cascading iterations in \myonline.}
    \vspace{-3mm}
    \label{fig:cmia_iter}
\end{figure}

\mypara{Improvement per Cascading Iteration}
We analyze the performance of each iteration in \myonline. 
Specifically, we use LiRA as the base attack and examine its performance across each iteration, as shown in~\Cref{fig:cmia_iter}.
The results demonstrate a gradual improvement in performance with each additional iteration, while the number of identified anchors increases until reaching the stopping criterion.
Notably, the most significant improvements occur during the first few iterations.
For example, adding just one cascading iteration increases the true-positive rate from 0.12\% to 0.41\% on MNIST, with similar improvements observed for CIFAR-100. 
This pattern likely emerges because identifying anchors is easier in early iterations, whereas it is increasingly difficult to reliably predict memberships for remaining instances.



\mypara{Efficiency Analysis}
While \myonline achieves impressive attack performance,~\Cref{tab:runtime} shows that it incurs a significantly higher computational cost than the base attack (\ie, LiRA).
To improve this trade-off, we explore three strategies:
\begin{itemize}
    \item \textit{Execute with Fewer Iterations.} 
    Most of \myonline's performance gains occur in the early iterations. Thus, reducing the cascading iterations can still achieve strong performance. 
    \item \textit{Optimize the Computational Budget.} 
    The intermediate shadow models of \myonline are primarily used for identifying anchors rather than full membership inference.
    Thus, we can reduce the number of shadow models trained per iteration and increase the iterations to take advantage of the cascading framework (denoted as \myonlineopt).
    \item \textit{Lightweight Attacks for Anchor Selection.} 
    A lightweight attack (\ie LOSS attack~\cite{csf18privacy}) can be used for anchor selection, followed by a stronger attack (\ie LiRA) for inference (denoted as \myonlineloss).
\end{itemize}

\begin{figure}[t]
    \centering
    \vspace{-5mm}
    \subfigure[MNIST]
    {
    \includegraphics[width=0.447\linewidth]{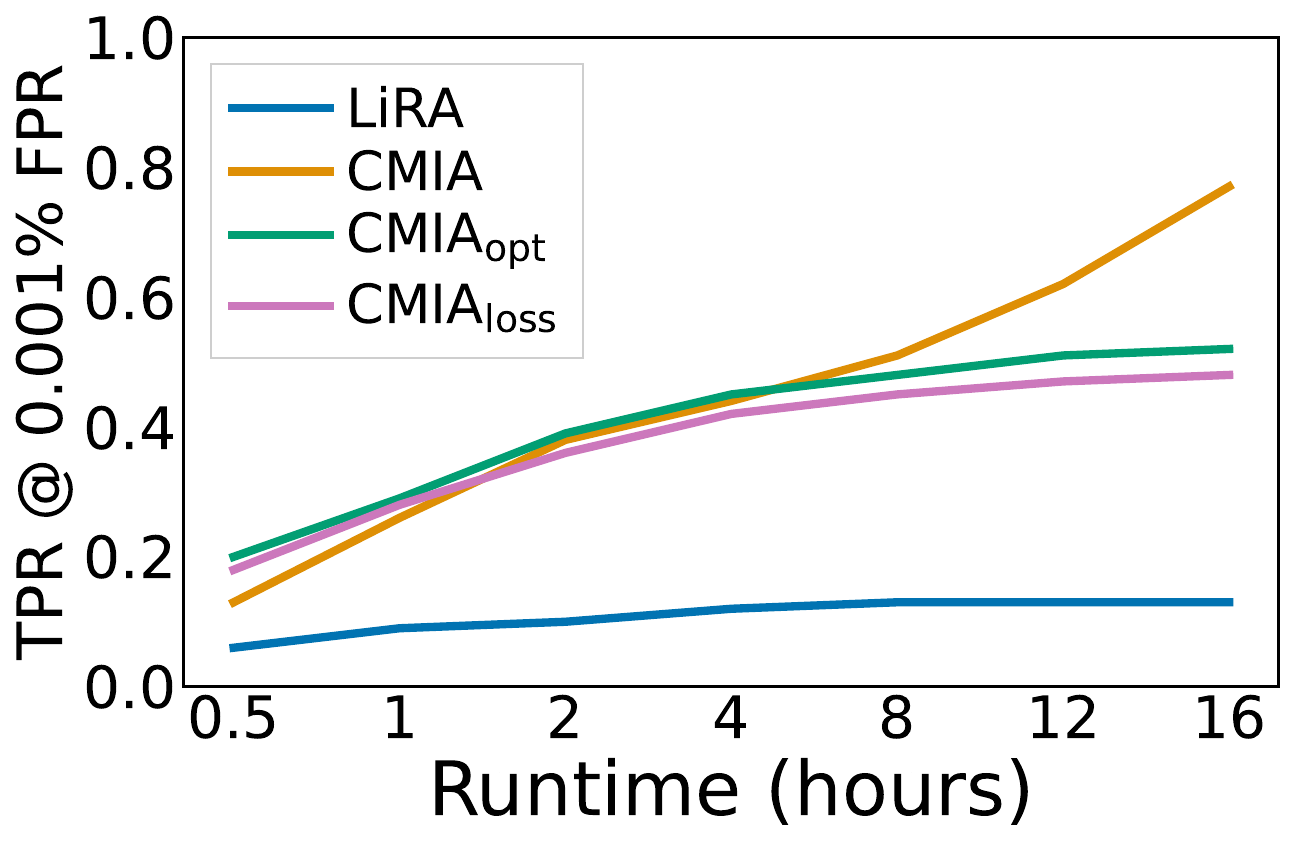}
    \label{fig:cmia_runtime_mnist}
    }
    \subfigure[CIFAR-10]
    {
    \includegraphics[width=0.447\linewidth]{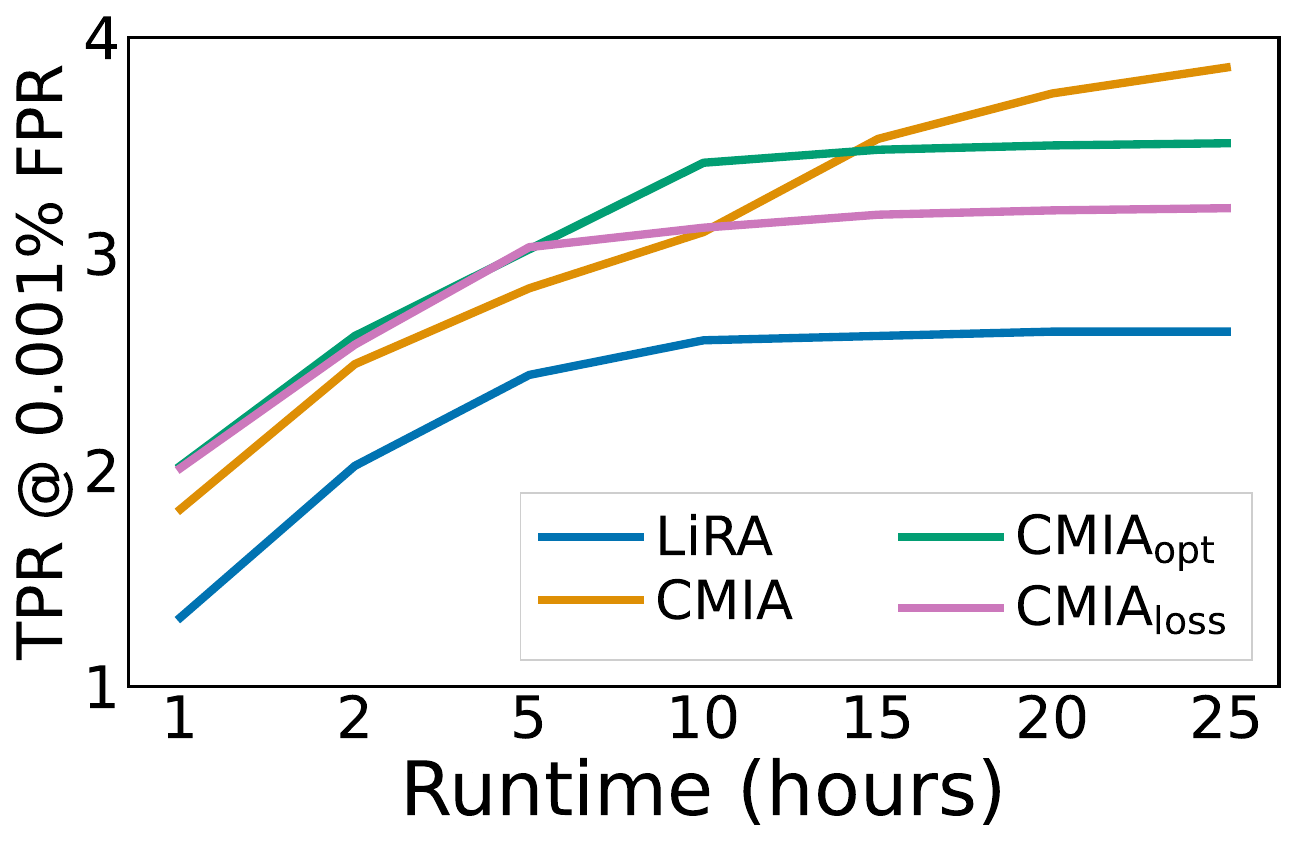}
    \label{fig:cmia_runtime_cifar10}
    }
    \vspace{-3mm}
    \caption{Efficiency analysis of \myonline with LiRA as the base attack. \myonlineopt optimize the computation efficiency by training 64 shadow models per iteration. \myonlineloss utilizes the LOSS attack~\cite{csf18privacy} to identify anchors for cascading attack.}
    \vspace{-3mm}
    \label{fig:cmia_efficiency}
\end{figure}

We evaluate these strategies on MNIST and CIFAR-10 using a cluster with 8 A100 GPUs.
As shown in~\Cref{fig:cmia_efficiency}, LiRA saturates after 2 and 5 hours (corresponding to 256 shadow models), whereas our proposed methods achieve higher accuracy within the same budget.
Notably, \myonlineopt surpasses LiRA while training only 64 shadow models per iteration.
Further, \myonlineloss proves effective without requiring any additional model training.
Overall, these enhancements allow \myonline to offer a better efficiency and effectiveness trade-off.

\subsection{Evaluation of \myoffline}
\label{sec:exp_offline}

\mypara{Main Results}
We compare \myoffline with SOTA MIAs in the non-adaptive setting. 
The results for ResNet50 are shown in \Cref{tab:pmia_main_resnet}, while performance on other model architectures and the ROC curves are provided in~\Cref{appendix:pmia_other_models}. 
It is observed that \myoffline achieves the best attack performance across all datasets.
For example, it achieves a true-positive rate of 5.90\% at a false-positive rate of 0.001\% on CIFAR-100, which is at least twice as high as the best-performing baseline (\ie RMIA). 
Despite using the same likelihood estimator as LiRA, \myoffline significantly outperforms it. 
This improvement stems from \myoffline's use of the approximated likelihood ratio for membership prediction, while LiRA offline relies on a one-sided hypothesis.



\mypara{Efficiency Analysis}
We analyze its computational cost in two phases: preparation and inference.
In the preparation phase, \myoffline trains the same number of shadow models (\eg 256) as LiRA. 
While this step is quite expensive, it only needs to be performed once. 
The more critical phase is inference, where the adversary must respond quickly to any queries. 
In~\Cref{tab:pmia_efficiency}, we report the total inference time cost for all attacks on MNIST.
\myoffline responds faster than RMIA and RAPID, demonstrating strong inference efficiency.
As a result, the overall runtime of \myoffline (prepare and inference) is only marginally greater than that of LiRA, as shown in~\Cref{tab:runtime}.


\subsection{Ablation Study}
\label{sec:exp_ablation}


\begin{table}[t]
\centering
\caption{Runtime comparison of the proposed methods, measured in hours on a cluster of 8 A100 GPUs. \myonline is measured using LiRA as the base attack.}
\vspace{-1mm}
\label{tab:runtime}
\resizebox{0.47\textwidth}{!}{
\begin{tabular}{c|r|r|r|r|r|r}
\toprule
 & \textbf{MNIST} & \textbf{FMNIST} & \textbf{C-10} & \textbf{C-100} & \textbf{Purchase} & \textbf{Texas}\\
\midrule
LiRA & 2.25 & 3.52 & 5.82 & 10.20 & 1.02 & 1.15 \\ 
\myonline & 13.30 & 17.53 & 34.83 & 71.48 & 5.14 & 5.93 \\ 
\myoffline & 2.30 & 3.58 & 5.85 & 10.40 & 1.09 & 1.18 \\ 
\bottomrule
\end{tabular}}
\vspace{-1mm}
\end{table}

\begin{table}[t]
\centering
\caption{Impact of the hyperparameters $K$ and $\delta$ in \myonline.}
\vspace{-1mm}
\label{tab:cmia_hyperparameter}
\resizebox{0.48\textwidth}{!}{
\begin{tabular}{c|c|c|c|c}
\toprule
 $K$ & $\delta$& \textbf{TPR @ 0.001\%FPR} & \textbf{TPR @ 0.1\%FPR} & \textbf{Balanced Accuracy} \\
\midrule
5 & 15 & 3.10\% & 8.90\% & 62.71\%  \\ 
5 & 30 & 3.18\% & 8.92\% & 62.84\%  \\ 
5 & 60 & 2.97\% & 8.65\% & 62.58\%  \\ 
10 & 15 & \textbf{3.88\%} & 9.67\% & 63.80\%  \\ 
10 & 30 & 3.86\% & \textbf{9.71\%} & \textbf{63.83\%}  \\ 
10 & 60 & 3.12\% & 9.03\% & 62.77\%  \\ 
\bottomrule
\end{tabular}}
\end{table}

\mypara{Impact of Hyperparameters for \myonline}
We vary the number of cascading iterations $K$ and stopping criterion $\delta$ in \myonline to investigate its impact. 
The results on CIFAR-10 are shown in~\Cref{tab:cmia_hyperparameter}. As illustrated, increasing the number of iterations $K$ generally improves performance. 
A strict stopping criterion (\eg $\delta=60$) tends to halt the cascading process prematurely, leading to degraded results. On the other hand, a relaxed criterion (\eg $\delta=15$) would increase cascading iterations with marginal gains. 
Overall, we find that  $K=10$ and $\delta=30$ provide a good trade-off between effectiveness and efficiency.

\mypara{Impact of Thresholds Selection for \myonline}
We vary the threshold $\tau_\text{out}$ and investigate its impact.
Specifically, we rank the membership scores and select the $r$-th lowest score among the members as $\tau_\text{out}$, where $r$ presents the tolerance level. 
We then vary $r$ and report the performance and anchor size for MNIST and CIFAR-100.
As shown in~\Cref{tab:cmia_threshold}, performance initially improves as $r$ increases, but drops as it increases further.
This happens because increasing the tolerance allows for more anchors for generating conditional shadow models, but also increases false negatives. 
The optimal trade-off occurs around $r = 10$, where sufficient anchor samples are selected for shadow training without introducing too many errors.
We use the strictest possible selection for $\tau_\text{in}$ (\ie no false positives are allowed) because allowing false positives would directly degrade the performance on metrics like TPR at low FPR.


\begin{table}[t]
\centering
\caption{Impact of the threshold parameter $r$ in \myonline. A larger $r$ introduces more false negatives for anchor selections.}
\vspace{-1mm}
\label{tab:cmia_threshold}
\resizebox{0.49\textwidth}{!}{
\begin{tabular}{r|rr|cc|cc}
\toprule
\multirow{2}{*}{{$r$}} & \multicolumn{2}{c|}{\textbf{\# Identified Anchors}} & \multicolumn{2}{c|}{\textbf{TPR @ 0.1\%FPR}} & \multicolumn{2}{c}{\textbf{TPR @ 0.001\%FPR}}\\
 & MNIST & CIFAR-100 & MNIST & CIFAR-100 & MNIST & CIFAR-100 \\
\midrule
1 & 245 & 1851 & 0.22\% & 23.68\% & 1.23\% & 38.95\% \\ 
5 & 573 & 3124 & 0.50\% & 26.88\% & 1.82\% & 41.57\% \\ 
10 & 1310 & 9512 & \textbf{0.77\%} & 36.74\% & \textbf{2.10\%} & 45.96\% \\
15 & 1507 & 13435 & 0.69\% & \textbf{38.12\%} & 1.74\% & \textbf{48.41\%} \\
20 & 1863 & 15941 & 0.51\% & 34.31\% & 1.45\% & 43.59\% \\
25 & 2437 & 17064 & 0.43\% & 31.51\% & 1.20\% & 41.14\% \\
\bottomrule
\end{tabular}}
\vspace{-3mm}
\end{table}

\begin{table*}[t]
\centering
\caption{Membership inference time cost of non-adaptive attacks against a ResNet50 model on MNIST. }
\vspace{-1mm}
\label{tab:pmia_efficiency}
\resizebox{0.88\textwidth}{!}{
\begin{tabular}{l|*{8}{c}|c}
\toprule
Attack Method&  LOSS & Entropy & Calibration & Attack-R & LiRA & Canary & RMIA & RAPID & \myoffline \\
\midrule
Inference Cost/seconds  & $1.23$ & $2.52$ & $1.85$ & $3.03$ & $10.47$ & $> 400,\!000$ & $49.5$ & $31.5$ & $\mathbf{15.8}$   \\
\bottomrule
\end{tabular}}
\end{table*}

\begin{figure}[t]
    \centering
    \subfigure[MNIST]
    {
    \includegraphics[width=0.447\linewidth]{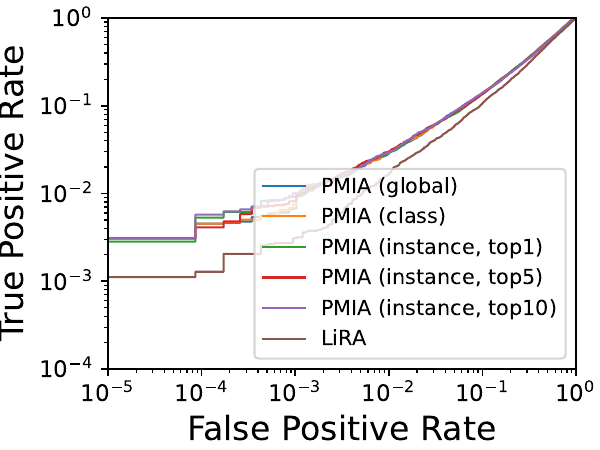}
    \label{fig:pmia_proxy_mnist}
    }
    \subfigure[CIFAR-100]
    {
    \includegraphics[width=0.447\linewidth]{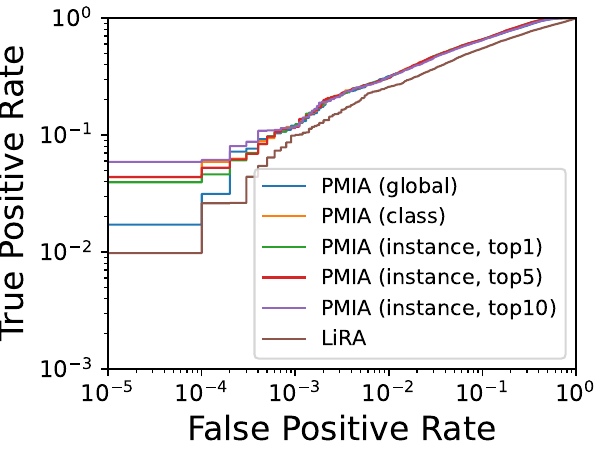}
    \label{fig:pmia_proxy_cifar100}
    }
    \vspace{-3mm}
    \caption{The impact of selecting different proxy data in \myoffline. The y-axis range is adjusted to enhance visibility. }
    \vspace{-3mm}
    \label{fig:pmia_proxy}
\end{figure}

\mypara{Impact of Selecting Proxy Data}
In \myoffline, we consider three proxy selection strategies at the global, class, and instance levels. 
\Cref{fig:pmia_proxy} illustrates the performance of these proxy selection strategies on the MNIST and CIFAR-100 datasets. 
All approaches significantly outperform LiRA, highlighting the importance of selecting proxies for attacks.
We observe a notable performance improvement when using proxy data from the same class, compared to using the entire adversary's dataset as proxies.
In CIFAR-100, performance improves when using similar images as proxy data; however, this improvement is not observed in MNIST.
We attribute this discrepancy to the lack of diversity in the MNIST dataset, making the instance-level proxy data less distinct and offering minimal improvement over the class-level proxy.
Additionally, we find that instance-level proxying is robust to the number of similar images selected, with comparable performance observed when using the top 1, 5, or 10 most similar images as proxies.

\begin{figure}[t]
    \centering
    \vspace{-5mm}
    \subfigure[\myonline with LiRA as base attack]
    {
    \includegraphics[width=0.44\linewidth]{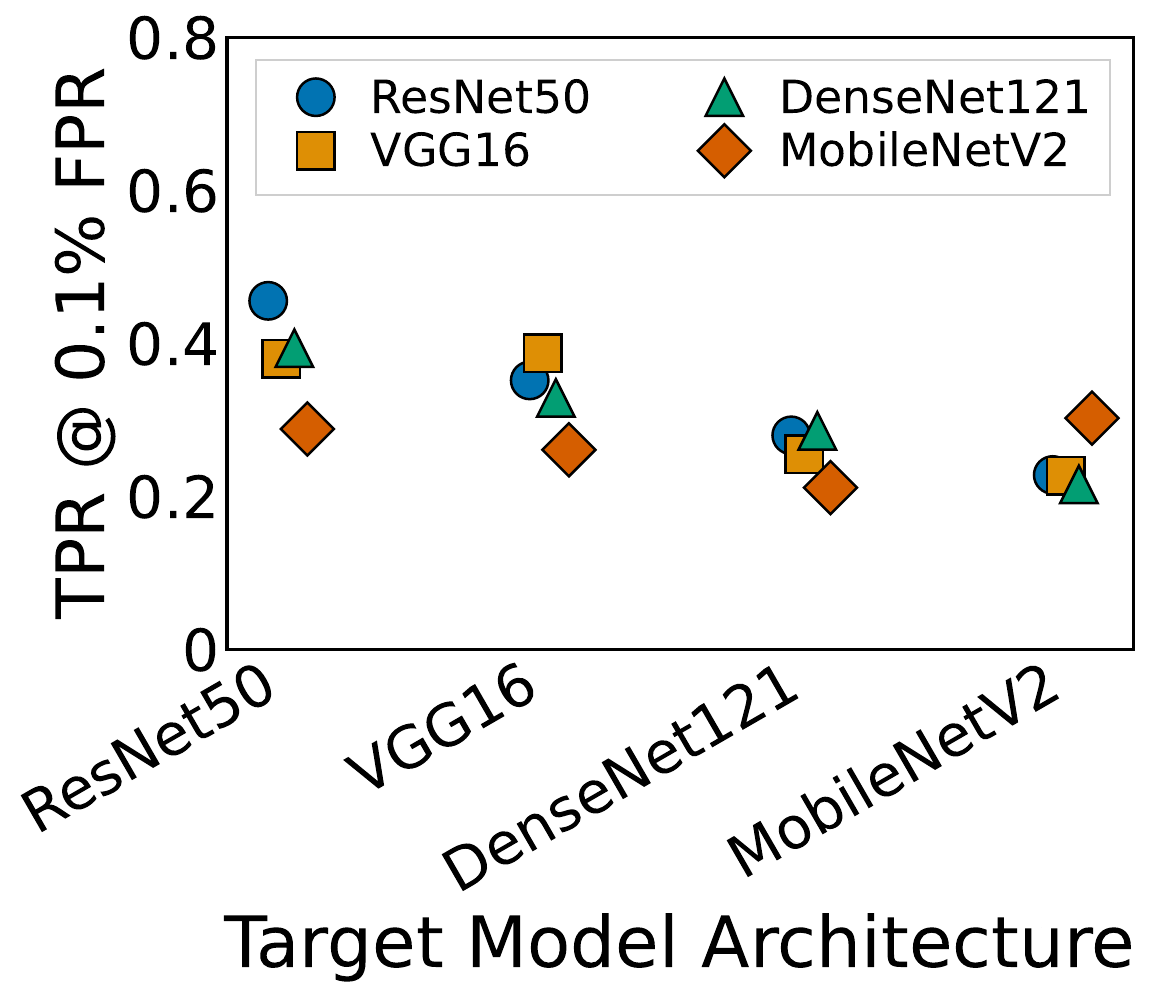}
    \label{fig:cmia_arch}
    }
    \subfigure[\myoffline]
    {
    \includegraphics[width=0.44\linewidth]{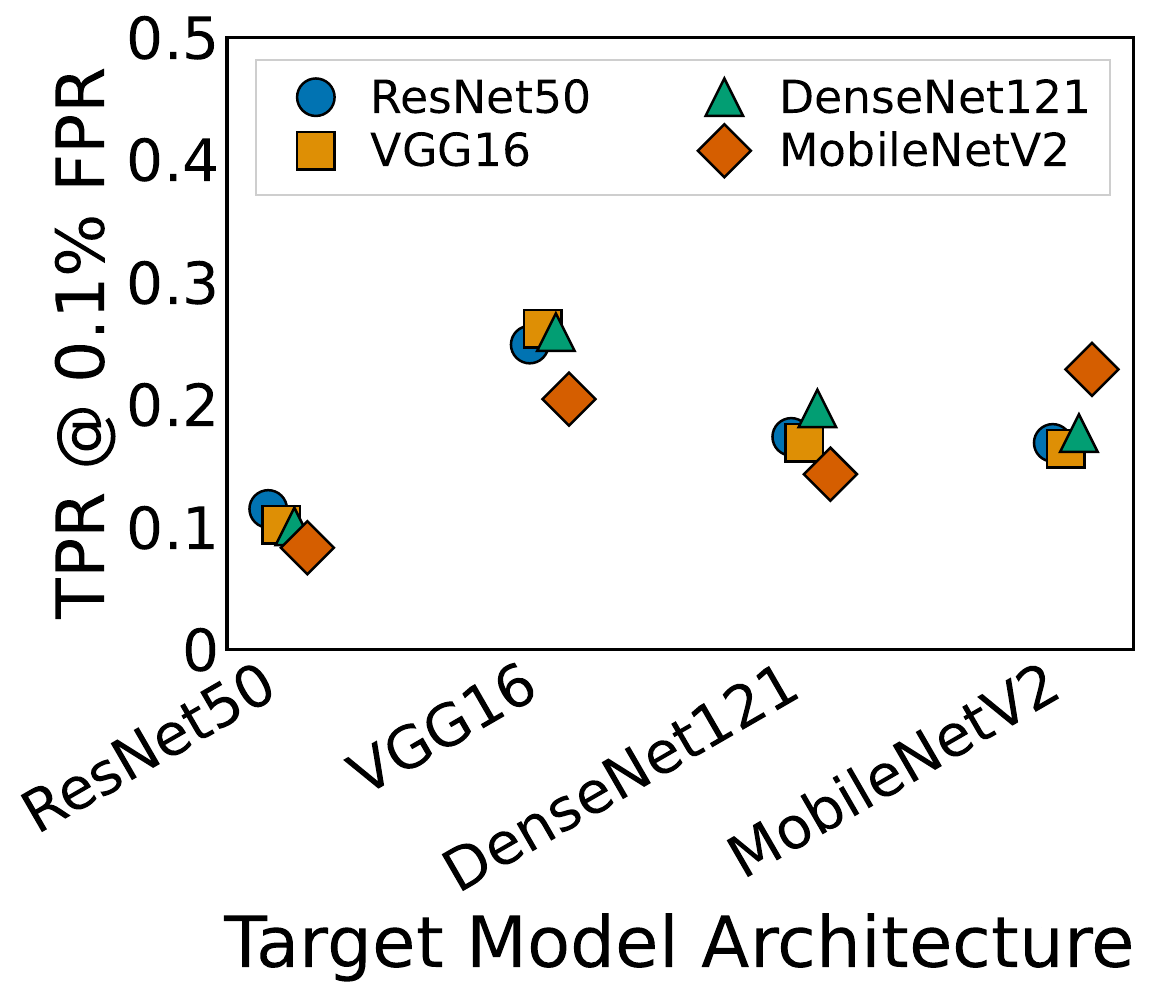}
    \label{fig:pmia_arch}
    }
    \vspace{-3mm}
    \caption{The impact of architecture differences between the target model and the shadow models trained on CIFAR-100.}
    \label{fig:impact_arch}
\end{figure}

\mypara{Mismatched Model Architecture}
We examine how our attack is affected when the attacker is unaware of the exact training procedure used for the target model. 
Specifically, we explore the impact of varying the target model's architecture on the performance of both \myonline and \myoffline attacks.
As shown in \Cref{fig:impact_arch}, our attack performs best when the attacker trains shadow models with the same architecture as the target model, and using a different model (\eg VGG16 instead of ResNet50) has only a minimal effect on the attack’s effectiveness. 
However, we observe a notable decrease in performance when using MobileNetV2. 
This decrease can be attributed to the inherent differences in the architectures: in MobileNetV2, the number of channels in the feature map increases and then decreases, which contrasts with other model architectures.
Similar results have been reported in previous studies~\cite{sp22lira,ccs24rapid}.

\mypara{Attack with Distribution Shift}
In our experiments so far, we assume that the adversary has access to the same underlying distribution as the target model's training datasets. 
However, in a real attack, the adversary's data is likely not perfectly aligned with the target's training data. 
We now explore this more realistic scenario to assess its impact. 
Specifically, we follow prior work~\cite{sp22lira,ccs24rapid} and conduct following experiments:
\begin{itemize}
    \item $\mathbb{D}_\text{target} = \mathbb{D}_\text{attack}$. 
    Both the target and shadow models are trained using disjoint subsets of the CIFAR-10 dataset. 
    This follows the non-adaptive setting in our main experiments.
    \item $\mathbb{D}_\text{target} \neq \mathbb{D}_\text{attack}$. 
    The target model is trained on a subset of CIFAR-10, while the shadow models use the ImageNet~\cite{cvpr09imagenet} portion of the CINIC-10~\cite{arxiv18cinic}. This creates a distribution shift between the target’s data and the adversary’s data.
\end{itemize}

\begin{figure}[t]
    \centering
    \vspace{-3mm}
    \subfigure[Balanced Accuracy]
    {
    \includegraphics[width=0.44\linewidth]{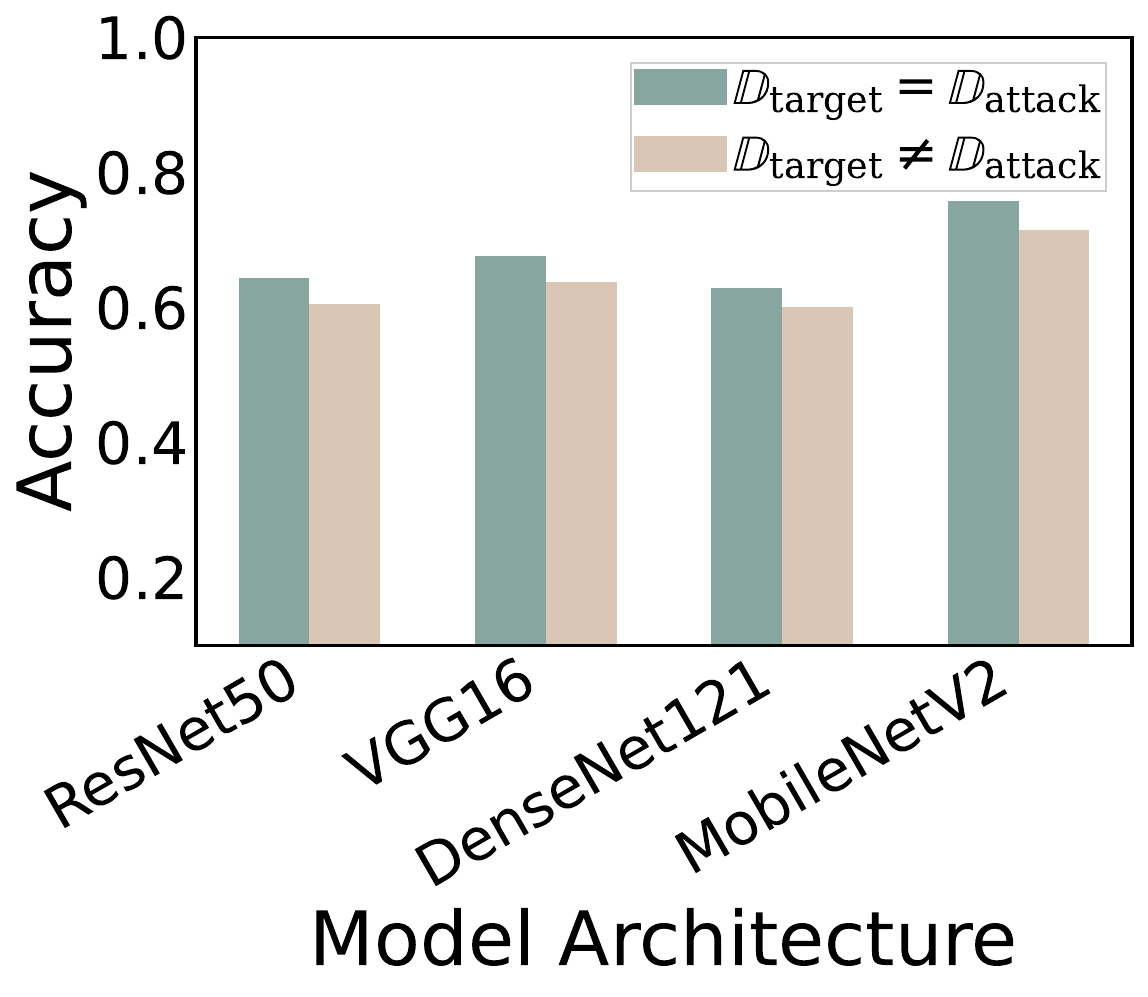}
    \label{fig:cinic_acc}
    }
    \subfigure[TPR @ 0.1\% FPR]
    {
    \includegraphics[width=0.44\linewidth]{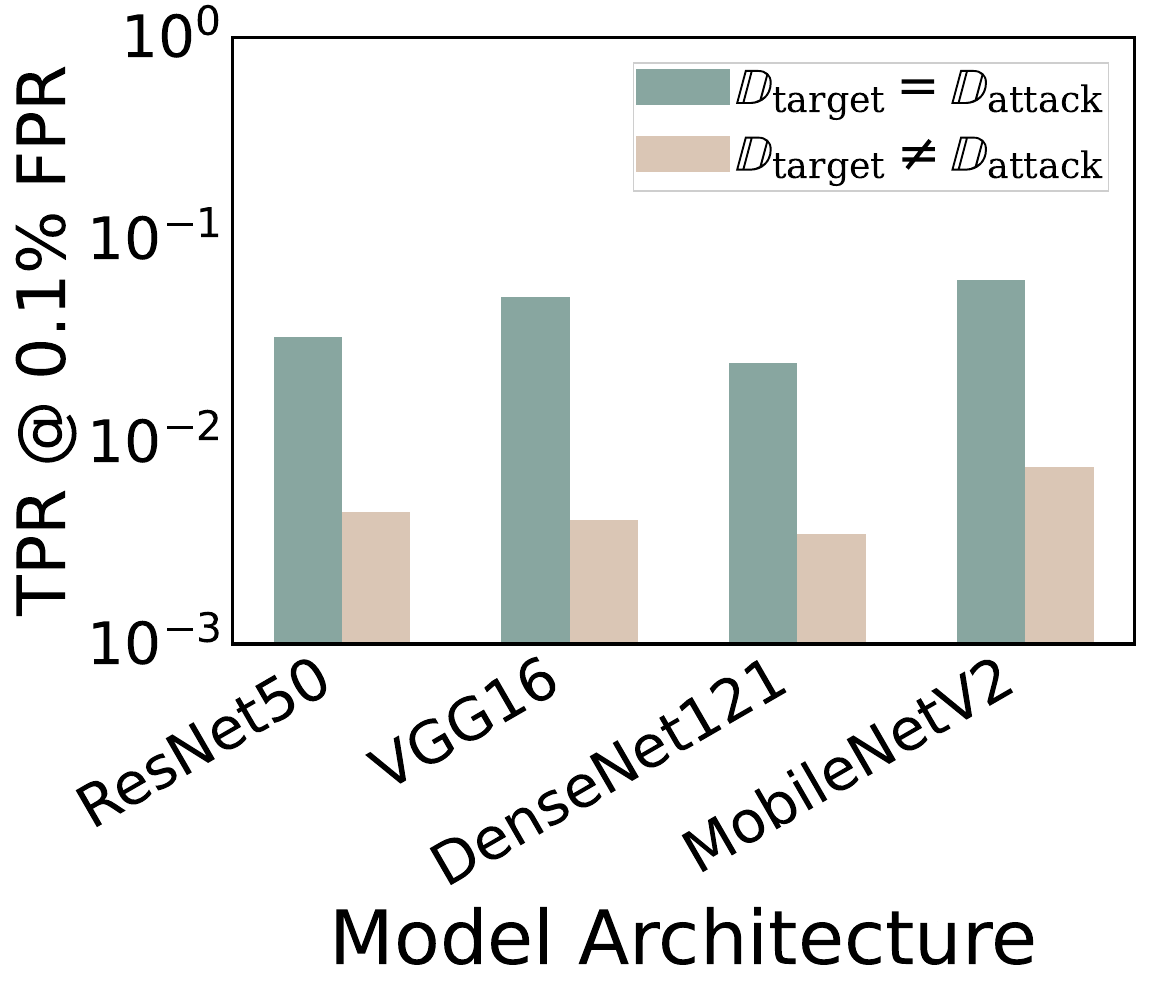}
    \label{fig:cinic_tpr}
    }
    \vspace{-3mm}
    \caption{The impact of distribution shift between the target model training dataset and the attacker's dataset on \myoffline.}
    \label{fig:impact_shift}
\end{figure}

We train the same number of shadow models in both settings, and apply the proposed \myoffline for attack.
\Cref{fig:impact_shift} shows that the distribution shift between the attacker’s and target’s training data leads to a noticeable decrease in performance, particularly for TPR at 0.1\% FPR.
This decrease can be attributed to errors in proxy data selection: when instances in the shadow dataset differ significantly from the query instances, using them as proxies introduces more approximation errors.
However, even in this more challenging setting, \myoffline still outperforms most baselines on balanced accuracy.


\subsection{Additional Investigations}
\label{sec:exp_add}

\begin{table}[t]
\centering
\vspace{-3mm}
\caption{Performance comparison on non-image datasets. \myonline employs LiRA as the base attack. \myonline and \myoffline outperform state-of-the-art MIAs on both attack settings.}
\vspace{-1mm}
\label{tab:nonimage_results}
\resizebox{0.49\textwidth}{!}{
\begin{tabular}{c|cc|cc|cc}
\toprule
\multirow{2}{*}{} & \multicolumn{2}{c|}{\textbf{TPR @ 0.001\%FPR}} & \multicolumn{2}{c|}{\textbf{TPR @ 0.1\%FPR}} & \multicolumn{2}{c}{\textbf{Balanced Accuracy}} \\
 & Purchase & Texas & Purchase & Texas & Purchase & Texas \\
\midrule
\multicolumn{7}{c}{\textbf{Adaptive Setting}} \\
\midrule
LiRA & 1.26\% & 10.28\% & 11.33\% & 24.24\% & 85.95\% & 90.10\% \\ 
RMIA & 0.41\% & 5.37\% & 4.73\% & 10.63\% & 84.62\% & 89.68\% \\ 
RAPID & 0.33\% & 9.64\% & 4.16\% & 15.73\% & 84.05\% & 90.05\% \\
\midrule
\myonline & \textbf{2.08\%} & \textbf{17.34\%} & \textbf{15.64\%} & \textbf{27.38\%} & \textbf{86.36\%} & \textbf{90.92\%} \\
\midrule
\multicolumn{7}{c}{\textbf{Non-Adaptive Setting}} \\
\midrule
LiRA & 0.01\% & 0.04\% & 0.06\% & 0.17\% & 62.12\% & 65.03\% \\ 
RMIA & 0.03\% & 0.12\% & 0.14\% & 0.46\% & 72.36\% & 80.52\% \\ 
RAPID & 0.03\% & 0.10\% & 0.17\% & 0.49\% & 73.72\% & 81.31\% \\
\midrule
\myoffline & \textbf{0.05\%} & \textbf{0.42\%} & \textbf{2.28\%} & \textbf{5.72\%} & \textbf{78.38}\% & \textbf{87.10\%} \\
\bottomrule
\end{tabular}}
\end{table}

\begin{table*}[t]
\centering
\caption{Performance comparison of \myonline and \myonlinegibbs (detailed in~\Cref{sec:exp_add}) on a ResNet50 trained on four benchmark datasets. LiRA is used as the base attack for both methods. The maximum number of iterations is set to 10.}
\vspace{-1mm}
\label{tab:cmia_enhance}
\resizebox{0.9\textwidth}{!}{
\begin{tabular}{l|*{12}{c}}
\toprule
 & \multicolumn{4}{c}{\textbf{TPR @ 0.001\% FPR}} & \multicolumn{4}{c}{\textbf{TPR @ 0.1\% FPR}} & \multicolumn{4}{c}{\textbf{Balanced Accuracy}} \\
\cmidrule(lr){2-5}\cmidrule(lr){6-9}\cmidrule(lr){10-13}
&  MNIST & FMNIST & C-10 & C-100 & MNIST & FMNIST & C-10 & C-100 & MNIST & FMNIST & C-10 & C-100 \\
\midrule
\myonline & 0.77\% & 4.42\% & 3.86\% & 36.74\% & 2.10\% & 8.34\% & 9.71\% & 45.37\% & 52.67\% & 60.91\% & 63.83\% & 84.89\% \\
\myonlinegibbs & 0.35\% & 3.13\% & 2.85\% & 29.52\% & 1.80\% & 7.82\% & 8.98\% & 40.05\% & 52.04\% & 59.58\% & 63.29\% & 84.02\% \\
\bottomrule
\end{tabular}}
\end{table*}

\begin{table}[t]
\centering
\vspace{-3mm}
\caption{Impact of joint membership inference on MNIST.}
\vspace{-1mm}
\label{tab:query_set_change}
\resizebox{0.5\textwidth}{!}{
\begin{tabular}{c c|c c}
\toprule
\textbf{\# Queries} & \textbf{\#Instance per query} & \textbf{TPR @ 0.001\% FPR} & \textbf{TPR @ 0.1\% FPR} \\
\midrule
1 & 60,000 & 0.77\% & 2.10\%  \\
6 & 10,000 & 0.69\% & 2.05\%  \\
60 & 1,000 & 0.45\% & 1.70\%  \\
600 & 100 & 0.27\% & 1.45\%  \\
60,000 & 1 & 0.12\% & 1.23\%  \\
\bottomrule
\end{tabular}}
\end{table}

\begin{figure}[t]
    \centering
    \vspace{-3mm}
    \subfigure[MNIST]
    {
    \includegraphics[width=0.44\linewidth]{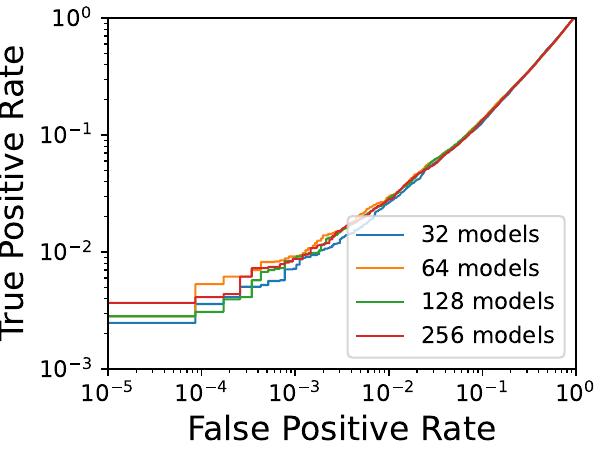}
    \label{fig:pmia_shadow_curve_resnet_mnist}
    }
    \subfigure[CIFAR-100]
    {
    \includegraphics[width=0.44\linewidth]{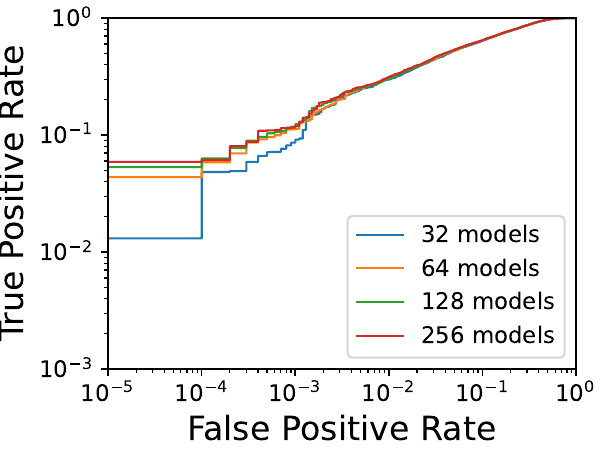}
    \label{fig:pmia_shadow_curve_resnet_cifar100}
    }
    \vspace{-3mm}
    \caption{The impact of the number of shadow models for \myoffline. The y-axis range is adjusted to enhance visibility.}
    \label{fig:impact_pmia_shadow}
    \vspace{-3mm}
\end{figure}

\begin{table}[t]
\centering
\vspace{-1mm}
\caption{Effectiveness of using DP-SGD against \myonline with different privacy budgets trained on CIFAR-10. $\sigma$ is the noise multiplier and $\varepsilon$ is the privacy budget for DPSGD.}
\vspace{-1mm}
\label{tab:cmia_dpsgd}
\resizebox{0.49\textwidth}{!}{
\begin{tabular}{ccccccc}
\toprule
& \multicolumn{2}{c}{Clipping norm $C = 10$} & \multirow{2}{*}{\text{Model Acc}} & \multicolumn{2}{c}{\text{TPR @ 0.1\% FPR (\%)}} \\
 \cmidrule(lr){2-3}\cmidrule(lr){5-6}
& $\sigma$ & $\varepsilon$ & & \text{Base (LiRA)} & \text{CMIA} \\
\midrule
\text{No defense} & - & - & 90.05\% & 8.45\% & \textbf{9.71\%} \\
\midrule
 & 0 & $\infty$ & 81.86\% & 1.36\% & \textbf{2.80\%} \\
 & 0.2 & $>1000$ & 48.62\% & 0.29\% & \textbf{0.35\%} \\
\text{DP-SGD} & 0.5 & 31 & 36.45\% & 0.21\% & \textbf{0.29\%} \\
 & 1.0 & 4 & 30.48\% & 0.14\% & \textbf{0.18\%} \\
\bottomrule
\end{tabular}}
\vspace{-3mm}
\end{table}

\mypara{Attack on Non-image Datasets}
We also evaluate the attack performance on two non-image datasets to demonstrate the generality of the proposed attacks. 
Specifically, we train multilayer perceptrons (MLPs) on the Purchase and Texas datasets~\cite{sp17miashokri}. 
For \myonline, we use LiRA as the base model and apply the cascading framework. 
For \myoffline, we employ the Wasserstein distance to select the top 10 proxy models for each query instance. 
The attack performance under both adaptive and non-adaptive settings is presented in Table~\ref{tab:nonimage_results}. 
As shown, both \myonline and \myoffline consistently outperform existing attacks, particularly in TPR at low FPR.
It is worth mentioning that the performance of all evaluated MIAs is notably lower on these non-image datasets compared to image datasets such as CIFAR-100, despite both having 100 classes. 
We attribute this to the smaller generalization gap in models trained on non-image datasets. 
For instance, as shown in~\Cref{tab:dataset_acc_offline}, the Purchase model achieves training and validation accuracies of 0.96 and 0.80, respectively. 
In contrast, the CIFAR-100 model shows a much larger gap, with a training accuracy of 0.92 and a validation accuracy of only 0.66.
Since MIAs typically exploit overfitting~\cite{sp22lira}, they are less effective against better-generalized models, as observed in the non-image datasets.

\mypara{Impact of Join MIA}
We experiment to illustrate the impact of joint membership dependence on attack performance. Specifically, we use the MNIST dataset and partition the 60,000 membership queries into the following scenarios:
(a) a single set of 60,000 instances,
(b) 6 sets with 10,000 instances each,
(c) 60 sets,
(d) 600 sets, and
(e) 60,000 individual queries (\ie no dependence is exploited).
We employ \myonline with LiRA as the base attack and evaluate the overall attack performance under each scenario. As shown in~\Cref{tab:query_set_change}, the attack performance consistently degrades from (a) to (e), demonstrating the importance of modeling joint membership dependence for achieving strong performance.

\mypara{Potential Improvements of \myonline}
One potential improvement of \myonline is to adopt a sampling strategy that better aligns with Gibbs sampling. 
Specifically, for each iteration, we generate conditional shadow datasets by sampling each instance with a probability proportional to its membership score. 
We implemented this approach (called \myonlinegibbs) and compared it with \myonline using the same computational budget (\ie a maximum of 10 iterations). 
As shown in~\Cref{tab:cmia_enhance}, \myonline consistently outperforms \myonlinegibbs across four image datasets.
We attribute this to the limited computations: while \myonlinegibbs aligns more closely with the spirit of Gibbs sampling, it requires significantly more iterations to be effective. 

\mypara{Impact of Number of Shadow Models for \myoffline}
In~\Cref{fig:impact_pmia_shadow}, we vary the number of shadow models from 32 to 256 and find that \myoffline is quite robust: the performance drop is less significant compared to LiRA. 
This robustness stems from using a set of proxy data to approximate the likelihood. 
By leveraging more data to approximate the confidence score distribution for normal distribution estimation, \myoffline remains effective even when trained with fewer shadow models.

\begin{table}[t]
\centering
\vspace{-3mm}
\caption{Attack performance of $\myoffline$ against a ResNet50 model trained on CIFAR-10 using DP-SGD.}
\vspace{-1mm}
\label{tab:pmia_dpsgd}
\resizebox{0.43\textwidth}{!}{
\begin{tabular}{ccccccc}
\toprule
 & \multicolumn{2}{c}{\text{TPR @ 0.1\% FPR (\%)}}  & \multicolumn{2}{c}{Balanced Accuracy} \\
 \cmidrule(lr){2-3}\cmidrule(lr){4-5}
 & $\sigma=0.2$ & $\sigma=0.5$ & $\sigma=0.2$ & $\sigma=0.5$ \\
\midrule
 LOSS & 0.02\% & 0.01\% &  51.36\% & 51.20\% \\
 Entropy & 0.05\% & 0.04\% &  51.77\% & 51.62\% \\
Calibration & 0.10\% & 0.12\% & 51.72\% & 51.64\% \\
Attack-R & 0.00\% & 0.00\% &  51.79\% & 51.60\% \\
LiRA & 0.13\% & 0.11\% &  50.13\% & 50.42\% \\
Canary & 0.12\% & 0.10\% &  50.34\% & 50.38\% \\
RMIA & 0.16\% & 0.12\% &  51.77\% & \textbf{52.10\%} \\
RAPID & 0.11\% & 0.13\% &  \textbf{51.81\%} & 51.96\% \\
 \midrule
\myoffline & \textbf{0.24\%} & \textbf{0.20\%}  & 51.55\% & 51.83\% \\
\bottomrule
\end{tabular}}
\vspace{-3mm}
\end{table}

\mypara{Attack Against DP-SGD}
Machine learning with differential privacy~\cite{ccs16dpsgd} is an effective defense mechanism against privacy attacks, including MIAs. 
We follow~\cite{sp22lira,ccs24rapid} and assess the effectiveness of the DP-SGD against our attack. 
Specifically, we fix the clipping norm $C$ as 10 and test \myonline and \myoffline on a ResNet50 model trained on CIFAR-10.
As shown in~\Cref{tab:cmia_dpsgd}, even when only the gradient norm is applied without adding noise, the model's accuracy and the effectiveness of our attack are significantly reduced. 
We observe that as the noise level increases, the improvement of \myonline compared to the base attack diminishes. 
This is because \myonline relies on identifying highly probable samples to perform cascading attacks on remaining instances.
When selecting reliable anchors becomes difficult, the benefit of the cascading framework decreases. 
We also evaluate the impact of DP-SGD in the non-adaptive setting in~\Cref{tab:pmia_dpsgd} and observe similar patterns: as the privacy level increases, attack performance drops for all methods, and the gap between attacks narrows.
Nevertheless, \myoffline still outperforms existing MIAs in the low false-positive regime.

\section{Related Work}
\label{sec:related}

Neural networks have been shown to be vulnerable to leaking sensitive information about their training data.
A variety of attacks~\cite{usenix21llm_extract, ccs15inversion,ccs18property} have been proposed to quantify the extent of data leakage and assess the associated privacy risks. 
In this paper, we focus on the membership inference attack (MIA)\cite{sp17miashokri}, which predicts whether specific instances were included in the target model's training set. 
Closedly connected to Differential Privacy~\cite{dwork_dp,ccs13membership},
MIA has become a widely used tool to audit training data privacy of ML models~\cite{arxiv20privacy_meter,tensorflow_mia,sp21auditing,nips20auditing}.
The first MIA against ML models was introduced by~\cite{sp17miashokri}, which also proposed the technique of shadow training. 
MIAs and shadow training have since been extended to various scenarios, including white-box~\cite{usenix20stolen, sp19comprehensive, icml19white}, black-box~\cite{long2018understanding,liu2022ml,usenix21systematic, ndss19ml-leak, arixv20revisiting, codaspy21mia}, label-only access~\cite{ccs21label, icml21label}, and model distillation~\cite{nips23distill}. 
Depending on when the adversary trains shadow models, MIAs can be categorized into two groups:

\mypara{Adaptive MIAs}
In the adaptive setting, the adversary is allowed to perform instance-by-instance analysis of the model’s behavior by training additional predictors~\cite{sp19comprehensive} or conducting hypothesis testing~\cite{sp22lira,iclr23canary}, achieving state-of-the-art attack performance. 
However, these MIAs typically determine the membership of instances independently, which fails to consider the inherent dependencies between instances.

\mypara{Non-Adaptive MIAs}
Some studies~\cite{icml24rmia, nips23quantile} argue that adaptive attacks are inefficient due to the need to train shadow models for every batch of queries. 
A series of non-adaptive MIAs have been proposed to relax this assumption by training shadow models (or forgoing shadow training entirely) before accessing the query set.
Several techniques have been introduced for non-adaptive MIAs, such as score functions~\cite{csf18privacy,usenix21systematic}, difficulty calibration~\cite{icml19white,iclr22calibrate, ccs24rapid}, loss trajectory~\cite{ccs24seqmia, ccs22trajectory}, quantile regression~\cite{nips23quantile}, and hypothesis testing~\cite{ccs22enhanced, icml24rmia}. 
However, these attacks deviate from the membership posterior odds ratio test, leading to suboptimal performance.

Recent research~\cite{sp22lira,eurosp20pragmatic} has emphasized evaluating attacks by calculating their true positive rate at a significantly low false positive rate, and most existing MIAs perform poorly under this evaluation paradigm. 
In~\Cref{sec:nonadaptive_theory} we provide theoretical insights into the rationale behind these evaluation metrics.
While some literature~\cite{nips22onion} explores instance vulnerabilities to MIAs, it does not propose an attack that leverages them. 
In contrast, we theoretically analyze membership dependencies and introduce a framework that exploits these vulnerabilities and dependencies to mount a more powerful attack.
The dependencies we investigate differ from group-level MIAs \cite{li2022user, usenix24llm}, which consider the inference target as a set of instances. 
Instead, our focus lies in exploring the membership dependencies between inference targets.






\section{Discussion \& Conclusion}
\label{sec:conclusion}

In this paper, we provide a new formulation of the MIA game and categorize existing MIAs into two categories: adaptive and non-adaptive.
Guided by theoretical analyses of joint membership estimation and the posterior odds test, we propose two attacks (\ie \myonline and \myoffline), one for each setting. 
Extensive experiments demonstrate the efficacy of the proposed attacks.
Our work has several limitations. 
First, the conditional shadow training strategy of \myonline limits its applicability for attacks that do not rely on shadow training. 
Second, our evaluations mainly focus on classification models, leaving the performance on generative models~\cite{arxiv17logan,icml23diffusion,arxiv24llmmia} unexamined. 
Nevertheless, our work provides both theoretical insights and practical approaches, offering new directions for MIAs.



\section*{Ethics Considerations}

This paper focuses on membership inference attacks using publicly available datasets that do not contain personally identifiable information. 
Although our proposed attacks demonstrate improved attack performance, we show that established defenses like DP-SGD remain an effective mitigation strategy.
We encourage the research community to use our findings in a manner that promotes ethical research and enhances privacy protections for users and data subjects.

\section*{Acknowledgment}
We thank the anonymous reviewers for their constructive comments.  This work was funded in part by the National Science Foundation (NSF) under awards CNS-2207204, CAREER IIS-1943364, CNS-2212160, and IIS-2229876, as well as by Cisco. Any opinions, findings, and conclusions or recommendations expressed in this material are those of the authors and do not necessarily reflect the views of the sponsors.



%
\bibliographystyle{IEEEtran}
\bibliography{ref}

\begin{thebibliography}{10}
\providecommand{\url}[1]{#1}
\csname url@samestyle\endcsname
\providecommand{\newblock}{\relax}
\providecommand{\bibinfo}[2]{#2}
\providecommand{\BIBentrySTDinterwordspacing}{\spaceskip=0pt\relax}
\providecommand{\BIBentryALTinterwordstretchfactor}{4}
\providecommand{\BIBentryALTinterwordspacing}{\spaceskip=\fontdimen2\font plus
\BIBentryALTinterwordstretchfactor\fontdimen3\font minus \fontdimen4\font\relax}
\providecommand{\BIBforeignlanguage}[2]{{%
\expandafter\ifx\csname l@#1\endcsname\relax
\typeout{** WARNING: IEEEtran.bst: No hyphenation pattern has been}%
\typeout{** loaded for the language `#1'. Using the pattern for}%
\typeout{** the default language instead.}%
\else
\language=\csname l@#1\endcsname
\fi
#2}}
\providecommand{\BIBdecl}{\relax}
\BIBdecl

\bibitem{sp17miashokri}
R.~Shokri, M.~Stronati, C.~Song, and V.~Shmatikov, ``{Membership inference attacks against machine learning models},'' in \emph{2017 IEEE symposium on security and privacy (SP)}, 2017, pp. 3--18.

\bibitem{dwork_dp}
C.~Dwork, ``{Differential Privacy},'' in \emph{Automata, Languages and Programming}, 2006, pp. 1--12.

\bibitem{arxiv20privacy_meter}
S.~K. Murakonda and R.~Shokri, ``{{ML} privacy meter: Aiding regulatory compliance by quantifying the privacy risks of machine learning},'' \emph{arXiv preprint arXiv:2007.09339}, 2020.

\bibitem{tensorflow_mia}
\BIBentryALTinterwordspacing
S.~Song and D.~Marn, ``{Introducing a New Privacy Testing Library in TensorFlow},'' 2022. [Online]. Available: \url{https://blog.tensorflow.org/2020/06/introducing-new-privacy-testing-library.html}
\BIBentrySTDinterwordspacing

\bibitem{usenix21llm_extract}
N.~Carlini, F.~Tramer, E.~Wallace, M.~Jagielski, A.~Herbert-Voss, K.~Lee, A.~Roberts, T.~Brown, D.~Song, U.~Erlingsson \emph{et~al.}, ``{Extracting training data from large language models},'' in \emph{30th USENIX security symposium (USENIX Security 21)}, 2021, pp. 2633--2650.

\bibitem{usenix23diffusion_extract}
N.~Carlini, J.~Hayes, M.~Nasr, M.~Jagielski, V.~Sehwag, F.~Tramer, B.~Balle, D.~Ippolito, and E.~Wallace, ``{Extracting training data from diffusion models},'' in \emph{32nd USENIX Security Symposium (USENIX Security 23)}, 2023, pp. 5253--5270.

\bibitem{iclr23canary}
Y.~Wen, A.~Bansal, H.~Kazemi, E.~Borgnia, M.~Goldblum, J.~Geiping, and T.~Goldstein, ``{Canary in a Coalmine: Better Membership Inference with Ensembled Adversarial Queries},'' in \emph{The Eleventh International Conference on Learning Representations}, 2023.

\bibitem{ccs24rapid}
Y.~He, B.~Li, Y.~Wang, M.~Yang, J.~Wang, H.~Hu, and X.~Zhao, ``{Is Difficulty Calibration All We Need? Towards More Practical Membership Inference Attacks},'' in \emph{Proceedings of the 2024 on ACM SIGSAC Conference on Computer and Communications Security}, 2024, pp. 1226--1240.

\bibitem{sp22lira}
N.~Carlini, S.~Chien, M.~Nasr, S.~Song, A.~Terzis, and F.~Tramer, ``{Membership inference attacks from first principles},'' in \emph{2022 IEEE symposium on security and privacy (SP)}.\hskip 1em plus 0.5em minus 0.4em\relax IEEE, 2022, pp. 1897--1914.

\bibitem{sp19comprehensive}
M.~Nasr, R.~Shokri, and A.~Houmansadr, ``{Comprehensive privacy analysis of deep learning: Passive and active white-box inference attacks against centralized and federated learning},'' in \emph{2019 IEEE symposium on security and privacy (SP)}, 2019, pp. 739--753.

\bibitem{codaspy21mia}
J.~Li, N.~Li, and B.~Ribeiro, ``{Membership Inference Attacks and Defenses in Classification Models},'' in \emph{Eleventh {ACM} Conference on Data and Application Security and Privacy}, 2021, pp. 5--16.

\bibitem{nips23quantile}
M.~Bertran, S.~Tang, A.~Roth, M.~Kearns, J.~H. Morgenstern, and S.~Z. Wu, ``{Scalable membership inference attacks via quantile regression},'' in \emph{Advances in Neural Information Processing Systems}, 2023, pp. 314--330.

\bibitem{ccs24seqmia}
H.~Li, Z.~Li, S.~Wu, C.~Hu, Y.~Ye, M.~Zhang, D.~Feng, and Y.~Zhang, ``{SeqMIA: Sequential-metric based membership inference attack},'' in \emph{Proceedings of the 2024 on ACM SIGSAC Conference on Computer and Communications Security}, 2024, pp. 3496--3510.

\bibitem{icml24rmia}
S.~Zarifzadeh, P.~Liu, and R.~Shokri, ``{Low-Cost High-Power Membership Inference Attacks},'' in \emph{International Conference on Machine Learning}, 2024, pp. 58\,244--58\,282.

\bibitem{csf18privacy}
S.~Yeom, I.~Giacomelli, M.~Fredrikson, and S.~Jha, ``{Privacy risk in machine learning: Analyzing the connection to overfitting},'' in \emph{2018 IEEE 31st computer security foundations symposium (CSF)}, 2018, pp. 268--282.

\bibitem{arixv20revisiting}
B.~Jayaraman, L.~Wang, K.~Knipmeyer, Q.~Gu, and D.~Evans, ``{Revisiting membership inference under realistic assumptions},'' \emph{arXiv preprint arXiv:2005.10881}, 2020.

\bibitem{albers2003online}
S.~Albers, ``Online algorithms: a survey,'' \emph{Mathematical Programming}, vol.~97, no.~1, pp. 3--26, 2003.

\bibitem{luby96pseudorandomness}
M.~Luby, \emph{{Pseudorandomness and cryptographic applications}}.\hskip 1em plus 0.5em minus 0.4em\relax Princeton University Press, 1996, vol.~1.

\bibitem{98adaptive}
D.~Bleichenbacher, ``{Chosen ciphertext attacks against protocols based on the RSA encryption standard PKCS\# 1},'' in \emph{Advances in Cryptology—CRYPTO'98: 18th Annual International Cryptology Conference Santa Barbara, California, USA August 23--27, 1998 Proceedings 18}.\hskip 1em plus 0.5em minus 0.4em\relax Springer, 1998, pp. 1--12.

\bibitem{gelfand1990sampling}
A.~E. Gelfand and A.~F. Smith, ``{Sampling-based approaches to calculating marginal densities},'' \emph{Journal of the American statistical association}, vol.~85, no. 410, pp. 398--409, 1990.

\bibitem{roberts2015surprising}
G.~O. Roberts and J.~S. Rosenthal, ``{Surprising convergence properties of some simple Gibbs samplers under various scans},'' \emph{International Journal of Statistics and Probability}, vol.~5, no.~1, pp. 51--60, 2015.

\bibitem{he2016scan}
B.~D. He, C.~M. De~Sa, I.~Mitliagkas, and C.~R{\'e}, ``{Scan order in Gibbs sampling: Models in which it matters and bounds on how much},'' \emph{Advances in neural information processing systems}, vol.~29, 2016.

\bibitem{latuszynski2013adaptive}
K.~{\L}atuszy{\'n}ski, G.~O. Roberts, and J.~S. Rosenthal, ``{Adaptive {G}ibbs samplers and related {MCMC} methods},'' \emph{The Annals of Applied Probability}, vol.~23, no.~1, pp. 66--99, 2013.

\bibitem{icml19white}
A.~Sablayrolles, M.~Douze, C.~Schmid, Y.~Ollivier, and H.~J{\'e}gou, ``{White-box vs black-box: Bayes optimal strategies for membership inference},'' in \emph{International Conference on Machine Learning}, 2019, pp. 5558--5567.

\bibitem{icml21clip}
A.~Radford, J.~W. Kim, C.~Hallacy, A.~Ramesh, G.~Goh, S.~Agarwal, G.~Sastry, A.~Askell, P.~Mishkin, J.~Clark \emph{et~al.}, ``{Learning transferable visual models from natural language supervision},'' in \emph{International conference on machine learning}, 2021, pp. 8748--8763.

\bibitem{24faiss}
M.~Douze, A.~Guzhva, C.~Deng, J.~Johnson, G.~Szilvasy, P.-E. Mazaré, M.~Lomeli, L.~Hosseini, and H.~Jégou, ``{The Faiss library},'' \emph{arXiv preprint arXiv:2401.08281}, 2024.

\bibitem{mnist}
Y.~LeCun, C.~Cortes, and C.~J. Burges, ``{The MNIST database of handwritten digits},'' \url{http://yann.lecun.com/exdb/mnist/}, 1998.

\bibitem{fmnist}
H.~Xiao, K.~Rasul, and R.~Vollgraf, ``{Fashion-mnist: a novel image dataset for benchmarking machine learning algorithms},'' \emph{arXiv preprint arXiv:1708.07747}, 2017.

\bibitem{cifar10}
A.~Krizhevsky and G.~Hinton, ``{Learning multiple layers of features from tiny images},'' 2009.

\bibitem{cvpr16resnet}
K.~He, X.~Zhang, S.~Ren, and J.~Sun, ``{Deep residual learning for image recognition},'' in \emph{Proceedings of the IEEE conference on computer vision and pattern recognition}, 2016, pp. 770--778.

\bibitem{vgg}
K.~Simonyan and A.~Zisserman, ``{Very deep convolutional networks for large-scale image recognition},'' \emph{arXiv preprint arXiv:1409.1556}, 2014.

\bibitem{densenet}
G.~Huang, Z.~Liu, L.~Van Der~Maaten, and K.~Q. Weinberger, ``{Densely connected convolutional networks},'' in \emph{Proceedings of the IEEE conference on computer vision and pattern recognition}, 2017, pp. 4700--4708.

\bibitem{mobilenetv2}
M.~Sandler, A.~Howard, M.~Zhu, A.~Zhmoginov, and L.-C. Chen, ``{Mobilenetv2: Inverted residuals and linear bottlenecks},'' in \emph{Proceedings of the IEEE conference on computer vision and pattern recognition}, 2018, pp. 4510--4520.

\bibitem{krogh1991simple}
A.~Krogh and J.~Hertz, ``{A simple weight decay can improve generalization},'' \emph{Advances in neural information processing systems}, vol.~4, 1991.

\bibitem{loshchilov16sgdr}
I.~Loshchilov and F.~Hutter, ``{SGDR: Stochastic gradient descent with warm restarts},'' \emph{arXiv preprint arXiv:1608.03983}, 2016.

\bibitem{aaai20random}
Z.~Zhong, L.~Zheng, G.~Kang, S.~Li, and Y.~Yang, ``{Random erasing data augmentation},'' in \emph{Proceedings of the AAAI conference on artificial intelligence}, 2020, pp. 13\,001--13\,008.

\bibitem{iclr22calibrate}
L.~Watson, C.~Guo, G.~Cormode, and A.~Sablayrolles, ``{On the Importance of Difficulty Calibration in Membership Inference Attacks},'' in \emph{The Tenth International Conference on Learning Representations}, 2022.

\bibitem{ccs22enhanced}
J.~Ye, A.~Maddi, S.~K. Murakonda, V.~Bindschaedler, and R.~Shokri, ``{Enhanced membership inference attacks against machine learning models},'' in \emph{Proceedings of the 2022 ACM SIGSAC Conference on Computer and Communications Security}, 2022, pp. 3093--3106.

\bibitem{usenix21systematic}
L.~Song and P.~Mittal, ``{Systematic evaluation of privacy risks of machine learning models},'' in \emph{30th USENIX Security Symposium (USENIX Security 21)}, 2021, pp. 2615--2632.

\bibitem{nature09genomic}
S.~Sankararaman, G.~Obozinski, M.~I. Jordan, and E.~Halperin, ``{Genomic privacy and limits of individual detection in a pool},'' \emph{Nature genetics}, vol.~41, no.~9, pp. 965--967, 2009.

\bibitem{cvpr09imagenet}
J.~Deng, W.~Dong, R.~Socher, L.-J. Li, K.~Li, and L.~Fei-Fei, ``{Imagenet: A large-scale hierarchical image database},'' in \emph{2009 IEEE conference on computer vision and pattern recognition}.\hskip 1em plus 0.5em minus 0.4em\relax Ieee, 2009, pp. 248--255.

\bibitem{arxiv18cinic}
L.~N. Darlow, E.~J. Crowley, A.~Antoniou, and A.~J. Storkey, ``{Cinic-10 is not imagenet or cifar-10},'' \emph{arXiv preprint arXiv:1810.03505}, 2018.

\bibitem{ccs16dpsgd}
M.~Abadi, A.~Chu, I.~Goodfellow, H.~B. McMahan, I.~Mironov, K.~Talwar, and L.~Zhang, ``{Deep learning with differential privacy},'' in \emph{Proceedings of the 2016 ACM SIGSAC conference on computer and communications security}, 2016, pp. 308--318.

\bibitem{ccs15inversion}
M.~Fredrikson, S.~Jha, and T.~Ristenpart, ``{Model inversion attacks that exploit confidence information and basic countermeasures},'' in \emph{Proceedings of the 22nd ACM SIGSAC conference on computer and communications security}, 2015, pp. 1322--1333.

\bibitem{ccs18property}
K.~Ganju, Q.~Wang, W.~Yang, C.~A. Gunter, and N.~Borisov, ``{Property inference attacks on fully connected neural networks using permutation invariant representations},'' in \emph{Proceedings of the 2018 ACM SIGSAC conference on computer and communications security}, 2018, pp. 619--633.

\bibitem{ccs13membership}
N.~Li, W.~Qardaji, D.~Su, Y.~Wu, and W.~Yang, ``{Membership privacy: A unifying framework for privacy definitions},'' in \emph{Proceedings of the 2013 ACM SIGSAC conference on Computer \& communications security}, 2013, pp. 889--900.

\bibitem{sp21auditing}
M.~Nasr, S.~Songi, A.~Thakurta, N.~Papernot, and N.~Carlin, ``{Adversary instantiation: Lower bounds for differentially private machine learning},'' in \emph{2021 IEEE Symposium on security and privacy (SP)}, 2021, pp. 866--882.

\bibitem{nips20auditing}
M.~Jagielski, J.~Ullman, and A.~Oprea, ``{Auditing differentially private machine learning: How private is private sgd?}'' in \emph{Advances in Neural Information Processing Systems}, vol.~33, 2020, pp. 22\,205--22\,216.

\bibitem{usenix20stolen}
K.~Leino and M.~Fredrikson, ``{Stolen memories: Leveraging model memorization for calibrated White-Box membership inference},'' in \emph{29th USENIX security symposium}, 2020, pp. 1605--1622.

\bibitem{long2018understanding}
Y.~Long, V.~Bindschaedler, L.~Wang, D.~Bu, X.~Wang, H.~Tang, C.~A. Gunter, and K.~Chen, ``{Understanding membership inferences on well-generalized learning models},'' \emph{arXiv preprint arXiv:1802.04889}, 2018.

\bibitem{liu2022ml}
Y.~Liu, R.~Wen, X.~He, A.~Salem, Z.~Zhang, M.~Backes, E.~De~Cristofaro, M.~Fritz, and Y.~Zhang, ``{ML-Doctor: Holistic risk assessment of inference attacks against machine learning models},'' in \emph{31st USENIX Security Symposium (USENIX Security 22)}, 2022, pp. 4525--4542.

\bibitem{ndss19ml-leak}
A.~Salem, Y.~Zhang, M.~Humbert, P.~Berrang, M.~Fritz, and M.~Backes, ``{ML-Leaks: Model and Data Independent Membership Inference Attacks and Defenses on Machine Learning Models},'' in \emph{26th Annual Network and Distributed System Security Symposium}, 2019.

\bibitem{ccs21label}
Z.~Li and Y.~Zhang, ``{Membership leakage in label-only exposures},'' in \emph{Proceedings of the 2021 ACM SIGSAC Conference on Computer and Communications Security}, 2021, pp. 880--895.

\bibitem{icml21label}
C.~A. Choquette-Choo, F.~Tramer, N.~Carlini, and N.~Papernot, ``{Label-only membership inference attacks},'' in \emph{International conference on machine learning}, 2021, pp. 1964--1974.

\bibitem{nips23distill}
M.~Jagielski, M.~Nasr, K.~Lee, C.~A. Choquette-Choo, N.~Carlini, and F.~Tramer, ``{Students parrot their teachers: Membership inference on model distillation},'' in \emph{Advances in Neural Information Processing Systems}, 2023, pp. 44\,382--44\,397.

\bibitem{ccs22trajectory}
Y.~Liu, Z.~Zhao, M.~Backes, and Y.~Zhang, ``{Membership Inference Attacks by Exploiting Loss Trajectory},'' in \emph{Proceedings of the 2022 {ACM} {SIGSAC} Conference on Computer and Communications Security}, H.~Yin, A.~Stavrou, C.~Cremers, and E.~Shi, Eds., 2022, pp. 2085--2098.

\bibitem{eurosp20pragmatic}
Y.~Long, L.~Wang, D.~Bu, V.~Bindschaedler, X.~Wang, H.~Tang, C.~A. Gunter, and K.~Chen, ``{A pragmatic approach to membership inferences on machine learning models},'' in \emph{2020 IEEE European Symposium on Security and Privacy (EuroS\&P)}, 2020, pp. 521--534.

\bibitem{nips22onion}
N.~Carlini, M.~Jagielski, C.~Zhang, N.~Papernot, A.~Terzis, and F.~Tramer, ``{The privacy onion effect: Memorization is relative},'' in \emph{Advances in Neural Information Processing Systems}, 2022, pp. 13\,263--13\,276.

\bibitem{li2022user}
G.~Li, S.~Rezaei, and X.~Liu, ``{User-Level Membership Inference Attack against Metric Embedding Learning},'' in \emph{ICLR 2022 Workshop on PAIR2Struct: Privacy, Accountability, Interpretability, Robustness, Reasoning on Structured Data}, 2022.

\bibitem{usenix24llm}
M.~Meeus, S.~Jain, M.~Rei, and Y.-A. de~Montjoye, ``Did the neurons read your book? document-level membership inference for large language models,'' in \emph{33rd USENIX Security Symposium}, 2024, pp. 2369--2385.

\bibitem{arxiv17logan}
J.~Hayes, L.~Melis, G.~Danezis, and E.~De~Cristofaro, ``{Logan: Membership inference attacks against generative models},'' \emph{arXiv preprint arXiv:1705.07663}, 2017.

\bibitem{icml23diffusion}
J.~Duan, F.~Kong, S.~Wang, X.~Shi, and K.~Xu, ``{Are diffusion models vulnerable to membership inference attacks?}'' in \emph{International Conference on Machine Learning}.\hskip 1em plus 0.5em minus 0.4em\relax PMLR, 2023, pp. 8717--8730.

\bibitem{arxiv24llmmia}
M.~Duan, A.~Suri, N.~Mireshghallah, S.~Min, W.~Shi, L.~Zettlemoyer, Y.~Tsvetkov, Y.~Choi, D.~Evans, and H.~Hajishirzi, ``{Do membership inference attacks work on large language models?}'' \emph{arXiv preprint arXiv:2402.07841}, 2024.

\bibitem{Williams1991}
D.~Williams, \emph{{Probability with martingales}}.\hskip 1em plus 0.5em minus 0.4em\relax Cambridge university press, 1991.

\bibitem{Tierney1994}
L.~Tierney, ``{Markov chains for exploring posterior distributions},'' \emph{the Annals of Statistics}, pp. 1701--1728, 1994.

\bibitem{Liu1996}
J.~S. Liu, ``{Peskun's theorem and a modified discrete-state Gibbs sampler.}'' \emph{Biometrika}, vol.~83, no.~3, 1996.

\bibitem{Roberts2004}
G.~O. Roberts and J.~S. Rosenthal, ``{General state space Markov chains and MCMC algorithms},'' \emph{Probability Surveys}, 2004.

\bibitem{MeynTweedie2012}
S.~P. Meyn and R.~L. Tweedie, \emph{{Markov chains and stochastic stability}}.\hskip 1em plus 0.5em minus 0.4em\relax Springer Science \& Business Media, 2012.

\bibitem{Douc2018}
R.~Douc, E.~Moulines, P.~Priouret, P.~Soulier, R.~Douc, E.~Moulines, P.~Priouret, and P.~Soulier, \emph{{Markov chains: Basic definitions}}.\hskip 1em plus 0.5em minus 0.4em\relax Springer, 2018.

\bibitem{Meyn2009}
S.~Meyn and R.~L. Tweedie, \emph{{Markov Chains and Stochastic Stability}}.\hskip 1em plus 0.5em minus 0.4em\relax Cambridge University Press, 2009.

\end{thebibliography}

\appendix
\section{Missing Proofs}

\subsection{Convergence of Joint MIA Gibbs Sampling}
\label{appendix:converge_gibbs}



\begin{restatable}[Convergence of Joint MIA Gibbs Sampling]{theorem}{MIAGibbsConvergence}
\label{thm:MIAGibbsConvergence}
Let $\mathbf{M} = (M_1, M_2, \ldots, M_n)$ be the vector of membership statuses, where $M_i = \mathbbm{1}[(x_i, y_i) \in D]$. Let $\pi(\mathbf{M}|o_\theta)= \Pr(\mathbf{M}|o_\theta)$ be the target joint distribution of membership statuses conditioned on the model output $o_\theta$. Consider the Gibbs sampling procedure that iteratively samples at step $t \geq 1$:
\begin{equation*}
M_i^{(t+1)} \sim \Pr(M_i | \mathbf{M}_{-i}^{(t+1,t)}, o_\theta), \: \forall (x_i,y_i) \in D_\text{query},
\end{equation*}
where $\mathbf{M}_{-i}^{(t+1,t)} = M_1^{(t+1)}, \ldots, M_{i-1}^{(t+1)}, M_{i+1}^{(t)}, \ldots, M_n^{(t)}$.
Then:
\begin{enumerate}
    \item The sequence of states $M^{(t)}$ forms a Markov chain with stationary distribution $\pi(\mathbf{M}|o_\theta) = \Pr(\mathbf{M}|o_\theta)$.
    \item For any measurable MIA performance metric $L$ with $\mathbb{E}_\pi[|L(\mathbf{M},D)|] < \infty$, the sequence 
   $$S_T = \frac{1}{T}\sum_{t=1}^T L(\mathbf{M}^{(t)},D)$$
   converges almost surely to $\mathbb{E}_\pi[L(\mathbf{M},D)]$ as $T \to \infty$.
\end{enumerate}
\end{restatable}


\begin{proof}
We prove the above theorem by establishing that the Gibbs sampling procedure forms a Markov chain that satisfies the conditions for the Martingale Convergence Theorem~\cite{Williams1991}.

\paragraph{Part 1: The Markov chain and its stationary distribution}

First, we establish that the Gibbs sampling procedure forms a Markov chain with stationary distribution $\pi(M|o_\theta)= \Pr(M|o_\theta)$.

By construction, the transition from $\mathbf{M}^{(t)}$ to $\mathbf{M}^{(t+1)}$ depends only on the current state $\mathbf{M}^{(t)}$ and not on the previous states, satisfying the Markov property.

Let $\Pr(\mathbf{M}'|\mathbf{M})$ denote the transition probability from state $\mathbf{M}$ to state $\mathbf{M}'$ in one complete iteration of Gibbs sampling. To prove that $\pi(\mathbf{M}|o_\theta)$ is a stationary distribution, we need to verify the detailed balance equation~\cite{Tierney1994}:
$$\pi(\mathbf{M}|o_\theta)\Pr(\mathbf{M}'|\mathbf{M}) = \pi(\mathbf{M}'|o_\theta)\Pr(\mathbf{M}|\mathbf{M}').$$
For Gibbs sampling, where we update one variable $M_i$ at a time, the transition probability is:
$$\Pr(\mathbf{M}'|\mathbf{M}) = \Pr(M'_i|\mathbf{M}_{-i}, o_\theta) \cdot \mathbbm{1}[\mathbf{M}'_{-i} = \mathbf{M}_{-i}]$$
where $\mathbf{M}_{-i}$ represents all components except $M_i$.

Following the results of Liu et al.~\cite{Liu1996}, the construction of Gibbs sampling ensures:
$$\Pr(M'_i|\mathbf{M}_{-i}, o_\theta) = \frac{\pi(M'_i, \mathbf{M}_{-i}|o_\theta)}{\sum_{m_i} \pi(m_i, \mathbf{M}_{-i}|o_\theta)} = \pi(M'_i | \mathbf{M}_{-i}, o_\theta)$$
Since the conditional distribution $\pi(M_i | M_{-i}, o_\theta)$ is used for sampling, the detailed balance condition is automatically satisfied~\cite{Roberts2004}, and $\pi(M|o_\theta)$ is indeed the stationary distribution of the Markov chain.

\paragraph{Part 2: Convergence via Martingale theory}

To establish convergence, we apply the Martingale Convergence Theorem as presented in Meryn and Tweedie~\cite{MeynTweedie2012}. Let the performance measure be such
$$|L(\mathbf{M},D)| < \infty,\quad \mathbf{M} \sim \pi(\mathbf{M}|o_\theta).$$
Let $\mathbf{M}^{
(1)} \sim \pi(\mathbf{M}|o_\theta)$ and define the sequence of $\sigma$-algebras $\mathcal{F}_t = \sigma(\mathbf{M}^{(1)}, \mathbf{M}^{(2)}, \ldots, \mathbf{M}^{(t)})$, representing the information available up to time $t$.
Following Douc et al.~\cite{Douc2018}, we define:
$$X_t = \mathbb{E}[L(\mathbf{M}^{(t)},D) | \mathcal{F}_t].$$
This defines a martingale, since:
\begin{align*}
\mathbb{E}[X_{t+1} | \mathcal{F}_t] &= \mathbb{E}[\mathbb{E}[L(\mathbf{M}^{(t+1)},D) | \mathcal{F}_{t+1}] | \mathcal{F}_t]\\
&= \mathbb{E}[L(\mathbf{M}^{(t+1)},D) | \mathcal{F}_t] = X_t,
\end{align*}
since the Markov chain is in steady state.
By the Martingale Convergence Theorem~\cite{Williams1991}, $X_t$ converges almost surely to a random variable $X_\infty$ as $t \to \infty$.

We now need to show that $X_\infty$ as $t \to \infty$ converges for any choice of $\mathbf{M}^{(1)}$. Under mild conditions on $\Pr(\mathbf{M}|o_\theta)$, where all query instances have some non-zero probability of being members and non-members, it ensures irreducibility and aperiodicity of the Gibbs sampler~\cite{Roberts2004, Tierney1994}, and then the Markov chain is ergodic. In this case, following the Ergodic Theorem for Markov chains~\cite{Meyn2009}, we have:
$$X_\infty = \mathbb{E}_\pi[L(\mathbf{M},D)].$$
And for the time average:
$$S_T = \frac{1}{T}\sum_{t=1}^T L(\mathbf{M}^{(t)},D),$$
the Law of Large Numbers for Markov chains~\cite{Meyn2009} ensures:
$$S_T \xrightarrow{a.s.} \mathbb{E}_\pi[L(\mathbf{M},D)]$$
as $T \to \infty$.
This establishes that the Gibbs sampling for joint MIA converges to the joint membership distribution, and empirical averages computed from the samples converge almost surely to expected values under the target distribution.
\end{proof}

\subsection{Proof of~\Cref{thm:marginalMIA}}
\label{appendix:proof_optimal}



\begin{proof}

From a Bayesian perspective, based on the adversary's observation $e_\theta$, we define the posterior probabilities as follows:
\begin{equation*}
\begin{aligned}
    p_{\text{in}} (x_i,y_i) &= \Pr\left(M_i = 1 \mid \mathcal{Q}(f_\theta)=e_\theta \right) \\
    &= \Pr\left((x_i,y_i) \in D \mid \mathcal{Q}(f_\theta)=e_\theta \right) \\
    & = \Pr\left(D \in \mathcal{S}^{+}_{(x_i,y_i)} \mid \mathcal{Q}(\mathcal{T}(D))=e_\theta \right), \\
    p_{\text{out}} (x_i,y_i) &= \Pr\left(M_i = 0 \mid \mathcal{Q}(f_\theta)=e_\theta \right) \\
   &=\Pr\left((x_i,y_i) \notin D \mid \mathcal{Q}(f_\theta) =e_\theta \right) \\
    &= \Pr\left(D \in \mathcal{S}^{-}_{(x_i,y_i)} \mid \mathcal{Q}(\mathcal{T}(D))=e_\theta \right),
\end{aligned}
\end{equation*}
where $p_{\text{in}
}(x_i,y_i)$ is the posterior probability that $(x_i,y_i)$ is in the training data and $p_{\text{out}}(x_i,y_i)$ is the posterior probability that it is not. 
The adversary should maximize the posterior probability of correctly identifying whether $(x_i,y_i)$ is part of the training set. 
Thus, the adversary will use the posterior odds to infer that $(x_i,y_i)$ is in the training set if:
\begin{equation}
\label{equ:posterior_ratio}
    \frac{p_{\text{in}} (x_i,y_i)}{p_{\text{out}} (x_i,y_i)} > 1,
\end{equation}
and it infers that $(x_i,y_i)$ is not in the training set otherwise. By the Bayesian rule, we can express the above ratio as follows:
\begin{equation}
\begin{aligned}
\label{equ:detailed_posterior_ratio}
    & \frac{\Pr\left(D \in \mathcal{S}^{+}_{(x_i,y_i)} \mid \mathcal{Q}(\mathcal{T}(D))=e_\theta \right)} {\Pr\left(D \in \mathcal{S}^{-}_{(x_i,y_i)} \mid \mathcal{Q}(\mathcal{T}(D))=e_\theta \right)} \\
    = & \frac{\Pr(\mathcal{Q}(\mathcal{T}(D))=e_\theta, D\in \mathcal{S}^+_{(x_i,y_i)} )}{\Pr(\mathcal{Q}(\mathcal{T}(D))=e_\theta, D\in \mathcal{S}^-_{(x_i,y_i)} )} \\
    = & \frac{\sum_{S \in \mathcal{S}^+_{(x_i,y_i)}} \Pr(D=S)\cdot \Pr(\mathcal{Q}(\mathcal{T}(S))=e_\theta)}{\sum_{S \in \mathcal{S}^-_{(x_i,y_i)}} \Pr(D=S)\cdot \Pr(\mathcal{Q}(\mathcal{T}(S))=e_\theta)
}\\
    = & \frac{\Pr (D\in \mathcal{S}^+_{(x_i,y_i)})}{\Pr (D\in \mathcal{S}^-_{(x_i,y_i)})} \times  \\
     &\frac{\sum_{S \in \mathcal{S}^+_{(x_i,y_i)}} \frac{\Pr(D=S)}{\Pr(D \in \mathcal{S}^+_{(x_i,y_i)})} \Pr(\mathcal{Q}(\mathcal{T}(S))=e_\theta)}{\sum_{S \in \mathcal{S}^-_{(x_i,y_i)}} \frac{\Pr(D=S)}{\Pr(D \in \mathcal{S}^-_{(x_i,y_i)})} \Pr(\mathcal{Q}(\mathcal{T}(S))=e_\theta)}.
\end{aligned}
\end{equation}

We define $\mathcal{L}(D,e_\theta)= \Pr(\mathcal{Q}(\mathcal{T}(D))=e_\theta)$ to denote the likelihood. 
Stemming from~\Cref{equ:posterior_ratio} and~\Cref{equ:detailed_posterior_ratio}, the adversary will infer $(x_i,y_i)$ is in the training set if:
\begin{equation*}
    \frac{\mathbb{E}_{D^\prime \sim \mathcal{S}^+_{(x_i,y_i)}}\mathcal{L}(D^\prime,e_\theta)}{\mathbb{E
}_{D^\prime \sim \mathcal{S}^-_{(x_i,y_i)}}\mathcal{L}(D^\prime,e_\theta)} > \frac{\Pr((x, y) \notin D)}{\Pr((x, y) \in D)}.
\end{equation*}
\end{proof}

\section{Details about Experimental Setups}

\begin{table}[t]
    \centering
    \caption{Data splits in the adaptive setting.}
    \vspace{-1mm}
    \label{tab:dataset_split_online}
    \resizebox{0.47\textwidth}{!}{
    \begin{tabular}{lccccccccccccc}
    \midrule
    \multirow{2}{*}{} & \multicolumn{2}{c}{MNIST} & \multicolumn{2}{c}{FMNIST} & \multicolumn{2}{c}{CIFAR-10} & \multicolumn{2}{c}{CIFAR-100} & \multicolumn{2}{c}{Purchase} & \multicolumn{2}{c}{Texas}\\
    \cmidrule(lr){2-3} \cmidrule(lr){4-5} \cmidrule(lr){6-7} \cmidrule(lr){8-9} \cmidrule(lr){10-11} \cmidrule(lr){12-13}
    & $D_1$ & $D_2$ & $D_1$ & $D_2$ & $D_1$ & $D_2$ & $D_1$ & $D_2$ & $D_1$ & $D_2$ & $D_1$ & $D_2$\\
    \midrule
    Size & 60k  & 10k & 60k  & 10k & 50k & 10k & 50k & 10k & 160k & 37k & 67k & 10k\\
    \midrule
    \end{tabular}}
\end{table}

\begin{table}[t]
    \centering
    \vspace{-3mm}
    \caption{Data splits in the non-adaptive setting.}
    \label{tab:dataset_split_offline}
    \vspace{-1mm}
    \resizebox{0.45\textwidth}{!}{
    \begin{tabular}{lccccc}
    \midrule
    \multirow{2}{*}{Dataset} & \multicolumn{2}{c}{$D_\text{query}$} & \multicolumn{3}{c}{$D_\text{adv}^\text{non-adapt}$}  \\
    \cmidrule(lr){2-3} \cmidrule(lr){4-6} 
    & Train & Val & Train & Val & Reference  \\
    \midrule
    MNIST & 11,667  & 11,667 & 11,667  & 11,667 & 23,334  \\
    \midrule
    FMNIST & 11,667  & 11,667 & 11,667  & 11,667 & 23,334  \\
    \midrule
    CIFAR-10 & 10,000  & 10,000 & 10,000  & 10,000 & 20,000 \\
    \midrule
    CIFAR-100 & 10,000  & 10,000 & 10,000  & 10,000 & 20,000 \\
    \midrule
    Purchase & 32,887 & 32,887 & 32,887 & 32,887 & 65,774 \\ 
    \midrule
    Texas & 11,221 & 11,221 & 11,221 & 11,221 & 22,443 \\ 
    \midrule
    \end{tabular}}
\end{table}

\subsection{Dataset Description}
\label{appendix:data_description}


\noindent\textbf{MNIST}~\cite{mnist} contains 60,000 grayscale images for training and 10,000 images for testing. The dataset consists of 10 classes, representing digits from 0 to 9.


\noindent\textbf{Fashion-MNIST}~\cite{fmnist} contains 60,000 grayscale images of size 28×28 pixels for training and 10,000 images for testing. It consists of 10 classes representing different fashion items.

\noindent\textbf{CIFAR10}~\cite{cifar10} is a benchmark dataset for general image classification tasks, containing 60,000 color images, equally distributed across 10 classes.

\noindent\textbf{CIFAR-100}~\cite{cifar10} is a dataset similar to CIFAR-10 but with a greater level of complexity, as it contains 100 classes. 

\noindent\textbf{Purchase}~\cite{sp17miashokri} is a dataset of shopping records with 197,324 samples of 600 dimensions. 

\noindent\textbf{Texas}~\cite{sp17miashokri} 
comprises records from 67,330 patients, with the 100 most frequent procedures used as classification labels.

\subsection{Accuracy of the Trained Target Model}
\label{appendix:acc_model}

The training and validation accuracies of the target model are in Table~\ref{tab:dataset_acc_online} and Table~\ref{tab:dataset_acc_offline}, respectively.

\section{Additional Experimental Results}

\begin{table}[t]
    \centering
    \caption{Prediction accuracy in the adaptive setting.}
    \vspace{-1mm}
    \label{tab:dataset_acc_online}
    \resizebox{0.51\textwidth}{!}{
    \begin{tabular}{lcccccccccccc}
    \midrule
    \multirow{2}{*}{Model} & \multicolumn{2}{c}{MNIST} & \multicolumn{2}{c}{FMNIST} & \multicolumn{2}{c}{C-10} & \multicolumn{2}{c}{C-100} & \multicolumn{2}{c}{Purchase} & \multicolumn{2}{c}{Texas} \\
    \cmidrule(lr){2-3} \cmidrule(lr){4-5} \cmidrule(lr){6-7} \cmidrule(lr){8-9} \cmidrule(lr){10-11} \cmidrule(lr){12-13}
    & Train & Val &  Train & Val & Train & Val & Train & Val & Train & Val & Train & Val \\
    \midrule
    ResNet50 & 1.00 & 0.99 & 1.00 & 0.91 & 0.99 & 0.90 & 0.99 & 0.67 & - & - & - & - \\
    VGG16 & 1.00 & 0.99 & 1.00 & 0.92 & 0.99 & 0.87 & 0.99 & 0.60 & - & - & - & - \\
    DenseNet121 & 1.00 & 0.99 & 0.99 & 0.93 & 0.99 & 0.89 & 0.99 & 0.63 & - & - & - & - \\
    MobileNetV2 & 1.00 & 0.99 & 1.00 & 0.92 & 0.99 & 0.88 & 0.99 & 0.61 & - & - & - & - \\
    MLP & - & - & - & - & - & - & - & - & 0.97 & 0.79 & 0.96 & 0.78 \\
    \midrule
    \end{tabular}}
\end{table}
\begin{table}[t]
    \centering
    \caption{Prediction accuracy in the non-adaptive setting.}
    \vspace{-1mm}
    \label{tab:dataset_acc_offline}
    \resizebox{0.51\textwidth}{!}{
    \begin{tabular}{lcccccccccccc}
    \midrule
    \multirow{2}{*}{Model} & \multicolumn{2}{c}{MNIST} & \multicolumn{2}{c}{FMNIST} & \multicolumn{2}{c}{C-10} & \multicolumn{2}{c}{C-100} & \multicolumn{2}{c}{Purchase} & \multicolumn{2}{c}{Texas} \\
    \cmidrule(lr){2-3} \cmidrule(lr){4-5} \cmidrule(lr){6-7} \cmidrule(lr){8-9} \cmidrule(lr){10-11} \cmidrule(lr){12-13}
    & Train & Val & Train & Val & Train & Val & Train  & Val & Train & Val & Train & Val \\
    \midrule
    ResNet50 & 1.00 & 0.99 & 1.00 & 0.94 & 0.99 & 0.90 & 0.92 & 0.66 & - & - & - & - \\
    VGG16 & 1.00 & 0.99 & 1.00 & 0.95 & 0.99 & 0.91 & 0.99 & 0.71 & - & - & - & - \\
    DenseNet121 & 1.00 & 0.99 & 1.00 & 0.96 & 0.99 & 0.90 & 0.99 & 0.72 & - & - & - & - \\
    MobileNetV2 & 1.00 & 0.99 & 1.00 & 0.94 & 0.99 & 0.97 & 1.00 & 0.72 & - & - & - & - \\
    MLP & - & - & - & - & - & - & - & - & 0.96 & 0.80 & 0.96 & 0.79 \\
    \midrule
    \end{tabular}}
\end{table}

\begin{table*}[th]
\centering
\vspace{-5mm}
\caption{Performance comparison of \textit{adaptive} attacks using \myonline on VGG16 trained on four image datasets.}
\vspace{-1mm}
\label{tab:cmia_main_vgg16}
\resizebox{0.99\textwidth}{!}{
\begin{tabular}{ll|*{12}{c}}
\toprule
\multirow{2}{*}{\textbf{Method}} & \multirow{2}{*}{} & \multicolumn{4}{c}{\textbf{TPR @ 0.001\% FPR}} & \multicolumn{4}{c}{\textbf{TPR @ 0.1\% FPR}} & \multicolumn{4}{c}{\textbf{Balanced Accuracy}} \\
\cmidrule(lr){3-6}\cmidrule(lr){7-10}\cmidrule(lr){11-14}
& & MNIST & FMNIST & C-10 & C-100 & MNIST & FMNIST & C-10 & C-100 & MNIST & FMNIST & C-10 & C-100 \\
\midrule
\multirow{3}{*}{Calibration} & Base & 0.01\% & 0.69\% & 0.46\% & 3.18\% & 0.40\% & 1.94\% & 1.67\% & 7.15\% & 50.61\% & 53.61\% & 55.17\% & 63.83\% \\
& \textbf{\myonline} & \textbf{0.04\%} & \textbf{0.94\%} & \textbf{0.77\%} & \textbf{5.02\%} & \textbf{0.69\%} & \textbf{2.58\%} & \textbf{1.95\%} & \textbf{8.82\%} & \textbf{51.37\%} & \textbf{53.94\%} & \textbf{56.52\%} & \textbf{65.93\%} \\
& \cellcolor[gray]{0.9}\%Imp.  & \cellcolor[gray]{0.9}300.00\% & \cellcolor[gray]{0.9}36.23\% & \cellcolor[gray]{0.9}67.39\% & \cellcolor[gray]{0.9}57.86\% & \cellcolor[gray]{0.9}72.50\% & \cellcolor[gray]{0.9}32.99\% & \cellcolor[gray]{0.9}16.77\% & \cellcolor[gray]{0.9}23.36\% & \cellcolor[gray]{0.9}1.50\% & \cellcolor[gray]{0.9}0.62\% & \cellcolor[gray]{0.9}2.45\% & \cellcolor[gray]{0.9}3.29\%  \\
\midrule
\multirow{3}{*}{Attack-R} & Base & 0.00\% & 0.00\% & 0.00\% & 0.26\% & 0.70\% & 4.08\% & 5.07\% & 1.06\% & 51.48\% & 57.03\% & 60.04\% & 74.79\% \\
& \textbf{\myonline} & \textbf{0.00\%} & \textbf{0.00\%} & \textbf{0.00\%} & \textbf{0.49\%} & \textbf{0.95\%} & \textbf{4.95\%} & \textbf{5.77\%} & \textbf{2.53\%} & \textbf{51.97\%} & \textbf{57.74\%} & \textbf{60.53\%} & \textbf{75.08\%}  \\
& \cellcolor[gray]{0.9}\%Imp.  & \cellcolor[gray]{0.9}- & \cellcolor[gray]{0.9}- & \cellcolor[gray]{0.9}- & \cellcolor[gray]{0.9}88.46\% & \cellcolor[gray]{0.9}35.71\% & \cellcolor[gray]{0.9}21.32\% & \cellcolor[gray]{0.9}13.81\% & \cellcolor[gray]{0.9}138.68\% & \cellcolor[gray]{0.9}0.95\% & \cellcolor[gray]{0.9}1.24\% & \cellcolor[gray]{0.9}0.82\% & \cellcolor[gray]{0.9}0.39\%  \\
\midrule
\multirow{3}{*}{LiRA} & Base & 0.11\% & 2.09\% & 2.01\% & 12.95\% & 0.94\% & 5.06\% & 7.21\% & 28.70\% & 51.24\% & 58.12\% & 64.02\% & 82.07\% \\
& \textbf{\myonline} & \textbf{0.79\%} & \textbf{3.63\%} & \textbf{3.02\%} & \textbf{17.88\%} & \textbf{1.47\%} & \textbf{5.98\%} & \textbf{9.18\%} & \textbf{32.02\%} & \textbf{52.05\%} & \textbf{58.88\%} & \textbf{65.47\%} & \textbf{82.55\%} \\
& \cellcolor[gray]{0.9}\%Imp.  & \cellcolor[gray]{0.9}618.18\% & \cellcolor[gray]{0.9}73.68\% & \cellcolor[gray]{0.9}50.25\% & \cellcolor[gray]{0.9}38.07\% & \cellcolor[gray]{0.9}56.38\% & \cellcolor[gray]{0.9}18.18\% & \cellcolor[gray]{0.9}27.32\% & \cellcolor[gray]{0.9}11.57\% & \cellcolor[gray]{0.9}1.58\% & \cellcolor[gray]{0.9}1.31\% & \cellcolor[gray]{0.9}2.26\% & \cellcolor[gray]{0.9}0.58\%  \\
\midrule
\multirow{3}{*}{Canary} & Base & 0.15\% & 2.16\% & 2.10\% & 12.82\% & 0.94\% & 5.38\% & 7.08\% & 27.53\% & 51.83\% & 58.49\% & 64.33\% & 81.44\% \\
& \textbf{\myonline} & \textbf{0.82\%} & \textbf{3.99\%} & \textbf{3.10\%} & \textbf{18.40\%} & \textbf{1.51\%} & \textbf{6.33\%} & \textbf{9.53\%} & \textbf{31.90\%} & \textbf{52.33\%} & \textbf{59.06\%} & \textbf{65.95\%} & \textbf{81.95\%} \\
& \cellcolor[gray]{0.9}\%Imp.  & \cellcolor[gray]{0.9}446.67\% & \cellcolor[gray]{0.9}84.72\% & \cellcolor[gray]{0.9}47.62\% & \cellcolor[gray]{0.9}43.53\% & \cellcolor[gray]{0.9}60.64\% & \cellcolor[gray]{0.9}17.66\% & \cellcolor[gray]{0.9}34.60\% & \cellcolor[gray]{0.9}15.87\% & \cellcolor[gray]{0.9}0.96\% & \cellcolor[gray]{0.9}0.97\% & \cellcolor[gray]{0.9}2.52\% & \cellcolor[gray]{0.9}0.63\%  \\
\midrule
\multirow{3}{*}{RMIA} & Base & 0.25\% & 2.07\% & 0.00\% & 0.27\% & 1.02\% & 3.81\% & 5.06\% & 20.49\% & 51.82\% & 59.02\% & 63.82\% & 80.74\% \\
& \textbf{\myonline} & \textbf{0.37\%} & \textbf{3.11\%} & \textbf{0.00\%} & \textbf{0.78\%} & \textbf{1.94\%} & \textbf{5.07\%} & \textbf{7.43\%} & \textbf{25.80\%} & \textbf{52.07\%} & \textbf{59.73\%} & \textbf{64.05\%} & \textbf{81.69\%} \\
& \cellcolor[gray]{0.9}\%Imp.  & \cellcolor[gray]{0.9}48.00\% & \cellcolor[gray]{0.9}50.24\% & \cellcolor[gray]{0.9}- & \cellcolor[gray]{0.9}188.89\% & \cellcolor[gray]{0.9}90.20\% & \cellcolor[gray]{0.9}33.07\% & \cellcolor[gray]{0.9}46.84\% & \cellcolor[gray]{0.9}25.92\% & \cellcolor[gray]{0.9}0.48\% & \cellcolor[gray]{0.9}1.20\% & \cellcolor[gray]{0.9}0.36\% & \cellcolor[gray]{0.9}1.18\%  \\
\midrule
\multirow{3}{*}{RAPID} & Base & 0.21\% & 1.10\% & 0.42\% & 3.02\% & 0.76\% & 3.02\% & 3.01\% & 14.03\% & 51.2\% & 56.89\% & 60.01\% & 76.82\% \\
& \textbf{\myonline} & \textbf{0.38\%} & \textbf{2.53\%} & \textbf{0.79\%} & \textbf{4.75\%} & \textbf{1.05\%} & \textbf{4.85\%} & \textbf{5.72\%} & \textbf{17.63\%} & \textbf{51.42\%} & \textbf{57.42\%} & \textbf{61.04\%} & \textbf{77.35\%} \\
& \cellcolor[gray]{0.9}\%Imp.  & \cellcolor[gray]{0.9}80.95\% & \cellcolor[gray]{0.9}130.00\% & \cellcolor[gray]{0.9}88.10\% & \cellcolor[gray]{0.9}57.28\% & \cellcolor[gray]{0.9}38.16\% & \cellcolor[gray]{0.9}60.60\% & \cellcolor[gray]{0.9}90.03\% & \cellcolor[gray]{0.9}25.66\% & \cellcolor[gray]{0.9}0.43\% & \cellcolor[gray]{0.9}0.93\% & \cellcolor[gray]{0.9}1.72\% & \cellcolor[gray]{0.9}0.69\%  \\
\bottomrule
\end{tabular}}
\end{table*}

\begin{table*}[t]
\centering
\vspace{-5mm}
\caption{Performance comparison of \textit{adaptive} attacks using \myonline on DenseNet121 trained on four image datasets. }
\label{tab:cmia_main_densenet}
\resizebox{0.99\textwidth}{!}{
\begin{tabular}{ll|*{12}{c}}
\toprule
\multirow{2}{*}{\textbf{Method}} & \multirow{2}{*}{} & \multicolumn{4}{c}{\textbf{TPR @ 0.001\% FPR}} & \multicolumn{4}{c}{\textbf{TPR @ 0.1\% FPR}} & \multicolumn{4}{c}{\textbf{Balanced Accuracy}} \\
\cmidrule(lr){3-6}\cmidrule(lr){7-10}\cmidrule(lr){11-14}
& & MNIST & FMNIST & C-10 & C-100 & MNIST & FMNIST & C-10 & C-100 & MNIST & FMNIST & C-10 & C-100 \\
\midrule
\multirow{3}{*}{Calibration} & Base & 0.09\% & 0.54\% & 0.53\% & 2.23\% & 0.28\% & 1.53\% & 1.84\% & 6.01\% & 50.45\% & 52.95\% & 53.94\% & 62.1\% \\
& \textbf{\myonline} & \textbf{0.18\%} & \textbf{1.05\%} & \textbf{0.68\%} & \textbf{2.64\%} & \textbf{0.39\%} & \textbf{2.03\%} & \textbf{2.07\%} & \textbf{6.85\%} & \textbf{51.42\%} & \textbf{53.49\%} & \textbf{54.90\%} & \textbf{62.5\%} \\
& \cellcolor[gray]{0.9}\%Imp.  & \cellcolor[gray]{0.9}100.00\% & \cellcolor[gray]{0.9}94.44\% & \cellcolor[gray]{0.9}28.30\% & \cellcolor[gray]{0.9}18.39\% & \cellcolor[gray]{0.9}39.29\% & \cellcolor[gray]{0.9}32.68\% & \cellcolor[gray]{0.9}12.50\% & \cellcolor[gray]{0.9}13.98\% & \cellcolor[gray]{0.9}1.92\% & \cellcolor[gray]{0.9}1.02\% & \cellcolor[gray]{0.9}1.78\% & \cellcolor[gray]{0.9}0.64\%  \\
\midrule
\multirow{3}{*}{Attack-R} & Base & 0.00\% & 0.00\% & 0.00\% & 0.00\% & 0.74\% & 3.51\% & 0.04\% & 0.93\% & 51.02\% & 56.14\% & 59.52\% & 74.88\% \\
& \textbf{\myonline} & \textbf{0.00\%} & \textbf{0.00\%} & \textbf{0.00\%} & \textbf{0.00\%} & \textbf{1.13\%} & \textbf{4.62\%} & \textbf{0.15\%} & \textbf{2.83\%} & \textbf{51.95\%} & \textbf{57.17\%} & \textbf{60.29\%} & \textbf{75.63\%} \\
& \cellcolor[gray]{0.9}\%Imp.  & \cellcolor[gray]{0.9}- & \cellcolor[gray]{0.9}- & \cellcolor[gray]{0.9}- & \cellcolor[gray]{0.9}- & \cellcolor[gray]{0.9}52.70\% & \cellcolor[gray]{0.9}31.62\% & \cellcolor[gray]{0.9}275.00\% & \cellcolor[gray]{0.9}204.30\% & \cellcolor[gray]{0.9}1.82\% & \cellcolor[gray]{0.9}1.83\% & \cellcolor[gray]{0.9}1.29\% & \cellcolor[gray]{0.9}1.00\%  \\
\midrule
\multirow{3}{*}{LiRA} & Base & 0.23\% & 2.01\% & 2.74\% & 10.05\% & 1.02\% & 6.05\% & 8.03\% & 24.60\% & 51.06\% & 58.46\% & 63.45\% & 80.09\% \\
& \textbf{\myonline} & \textbf{0.36\%} & \textbf{2.80\%} & \textbf{3.33\%} & \textbf{13.67\%} & \textbf{1.36\%} & \textbf{7.18\%} & \textbf{10.85\%} & \textbf{28.64\%} & \textbf{51.50\%} & \textbf{58.80\%} & \textbf{63.84\%} & \textbf{81.68\%} \\
& \cellcolor[gray]{0.9}\%Imp.  & \cellcolor[gray]{0.9}56.52\% & \cellcolor[gray]{0.9}39.30\% & \cellcolor[gray]{0.9}21.53\% & \cellcolor[gray]{0.9}36.02\% & \cellcolor[gray]{0.9}33.33\% & \cellcolor[gray]{0.9}18.68\% & \cellcolor[gray]{0.9}35.12\% & \cellcolor[gray]{0.9}16.42\% & \cellcolor[gray]{0.9}0.86\% & \cellcolor[gray]{0.9}0.58\% & \cellcolor[gray]{0.9}0.61\% & \cellcolor[gray]{0.9}1.99\%  \\
\midrule
\multirow{3}{*}{Canary} & Base & 0.31\% & 2.17\% & 2.77\% & 10.16\% & 0.97\% & 6.28\% & 8.16\% & 24.73\% & 51.53\% & 58.92\% & 63.58\% & 80.12\% \\
& \textbf{\myonline} & \textbf{0.39\%} & \textbf{3.04\%} & \textbf{3.51\%} & \textbf{13.48\%} & \textbf{1.29\%} & \textbf{7.25\%} & \textbf{11.01\%} & \textbf{28.41\%} & \textbf{51.73\%} & \textbf{59.00\%} & \textbf{63.75\%} & \textbf{81.50\%} \\
& \cellcolor[gray]{0.9}\%Imp.  & \cellcolor[gray]{0.9}25.81\% & \cellcolor[gray]{0.9}40.09\% & \cellcolor[gray]{0.9}26.71\% & \cellcolor[gray]{0.9}32.68\% & \cellcolor[gray]{0.9}32.99\% & \cellcolor[gray]{0.9}15.45\% & \cellcolor[gray]{0.9}34.93\% & \cellcolor[gray]{0.9}14.88\% & \cellcolor[gray]{0.9}0.39\% & \cellcolor[gray]{0.9}0.14\% & \cellcolor[gray]{0.9}0.27\% & \cellcolor[gray]{0.9}1.72\%  \\
\midrule
\multirow{3}{*}{RMIA} & Base & 0.28\% & 1.04\% & 0.54\% & 0.29\% & 0.94\% & 4.06\% & 3.60\% & 1.84\% & 51.04\% & 58.07\% & 62.04\% & 79.11\% \\
& \textbf{\myonline} & \textbf{0.33\%} & \textbf{1.97\%} & \textbf{0.66\%} & \textbf{0.77\%} & \textbf{1.42\%} & \textbf{4.66\%} & \textbf{4.15\%} & \textbf{3.63\%} & \textbf{51.99\%} & \textbf{58.86\%} & \textbf{62.38\%} & \textbf{79.57\%} \\
& \cellcolor[gray]{0.9}\%Imp.  & \cellcolor[gray]{0.9}17.86\% & \cellcolor[gray]{0.9}89.42\% & \cellcolor[gray]{0.9}22.22\% & \cellcolor[gray]{0.9}165.52\% & \cellcolor[gray]{0.9}51.06\% & \cellcolor[gray]{0.9}14.78\% & \cellcolor[gray]{0.9}15.28\% & \cellcolor[gray]{0.9}97.28\% & \cellcolor[gray]{0.9}1.86\% & \cellcolor[gray]{0.9}1.36\% & \cellcolor[gray]{0.9}0.55\% & \cellcolor[gray]{0.9}0.58\%  \\
\midrule
\multirow{3}{*}{RAPID} & Base & 0.12\% & 1.36\% & 1.05\% & 5.48\% & 0.65\% & 3.75\% & 3.81\% & 13.09\% & 51.39\% & 57.02\% & 59.20\% & 78.01\% \\
& \textbf{\myonline} & \textbf{0.37\%} & \textbf{2.04\%} & \textbf{1.98\%} & \textbf{5.96\%} & \textbf{1.33\%} & \textbf{4.40\%} & \textbf{4.63\%} & \textbf{17.73\%} & \textbf{51.96\%} & \textbf{57.77\%} & \textbf{60.12\%} & \textbf{79.04\%} \\
& \cellcolor[gray]{0.9}\%Imp.  & \cellcolor[gray]{0.9}208.33\% & \cellcolor[gray]{0.9}50.00\% & \cellcolor[gray]{0.9}88.57\% & \cellcolor[gray]{0.9}8.76\% & \cellcolor[gray]{0.9}104.62\% & \cellcolor[gray]{0.9}17.33\% & \cellcolor[gray]{0.9}21.52\% & \cellcolor[gray]{0.9}35.45\% & \cellcolor[gray]{0.9}1.11\% & \cellcolor[gray]{0.9}1.32\% & \cellcolor[gray]{0.9}1.55\% & \cellcolor[gray]{0.9}1.32\%  \\
\bottomrule
\end{tabular}}
\end{table*}

\begin{table*}[t]
\centering
\vspace{-5mm}
\caption{Performance comparison of \textit{adaptive} attacks using \myonline on MobileNetV2 trained on four image datasets.}
\vspace{-1mm}
\label{tab:cmia_main_mobilenet}
\resizebox{0.99\textwidth}{!}{
\begin{tabular}{ll|*{12}{c}}
\toprule
\multirow{2}{*}{\textbf{Method}} & \multirow{2}{*}{} & \multicolumn{4}{c}{\textbf{TPR @ 0.001\% FPR}} & \multicolumn{4}{c}{\textbf{TPR @ 0.1\% FPR}} & \multicolumn{4}{c}{\textbf{Balanced Accuracy}} \\
\cmidrule(lr){3-6}\cmidrule(lr){7-10}\cmidrule(lr){11-14}
& & MNIST & FMNIST & C-10 & C-100 & MNIST & FMNIST & C-10 & C-100 & MNIST & FMNIST & C-10 & C-100 \\
\midrule
\multirow{3}{*}{Calibration} & Base & 0.10\% & 0.00\% & 0.43\% & 1.73\% & 0.35\% & 1.62\% & 1.56\% & 6.32\% & 51.35\% & 54.21\% & 55.53\% & 64.3\% \\
& \textbf{\myonline} & \textbf{0.42\%} & \textbf{0.00\%} & \textbf{0.72\%} & \textbf{3.03\%} & \textbf{0.77\%} & \textbf{1.93\%} & \textbf{2.00\%} & \textbf{8.84\%} & \textbf{51.92\%} & \textbf{54.83\%} & \textbf{55.76\%} & \textbf{65.78\%} \\
& \cellcolor[gray]{0.9}\%Imp.  & \cellcolor[gray]{0.9}320.00\% & \cellcolor[gray]{0.9}- & \cellcolor[gray]{0.9}67.44\% & \cellcolor[gray]{0.9}75.14\% & \cellcolor[gray]{0.9}120.00\% & \cellcolor[gray]{0.9}19.14\% & \cellcolor[gray]{0.9}28.21\% & \cellcolor[gray]{0.9}39.87\% & \cellcolor[gray]{0.9}1.11\% & \cellcolor[gray]{0.9}1.14\% & \cellcolor[gray]{0.9}0.41\% & \cellcolor[gray]{0.9}2.30\%  \\
\midrule
\multirow{3}{*}{Attack-R} & Base & 0.00\% & 0.00\% & 0.00\% & 0.14\% & 0.00\% & 0.00\% & 0.92\% & 2.85\% & 51.52\% & 56.56\% & 58.69\% & 74.63\% \\
& \textbf{\myonline} & \textbf{0.00\%} & \textbf{0.00\%} & \textbf{0.00\%} & \textbf{0.43\%} & \textbf{0.00\%} & \textbf{0.00\%} & \textbf{1.26\%} & \textbf{4.48\%} & \textbf{51.96\%} & \textbf{56.84\%} & \textbf{58.88\%} & \textbf{75.12\%} \\
& \cellcolor[gray]{0.9}\%Imp.  & \cellcolor[gray]{0.9}- & \cellcolor[gray]{0.9}- & \cellcolor[gray]{0.9}- & \cellcolor[gray]{0.9}207.14\% & \cellcolor[gray]{0.9}- & \cellcolor[gray]{0.9}- & \cellcolor[gray]{0.9}36.96\% & \cellcolor[gray]{0.9}57.19\% & \cellcolor[gray]{0.9}0.85\% & \cellcolor[gray]{0.9}0.50\% & \cellcolor[gray]{0.9}0.32\% & \cellcolor[gray]{0.9}0.66\%  \\
\midrule
\multirow{3}{*}{LiRA} & Base & 0.18\% & 1.01\% & 2.33\% & 11.52\% & 1.02\% & 4.01\% & 6.01\% & 23.90\% & 51.73\% & 58.62\% & 62.04\% & 80.08\% \\
& \textbf{\myonline} & \textbf{0.64\%} & \textbf{2.17\%} & \textbf{3.02\%} & \textbf{16.90\%} & \textbf{1.72\%} & \textbf{5.36\%} & \textbf{7.23\%} & \textbf{30.65\%} & \textbf{52.24\%} & \textbf{59.61\%} & \textbf{62.65\%} & \textbf{81.69\%} \\
& \cellcolor[gray]{0.9}\%Imp.  & \cellcolor[gray]{0.9}255.56\% & \cellcolor[gray]{0.9}114.85\% & \cellcolor[gray]{0.9}29.61\% & \cellcolor[gray]{0.9}46.70\% & \cellcolor[gray]{0.9}68.63\% & \cellcolor[gray]{0.9}33.67\% & \cellcolor[gray]{0.9}20.30\% & \cellcolor[gray]{0.9}28.24\% & \cellcolor[gray]{0.9}0.99\% & \cellcolor[gray]{0.9}1.69\% & \cellcolor[gray]{0.9}0.98\% & \cellcolor[gray]{0.9}2.01\%  \\
\midrule
\multirow{3}{*}{Canary} & Base & 0.16\% & 1.25\% & 2.37\% & 12.07\% & 1.05\% & 4.04\% & 6.12\% & 24.07\% & 51.77\% & 58.61\% & 62.53\% & 80.17\% \\
& \textbf{\myonline} & \textbf{1.53\%} & \textbf{2.46\%} & \textbf{3.18\%} & \textbf{17.74\%} & \textbf{1.85\%} & \textbf{5.77\%} & \textbf{7.35\%} & \textbf{31.02\%} & \textbf{52.31\%} & \textbf{59.99\%} & \textbf{62.98\%} & \textbf{81.72\%} \\
& \cellcolor[gray]{0.9}\%Imp.  & \cellcolor[gray]{0.9}856.25\% & \cellcolor[gray]{0.9}96.80\% & \cellcolor[gray]{0.9}34.18\% & \cellcolor[gray]{0.9}46.98\% & \cellcolor[gray]{0.9}76.19\% & \cellcolor[gray]{0.9}42.82\% & \cellcolor[gray]{0.9}20.10\% & \cellcolor[gray]{0.9}28.87\% & \cellcolor[gray]{0.9}1.04\% & \cellcolor[gray]{0.9}2.35\% & \cellcolor[gray]{0.9}0.72\% & \cellcolor[gray]{0.9}1.93\%  \\
\midrule
\multirow{3}{*}{RMIA} & Base & 0.17\% & 1.83\% & 0.10\% & 3.74\% & 1.23\% & 3.05\% & 1.18\% & 6.38\% & 52.43\% & 59.07\% & 61.01\% & 79.64\% \\
& \textbf{\myonline} & \textbf{0.43\%} & \textbf{2.84\%} & \textbf{0.27\%} & \textbf{5.06\%} & \textbf{1.83\%} & \textbf{4.49\%} & \textbf{1.99\%} & \textbf{8.62\%} & \textbf{52.62\%} & \textbf{59.82\%} & \textbf{61.49\%} & \textbf{80.81\%} \\
& \cellcolor[gray]{0.9}\%Imp.  & \cellcolor[gray]{0.9}152.94\% & \cellcolor[gray]{0.9}55.19\% & \cellcolor[gray]{0.9}170.00\% & \cellcolor[gray]{0.9}35.29\% & \cellcolor[gray]{0.9}48.78\% & \cellcolor[gray]{0.9}47.21\% & \cellcolor[gray]{0.9}68.64\% & \cellcolor[gray]{0.9}35.11\% & \cellcolor[gray]{0.9}0.36\% & \cellcolor[gray]{0.9}1.27\% & \cellcolor[gray]{0.9}0.79\% & \cellcolor[gray]{0.9}1.47\%  \\
\midrule
\multirow{3}{*}{RAPID} & Base & 0.23\% & 0.64\% & 0.32\% & 5.31\% & 1.08\% & 3.01\% & 2.93\% & 14.73\% & 52.16\% & 58.57\% & 59.39\% & 76.28\% \\
& \textbf{\myonline} & \textbf{0.41\%} & \textbf{1.05\%} & \textbf{0.52\%} & \textbf{7.29\%} & \textbf{2.05\%} & \textbf{4.21\%} & \textbf{3.40\%} & \textbf{20.39\%} & \textbf{52.36\%} & \textbf{58.90\%} & \textbf{59.74\%} & \textbf{77.58\%} \\
& \cellcolor[gray]{0.9}\%Imp.  & \cellcolor[gray]{0.9}78.26\% & \cellcolor[gray]{0.9}64.06\% & \cellcolor[gray]{0.9}62.50\% & \cellcolor[gray]{0.9}37.29\% & \cellcolor[gray]{0.9}89.81\% & \cellcolor[gray]{0.9}39.87\% & \cellcolor[gray]{0.9}16.04\% & \cellcolor[gray]{0.9}38.42\% & \cellcolor[gray]{0.9}0.38\% & \cellcolor[gray]{0.9}0.56\% & \cellcolor[gray]{0.9}0.59\% & \cellcolor[gray]{0.9}1.70\%  \\
\bottomrule
\end{tabular}}
\end{table*}

\begin{figure*}[t]
    \centering
    \vspace{-5mm}
    \subfigure[MNIST]
    {
    \includegraphics[width=0.222\linewidth]{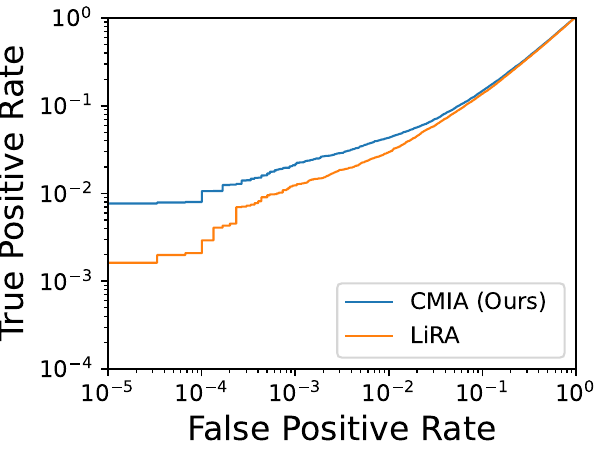}
    }
    \subfigure[Fashion-MNIST]
    {
    \includegraphics[width=0.222\linewidth]{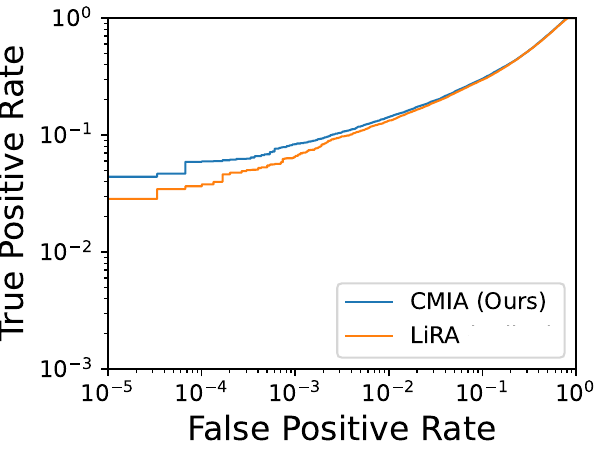}
    }
    \subfigure[CIFAR-10]
    {
    \includegraphics[width=0.222\linewidth]{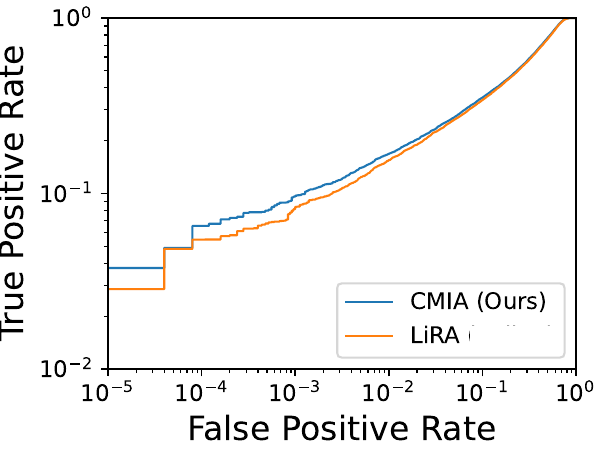}
    }
    \subfigure[CIFAR-100]
    {
    \includegraphics[width=0.222\linewidth]{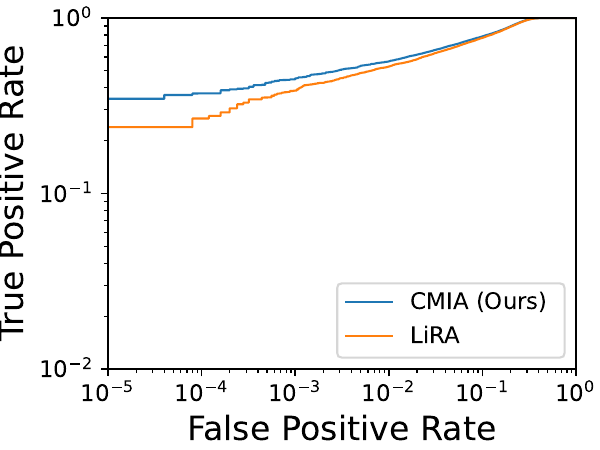}
    }
    \vspace{-3mm}
    \caption{The ROC curves of LIRA and \myonline (base: LiRA) on ResNet50 models trained on four image datasets.}
    \label{fig:cmia_curve_resnet}
\end{figure*}

\subsection{Experiments of \myonline on Other Models}
\label{appendix:cmia_other_models}

We present the results on the remaining three model architectures in the adaptive setting. Specifically, \Cref{tab:cmia_main_vgg16},~\Cref{tab:cmia_main_densenet} and \Cref{tab:cmia_main_mobilenet} displays the results for VGG16, DenseNet121, and MobileNetV2.
We also present the ROC curves of \myonline using LiRA as the base attack in~\Cref{fig:cmia_curve_resnet}.

\begin{figure*}[t]
    \centering
    \vspace{-7mm}
    \subfigure[MNIST]
    {
    \includegraphics[width=0.222\linewidth]{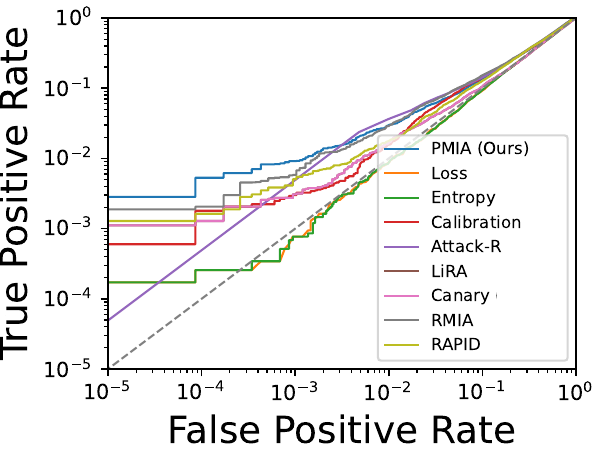}
    }
    \subfigure[Fashion-MNIST]
    {
    \includegraphics[width=0.222\linewidth]{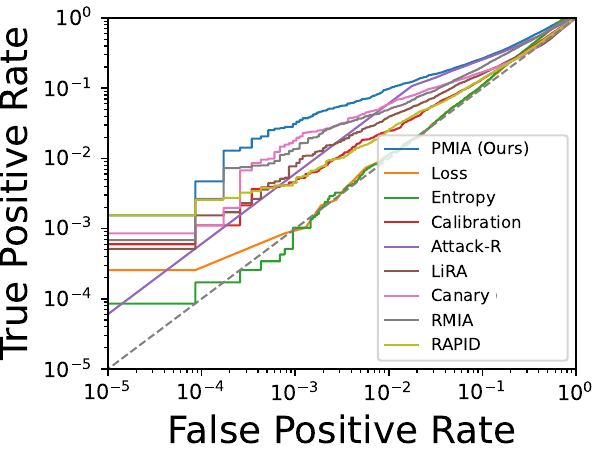}
    }
    \subfigure[CIFAR-10]
    {
    \includegraphics[width=0.222\linewidth]{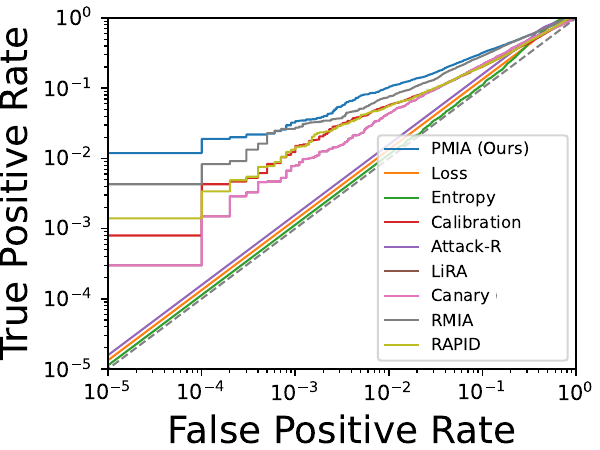}
    }
    \subfigure[CIFAR-100]
    {
    \includegraphics[width=0.222\linewidth]{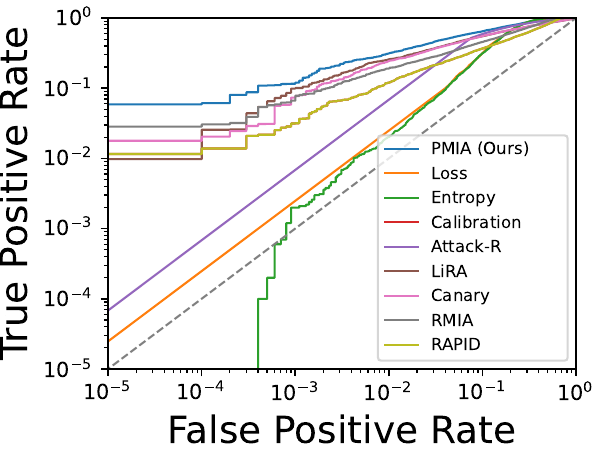}
    }
    \vspace{-3mm}
    \caption{The ROC curves of non-adaptive attack results on ResNet50 models trained on four image datasets.}
    \label{fig:pmia_curve_resnet}
\end{figure*}

\begin{table*}[t]
\centering
\vspace{-2mm}
\caption{Performance comparison of \textit{non-adaptive} attacks on VGG16 trained on four image datasets.
}
\vspace{-1mm}
\label{tab:pmia_main_vgg}
\resizebox{0.99\textwidth}{!}{
\begin{tabular}{l|*{12}{c}}
\toprule
\multirow{2}{*}{\textbf{Method}} & \multicolumn{4}{c}{\textbf{TPR @ 0.001\% FPR}} & \multicolumn{4}{c}{\textbf{TPR @ 0.1\% FPR}} & \multicolumn{4}{c}{\textbf{Balanced Accuracy}} \\
\cmidrule(lr){2-5}\cmidrule(lr){6-9}\cmidrule(lr){10-13}
&  MNIST & FMNIST & C-10 & C-100 & MNIST & FMNIST & C-10 & C-100 & MNIST & FMNIST & C-10 & C-100 \\
\midrule
LOSS & 0.00\% & 0.00\% & 0.00\% & 0.00\% & 0.00\% & 0.00\% & 0.01\% & 0.16\% & 52.18\% & 59.03\% & 64.86\% & 85.68\% \\
Entropy & 0.00\% & 0.00\% & 0.00\% & 0.05\% & 0.00\% & 0.01\% & 0.02\% & 0.23\% & \underline{52.02\%} & \underline{59.05\%} & \underline{67.56\%} & \underline{88.09\%} \\
Calibration & 0.07\% & 0.07\% & 0.01\% & 1.53\% & 0.27\% & 1.30\% & 1.60\% & 4.52\% & 52.15\% & 54.79\% & 58.61\% & 68.21\% \\
Attack-R & 0.00\% & 0.00\% & 0.00\% & 0.00\% & \underline{0.33\%} & 0.00\% & 0.00\% & 0.00\% & 52.01\% & 57.60\% & 64.19\% & 82.89\% \\
LiRA & 0.00\% & 0.00\% & 0.00\% & 2.95\% & 0.00\% & 0.29\% & 1.51\% & 9.83\% & 50.41\% & 51.98\% & 56.68\% & 78.51\% \\
Canary & 0.00\% & 0.00\% & 0.02\% & 3.58\% & 0.00\% & 0.37\% & 1.93\% & 10.15\% & 50.77\% & 51.34\% & 56.69\% & 80.04\% \\
RMIA & \underline{0.13\%} & \underline{0.08\%} & \underline{0.24\%} & \underline{5.74\%} & 0.32\% & \underline{1.78\%} & \underline{3.42\%} & 10.12\% & 51.94\% & 57.78\% & 64.46\% & 75.53\% \\
RAPID & 0.08\% & 0.04\% & 0.13\% & 0.28\% & 0.25\% & 0.15\% & 2.73\% & \underline{10.78\%} & 52.07\% & 58.49\% & 60.88\% & 82.52\% \\
\midrule
\textbf{\myoffline} & \textbf{0.16\%} & \textbf{0.12\%} & \textbf{0.65\%} & \textbf{9.00\%} & \textbf{0.46\%} & \textbf{3.09\%} & \textbf{5.22\%} & \textbf{26.21\%} & \textbf{52.27\%} & \textbf{59.76\%} & \textbf{67.68\%} & \textbf{88.93\%} \\
\cellcolor[gray]{0.9}\%Imp.  & \cellcolor[gray]{0.9}23.08\% & \cellcolor[gray]{0.9}50.00\% & \cellcolor[gray]{0.9}170.83\% & \cellcolor[gray]{0.9}56.79\% & \cellcolor[gray]{0.9}39.39\% & \cellcolor[gray]{0.9}73.60\% & \cellcolor[gray]{0.9}52.63\% & \cellcolor[gray]{0.9}143.14\% & \cellcolor[gray]{0.9}0.01\% & \cellcolor[gray]{0.9}1.20\% & \cellcolor[gray]{0.9}0.18\% & \cellcolor[gray]{0.9}0.95\%  \\
\bottomrule
\end{tabular}}
\end{table*}
\begin{table*}[t]
\centering
\caption{Performance comparison of \textit{non-adaptive} attacks on DenseNet121 trained on four images datasets.}
\vspace{-1mm}
\label{tab:pmia_main_densenet}
\resizebox{0.99\textwidth}{!}{
\begin{tabular}{l|*{12}{c}}
\toprule
\multirow{2}{*}{\textbf{Method}} & \multicolumn{4}{c}{\textbf{TPR @ 0.001\% FPR}} & \multicolumn{4}{c}{\textbf{TPR @ 0.1\% FPR}} & \multicolumn{4}{c}{\textbf{Balanced Accuracy}} \\
\cmidrule(lr){2-5}\cmidrule(lr){6-9}\cmidrule(lr){10-13}
&  MNIST & FMNIST & C-10 & C-100 & MNIST & FMNIST & C-10 & C-100 & MNIST & FMNIST & C-10 & C-100 \\
\midrule
LOSS & 0.00\% & 0.00\% & 0.00\% & 0.00\% & 0.00\% & 0.00\% & 0.00\% & 0.00\% & 52.15\% & 56.55\% & 61.57\% & 86.53\% \\
Entropy & 0.00\% & 0.00\% & 0.00\% & 0.00\% & 0.00\% & 0.15\% & 0.12\% & 0.13\% & 52.07\% & 56.43\% & 61.43\% & \underline{87.61\%} \\
Calibration & 0.00\% & 0.12\% & 0.10\% & 1.63\% & 0.51\% & 0.83\% & 0.78\% & 4.70\% & 52.36\% & 54.06\% & 59.78\% & 69.41\% \\
Attack-R & 0.00\% & 0.00\% & 0.00\% & 0.00\% & 0.00\% & 0.00\% & 0.00\% & 0.00\% & 52.15\% & \underline{56.06\%} & \underline{62.13\%} & 83.46\% \\
LiRA & 0.00\% & 0.13\% & 0.00\% & 5.73\% & 0.60\% & 0.42\% & 0.78\% & 12.05\% & 51.05\% & 52.96\% & 59.56\% & 79.41\% \\
Canary & 0.00\% & 0.21\% & 0.00\% & \underline{5.87\%} & 0.62\% & 0.51\% & 0.93\% & \underline{12.18\%} & 51.12\% & 53.18\% & 59.38\% & 79.80\% \\
RMIA & \underline{0.01\%} & \underline{0.23\%} & \underline{0.22\%} & 3.21\% & 0.82\% & \underline{1.05\%} & \underline{1.21\%} & 11.83\% & \underline{52.41\%} & 56.62\% & 62.06\% & 76.17\% \\
RAPID & 0.01\% & 0.20\% & 0.18\% & 3.05\% & \underline{0.84\%} & 0.93\% & 0.92\% & 10.84\% & 51.93\% & 55.17\% & 60.25\% & 77.89\% \\
\midrule
\textbf{\myoffline} & \textbf{0.19\%} & \textbf{0.27\%} & \textbf{0.26\%} & \textbf{9.75\%} & \textbf{1.23\%} & \textbf{1.25\%} & \textbf{2.45\%} & \textbf{19.70\%} & \textbf{52.64\%} & \textbf{56.70\%} & \textbf{62.94\%} & \textbf{89.07\%} \\
\cellcolor[gray]{0.9}\%Imp.  & \cellcolor[gray]{0.9}1800.00\% & \cellcolor[gray]{0.9}17.39\% & \cellcolor[gray]{0.9}18.19\% & \cellcolor[gray]{0.9}66.10\% & \cellcolor[gray]{0.9}48.80\% & \cellcolor[gray]{0.9}19.05\% & \cellcolor[gray]{0.9}102.48\% & \cellcolor[gray]{0.9}61.74\% & \cellcolor[gray]{0.9}0.44\% & \cellcolor[gray]{0.9}1.14\% & \cellcolor[gray]{0.9}1.30\% & \cellcolor[gray]{0.9}1.67\%  \\
\bottomrule
\end{tabular}}
\end{table*}
\begin{table*}[t]
\centering
\caption{Performance comparison of \textit{non-adaptive} attacks on MobileNetV2 trained on four image datasets.
}
\vspace{-1mm}
\label{tab:pmia_main_mobilenet}
\resizebox{0.99\textwidth}{!}{
\begin{tabular}{l|*{12}{c}}
\toprule
\multirow{2}{*}{\textbf{Method}} & \multicolumn{4}{c}{\textbf{TPR @ 0.001\% FPR}} & \multicolumn{4}{c}{\textbf{TPR @ 0.1\% FPR}} & \multicolumn{4}{c}{\textbf{Balanced Accuracy}} \\
\cmidrule(lr){2-5}\cmidrule(lr){6-9}\cmidrule(lr){10-13}
&  MNIST & FMNIST & C-10 & C-100 & MNIST & FMNIST & C-10 & C-100 & MNIST & FMNIST & C-10 & C-100 \\
\midrule
LOSS & 0.00\% & 0.00\% & 0.00\% & 0.00\% & 0.00\% & 0.00\% & 0.00\% & 0.00\% & 54.60\% & \underline{64.45\%} & 72.61\% & 88.55\% \\
Entropy & 0.00\% & 0.00\% & 0.00\% & 0.00\% & 0.00\% & 0.01\% & 0.00\% & 0.00\% & 54.61\% & 64.37\% & \underline{74.41\%} & \underline{89.92\%} \\
Calibration & 0.00\% & 0.16\% & 0.35\% & 1.53\% & 0.53\% & 0.92\% & 1.57\% & 4.81\% & 53.70\% & 57.21\% & 60.71\% & 69.85\% \\
Attack-R & 0.00\% & 0.00\% & 0.00\% & 0.00\% & 0.83\% & 0.00\% & 0.00\% & 0.00\% & 52.96\% & 59.94\% & 69.43\% & 84.20\% \\
LiRA & 0.00\% & 0.00\% & 0.47\% & 0.15\% & 0.32\% & 0.78\% & 3.71\% & 12.65\% & 50.22\% & 51.92\% & 62.63\% & 80.04\% \\
Canary & 0.00\% & 0.02\% & 0.53\% & 0.23\% & 0.35\% & 0.92\% & \underline{3.85\%} & \underline{13.73\%} & 50.89\% & 52.38\% & 63.61\% & 81.09\% \\
RMIA & \underline{0.13\%} & 0.25\% & \underline{0.80\%} & \underline{3.18\%} & \underline{0.86\%} & 1.24\% & 3.12\% & 13.12\% & 54.17\% & 61.62\% & 68.60\% & 78.57\% \\
RAPID & 0.07\% & \underline{0.38\%} & 0.52\% & 2.57\% & 0.75\% & \underline{2.01\%} & 2.96\% & 7.98\% & \underline{54.61\%} & 62.38\% & 62.41\% & 80.72\% \\
\midrule
\textbf{\myoffline} & \textbf{0.56\%} & \textbf{2.25\%} & \textbf{3.34\%} & \textbf{6.15\%} & \textbf{1.15\%} & \textbf{4.15\%} & \textbf{6.30\%} & \textbf{22.89\%} & \textbf{54.89\%} & \textbf{64.49\%} & \textbf{75.81\%} & \textbf{91.2\%} \\
\cellcolor[gray]{0.9}\%Imp.  & \cellcolor[gray]{0.9}330.77\% & \cellcolor[gray]{0.9}492.11\% & \cellcolor[gray]{0.9}317.50\% & \cellcolor[gray]{0.9}93.40\% & \cellcolor[gray]{0.9}33.72\% & \cellcolor[gray]{0.9}106.47\% & \cellcolor[gray]{0.9}63.64\% & \cellcolor[gray]{0.9}66.72\% & \cellcolor[gray]{0.9}0.51\% & \cellcolor[gray]{0.9}0.06\% & \cellcolor[gray]{0.9}1.88\% & \cellcolor[gray]{0.9}1.42\%  \\
\bottomrule
\end{tabular}}
\end{table*}

\subsection{Experiments of \myoffline on Other Models}
\label{appendix:pmia_other_models}

We present the results on the remaining three model architectures in the non-adaptive setting. Specifically, \Cref{tab:pmia_main_vgg},~\Cref{tab:pmia_main_densenet} and \Cref{tab:pmia_main_mobilenet} displays the results for VGG16, DenseNet121, and MobileNetV2.
The ROC curves of non-adaptive attacks are shown in~\Cref{fig:pmia_curve_resnet}.

\end{document}